\input{epsf}

\documentclass[journal]{IEEEtran}
\usepackage{epsf}
\usepackage{graphics}
\usepackage[cmex10]{amsmath}
\usepackage{amsthm}
\usepackage{amssymb}
\usepackage{epsfig,latexsym,amsmath,epsf,amssymb,amsfonts}
\usepackage{algorithm, algorithmic}
\usepackage{placeins}
\usepackage{cite}
\usepackage{comment} 
\usepackage{subfigure}
\usepackage{lipsum}

\usepackage[usenames,dvipsnames,svgnames,table]{xcolor}
\usepackage{dsfont}
\usepackage{multirow}
\usepackage{color}

\usepackage{pgf,tikz}
\usetikzlibrary{shapes,arrows,shadows}
\renewcommand{\theequation}{\arabic{equation}}
\newtheorem{Definition}{\hskip 0pt Definition}
\newtheorem{theorem}{\hskip 0pt Theorem}
\newtheorem{Proposition}{\hskip 0pt Proposition}

\theoremstyle{remark}

\begin{document}
%
\title{A Survey on Applications of Model-Free Strategy Learning in Cognitive Wireless Networks}

\author{
\IEEEauthorblockN{Wenbo Wang ~\IEEEmembership{Student Member,~IEEE,}
Andres Kwasinski ~\IEEEmembership{Senior Member,~IEEE,} 
Dusit Niyato ~\IEEEmembership{Senior Member,~IEEE,}  and
Zhu Han ~\IEEEmembership{Fellow,~IEEE}}

\thanks{Wenbo Wang and Andres Kwasinski are with the Department of Computer Engineering, Rochester Institute of Technology, Rochester, NY 
14623 USA (email: wxw4213@rit.edu, axkeec@rit.edu).}
\thanks{Dusit Niyato is with the School of Computer Engineering, Nanyang Technological University, Singapore 639798
(email: dniyato@ntu.edu.sg).}
\thanks{Zhu Han is with the Department of Electrical and Computer Engineering as well as the Department of Computer Science, University of Houston, TX 77004 USA
(email: zhan2@uh.edu).}}


\maketitle

\begin{abstract}
The framework of cognitive wireless radio is expected to endow the wireless devices with the cognition-intelligence ability, with which they can 
efficiently learn and respond to the dynamic wireless environment. In many practical scenarios, the complexity of network dynamics makes it difficult to determine the network
evolution model in advance. As a result, the wireless decision-making entities may face a black-box network control problem and the model-based network management
mechanisms will be no longer applicable. In contrast, 
model-free learning has been considered as an efficient tool for designing control mechanisms when the model of the system environment or the interaction between the 
decision-making entities is not available as a-priori knowledge. With model-free learning, the decision-making entities adapt their behaviors based on the reinforcement from 
their interaction with the environment and are able to (implicitly) build the understanding of the system through trial-and-error mechanisms. Such characteristics of model-free 
learning is highly in accordance with the requirement of cognition-based intelligence for devices in cognitive wireless networks. Recently, model-free learning has been 
considered as one key implementation approach to adaptive, self-organized network control in cognitive wireless networks. In this paper, we provide a comprehensive survey on the 
applications of the state-of-the-art model-free learning mechanisms in cognitive wireless networks. According to the system models that those applications are based on,  
a systematic overview of the learning algorithms in the domains of single-agent system, multi-agent systems and multi-player games is provided. Furthermore, the applications of 
model-free learning to various problems in cognitive wireless networks are discussed with the focus on how the learning 
mechanisms help to provide the solutions to these problems and improve the network performance over the existing model-based, non-adaptive methods. Finally, a broad spectrum of 
challenges and open issues is discussed to offer a guideline for the future research directions.
\end{abstract}

\IEEEpeerreviewmaketitle

\section{Introduction}
\label{sec_intro}
\index{Cognitive Radio Networks@\emph{Cognitive Radio Networks}}
\index{Cognitive Radio Networks@\emph{Cognitive Radio Networks}}%
\subsection{Cognitive Radio Networks}
\label{subsec_CRN}
The original concept of Cognitive Radio (CR) was first proposed a little over one decade ago \cite{2000mitola_cognitive_radio}. In a broad sense, CR is defined as a prototypical 
radio framework that adopts a radio-knowledge-representation language for the software-defined radio devices to autonomously learn about the dynamics of radio environments and
adapt to changes of application/protocol requirements. In recent years, Cognitive Radio Networks (CRNs) have been widely recognized from a high-level perspective as 
\emph{an intelligent wireless communication system}. A device in a CRN is expected to be aware of its surrounding environment and uses the methodology of understanding-by-building
to reconfigure the operational parameters in real-time, in order to achieve the optimal network performance \cite{SimonHaykin1391031,5639025}. In the framework of CRNs, the 
following abilities are typically emphasized:
\begin{itemize}
 \item radio-environment awareness by sensing (cognition) in a time-varying radio environment;
 \item autonomous, adaptive reconfigurability by learning (intelligence);
 \item cost-efficient and scalable network configuration. 
\end{itemize}

Many recent studies on CR technologies focus on radio-environment awareness in order to enhance spectrum efficiency. This leads to the concept of Dynamic Spectrum Access (DSA) 
networks \cite{haykin2012cognitive}, which are featured by a novel PHY-MAC architecture (namely, primary users vs. secondary users) for opportunistic spectrum access based on 
the detection of spectrum holes \cite{4481339}. It is worth noting that by emphasizing the network architecture of spectrum sharing between the licensed/primary networks and the 
unlicensed/secondary networks \cite{haykin2012cognitive}, ``DSA networks'' is frequently considered a terminology that is interchangeable with ``CR networks'' \cite{5639025}. The
rationale behind such a consideration is that a secondary network relies on spectrum cognition modules to make proper decisions for seamless spectrum access without 
interfering the primary transmissions. For this category of works in the literature, ``learning'' is mostly about the techniques of feature classification for primary signal 
identification \cite{6336689}. For an overview of the relevant techniques, the readers may refer to recent survey works in \cite{4796930, Akyildiz201140, 
Zeng:2010:RSS:1809168.1809170}. 

\begin{figure}[!t]
    \centering
    \includegraphics[width=0.499\textwidth]{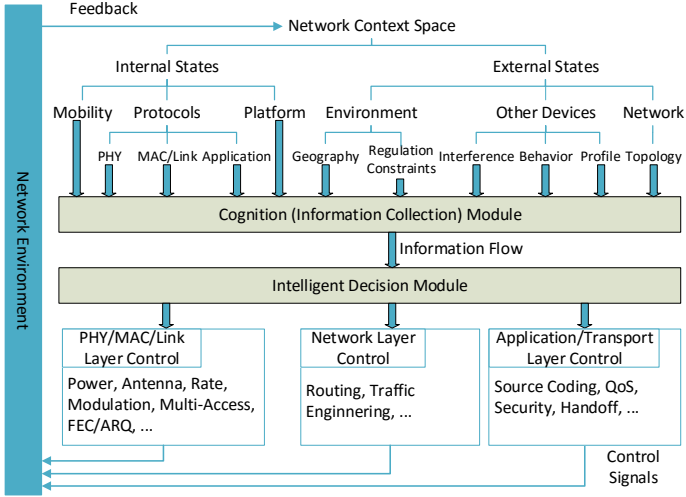}
    \caption{Relationship between the functionalities of cognition and intelligence in a cognitive wireless network.}
    \label{fig:Context_Graph}
\end{figure}

However, in order to achieve autonomous and cost-efficient network configuration, the functionalities of self-organized, adaptive reconfigurability also become fundamental for 
CRNs, since these functionalities shape the mechanisms of network control and transmission strategy acquisition. By emphasizing such an objective, the network management 
mechanism is required to
dynamically characterize the situation of the decision-making entities in the network and accordingly infer the proper transmission strategies. 
As the network management mechanisms in conventional wireless networks are acquiring more and more levels of such a cognition-intelligence ability, the border between a pure CRN 
(namely, a CRN in the sense of DSA networks) and a conventional wireless network is gradually diminishing \cite{6182559, Balamuralidhar_Prasad2008}. In recent years, 
the emerging networking technologies (e.g., CRNs and self-organized networks \cite{6507392,6157579}) emphasize more on autonomous, adaptive reconfigurability.
For these networks, the concept of ``intelligent network management'' 
based on ``cognition'' can be re-defined as providing the functionalities of autonomous transmission policy adaptation according to the radio-environment awareness capability of 
the CR devices in numerous dimensions across the networking protocol stacks \cite{Balamuralidhar_Prasad2008}. In Figure \ref{fig:Context_Graph}, we provide an overview of 
the perceivable network states for cognition and the cross-layer network functionalities for configuration in cognitive wireless networks. Interested readers are referred 
to recent surveys such as \cite{Cesana2011228, 4738464} for more details about the CR applications in different protocol layers.

Considering the distributed nature of wireless networks, a good CR-based framework of autonomous network configuration in time-varying environments needs to address the 
following questions: 
\begin{itemize}
 \item[1)] How to properly configure the transmission parameters with limited ability of network modeling or environment observation? 
 \item[2)] How to coordinate the distributed transmitting entities (e.g., end users and base stations) with limited resources for information exchange?
 \item[3)] How to guarantee the network convergence under the condition of interest conflicts among transmitting entities?
\end{itemize}
The need to address question 1) lies in the fact that in practical scenarios, the abilities of environment perception may be limited on different levels and/or 
for different devices. Therefore, the solution to the problems raised by question 1) requires that a decision making mechanism should be able to learn the transmission 
policies without explicitly knowing the accurate mathematical model of the networks beforehand. Meanwhile, questions 2) and 3) are raised by the basic requirement 
of a self-organized, distributed control system. Only by addressing questions 2) and 3) can the network configuration process be efficient in both information acquisition and 
policy computation. In summary, the key to answering questions 1), 2) and 3) lies in the prospect of enabling the devices in CRNs to distributively achieve their stable 
operation point under the condition of information incompleteness/locality.

\subsection{From Model-Based Network Management to Model-Free Strategy Learning}
\label{subsec_evo_2_CRN}
When the designer of (distributed) network-controlling mechanisms has complete and global information, the network control problem are frequently addressed in the 
model-base ways such as the optimization-decomposition-based formulation/solution \cite{4118456}. With a model-base design methodology, the network control 
algorithms are usually designed as a set of distributed computations by the network entities (also known as decision-making agents in the domain of control theory) to solve a 
global constrained optimization problem through decomposition. Under such a framework, since the model of the network dynamics is known in advance, there is no need for 
``learning'' anything about the network dynamics other than the time-varying network parameters. However,
in order to adopt such a design methodology, it is necessary to assume that the set of the network parameters (e.g., channel information and channel availability probabilities)
that determines the target network utilities is fully available or perfectly known to all the CR devices\footnote{More details about the common assumptions for
the model-based methods can be found in \cite{6407454}.}. 
If an equilibrium \cite{han2012game} of a multi-entity network is expected instead of the global optimality, the game theoretic approaches (e.g., for multiple access 
problems \cite{5692880} and network security problems \cite{6238282}) can also be adopted. Similar to the optimization-decomposition-based solutions, the game theoretic
approaches may still depend on a pre-known model of the network dynamics. In this case, the mathematical tools of optimization theory can also be used for the game theoretic 
approaches to achieve the goal of obtaining an equilibrium or locally optimal payoff, given that the strategies of the other network entities are accessible.

However, due to the practical limitation of information incompleteness/locality, directly applying the model-based solutions will face difficulties since a 
model of the network dynamics may even not be available in advance, or in most cases its details may be inaccurate or not instantaneously known to every device. Under the 
model-based framework, the attempts to conquer the obstacles of information incompleteness/inaccuracy are limited within a small scope by allowing more uncertainty/inaccuracy in 
the a-priori network model. Examples of these attempts include the introduction of robust control (e.g., variation inequality for spectrum sharing 
\cite{6802394}) and fuzzy logic (e.g., fuzzy logic for call admission control \cite{1390885}). Nevertheless, these techniques still lack the strength of fully addressing the 
three questions raised in Section \ref{subsec_CRN}.

\begin{figure}[!t]
    \centering
    \includegraphics[width=0.49\textwidth]{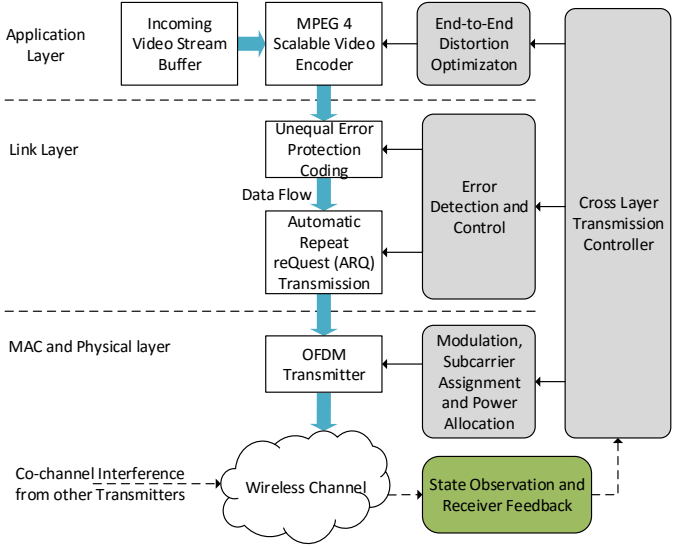}
    \caption{Scalable video transmission over a one-hop OFDM-based ad-hoc network.}
    \label{fig_example}
\end{figure}
The difficulty of obtaining an accurate model in advance for dynamic network control in practical scenarios can be illustrated by a multimedia transmission task over an 
one-hop OFDM-based ad-hoc network (Figure \ref{fig_example}). In the network illustrated by Figure \ref{fig_example}, the goal of the transmitter-receiver pairs is to achieve the 
minimized end-to-end distortion through joint power allocation, channel code adaptation and source coding control over dynamic channels. In the practical situation, the obstacle 
for obtaining an appropriate device behavior 
model first lies in the difficulty in constructing an accurate end-to-end rate-distortion model at the source codec level, since modeling the rate-distortion relationship 
for MPEG-4 Scalable Video Coding (SVC) mechanism is notoriously difficult \cite{5668507}. Moreover, one analytical model may only apply to a certain category of video sources \cite{5668507}. Meanwhile, 
the stochastic evolution of the channel condition makes it difficult to predict the transition of the states for the channel-coding/retransmission mechanism, which in return 
will result in uncertain error propagation at the video decoder of the receiver \cite{7248799}. Furthermore, when distributed power control and subcarrier allocation mechanism is
adopted, it is impractical 
for a transmitter-receiver pair to fully observe the transmission behaviors of the other pair of nodes, thus rendering the optimal power-channel allocation difficult with 
merely the local channel observation. As a result, without knowing the end-to-end distortion model, the channel evolution model and the information of peer-node behaviors, the 
wireless nodes are facing a black-box optimization problem with a limited level of coordination. In this situation, it will be difficult to apply the aforementioned model-based 
methods for the solution of video transmission control.

\begin{figure}[!t]
    \centering
    \includegraphics[width=0.40\textwidth]{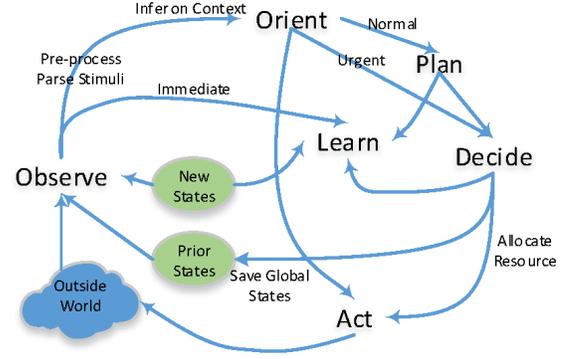}
    \caption{Cognition cycle of a single wireless device \cite{2000mitola_cognitive_radio}.}
    \label{fig_single_cog_cycle}
\end{figure}

In the scenarios of black-box network optimization/control with limited signaling, it is highly desirable that the network control mechanisms do not depend on the a-priori design 
of the devices' behavior model. As a result, the methods of controlling-by-learning without the need for the a-priori network model, namely, the model-free decision-making 
approaches \cite{Kaelbling:1996:RLS:1622737:1622748, 4445757}, are considered more proper, especially within the framework of CR technologies. In the context of adaptive control, 
controlling-by-learning in CRNs is usually described by the cognition-decision paradigm (Figure \ref{fig_single_cog_cycle}) \cite{2000mitola_cognitive_radio}. This paradigm 
describes the learning-based strategy-taking process of a single device from a high-level perspective and interprets it as a cognition cycle to present the information flow from 
environment cognition to the final network control decision. In the paradigm, the model-based decision making process is replaced by the observation-decision-action-learning loop. 
However, the paradigm itself does not provide any detail on how much information about the system model should be learned before a proper transmission strategy can be determined,
or in what way the information could be learned. 

Under the settings of not knowing a network model in advance, 
the strategy-learning process can be further divided into two categories according to the ways of using the model knowledge obtained from the learning process:
the ``model-dependent'' methods and the ``model-free'' methods \cite{Kaelbling:1996:RLS:1622737:1622748}. For model-dependent learning, an arbitrary division exists between the 
learning phase and the decision phase, and the goal of learning is to construct the network model first and then use it to derive the network control strategies. By contrast,
model-free learning directly learns the network controller without explicitly learning the network model in advance. Early research has pointed out that the model-dependent learning
methods are generally more computationally intensive, while model-free learning makes a trade-off of the time to reach controller convergence for reducing computational 
complexity \cite{Kaelbling:1996:RLS:1622737:1622748}. Although most of the existing research on strategy-learning methods in wireless networks focus on model-free learning due to
the limited computational resources in mobile devices, recent years have seen a tendency that the border between the two categories of strategy-learning methods keeps diminishing
\cite{4445757}. 

\subsection{A Brief Review of the Existing Survey Works on Learning in CRNs}
\label{subsec_summary_survey}
As indicated by our discussion in Sections \ref{subsec_CRN} and \ref{subsec_evo_2_CRN}, the problem domain of learning in cognitive wireless networks can be divided into 
two categories: the problems of wireless environment cognition (namely, spectrum sensing) \cite{4796930, Akyildiz201140, Zeng:2010:RSS:1809168.1809170} and the problems of
network management (namely, strategy learning). The solutions to the former problem sub-domain generally provide the information that works as the feed-in to the strategy 
managers of the latter  problem sub-domain. In the literature, the existing surveys on the network management problems are generally organized in accordance with the 
protocol layers of the OSI/ISO model. These problems include the DSA-based MAC protocol design in CRNs \cite{haykin2012cognitive, 6365154, 5639025, Akyildiz20062127}, routing 
protocol design in CRNs \cite{Cesana2011228, AlRawi2012} and cross-layer network control problems in CRNs such as self-organization \cite{6507392, 6827574} and network security 
problems \cite{6238282,liu2010cognitive}.
\begin{table*}[!t]
  \caption{Summary of existing survey works on CR networking problems and model-free learning methods}
  \centering
 \begin{tabular}{  p{75pt}|  p{95pt} | p{100pt} | p{120pt} }
  \hline
  Problem Domain of Cognitive Wireless Networks &  Sub-domain of CR Networking Problems & Category of Corresponding Machine-learning Methods & Sub-category of Learning Methods \\
  \hline
  Wireless environment cognition & Spectrum sensing \cite{4796930, Akyildiz201140, Zeng:2010:RSS:1809168.1809170} & Supervised learning (pattern classification) \cite{6336689, 
  5420026} & N/A \\
  \hline
  \multirow{4}{80pt} {Network management} & DSA-based MAC protocol design \cite{haykin2012cognitive, 6365154, 5639025, Akyildiz20062127} & \multirow{4}{80pt} 
  {Unsupervised learning (model-free learning) \cite{Kaelbling:1996:RLS:1622737:1622748, 4445757, fudenberg1998theory, Buşoniu2010, 1049606}} & Single-agent-based reinforcement learning
  \cite{Kaelbling:1996:RLS:1622737:1622748}
  \\
  \cline{2-2} \cline{4-4} & Spectrum-aware routing \cite{Cesana2011228, AlRawi2012} & & Multi-agent-based reinforcement learning \cite{4445757, Buşoniu2010}\\
  \cline{2-2} \cline{4-4} & Self-organization \cite{6507392, 6827574} & & Learning automata \cite{1049606} \\
  \cline{2-2} \cline{4-4} & Network security \cite{6238282,liu2010cognitive} & & Repeated-game-based learning \cite{fudenberg1998theory, lasaulce2011game, tembine2012distributed}\\
  \hline
\end{tabular}
  \label{table_confluence}
\end{table*}

With respect to different domains of networking problems, the pool of the potential machine-learning-based solutions can also be grouped into two major categories. For the 
problems of spectrum sensing, the survey on applications of signal-classification-oriented learning methods can be found in recent studies such as \cite{6336689, 5420026}. For 
the network management problems, the (model-free) strategy-learning-based solutions are generally identified as belonging to the category of unsupervised learning 
\cite{6336689}.
More specifically, the techniques of controlling-by-learning in CRNs are usually featured by the trial-and-error interactions with the dynamic wireless environment and thus 
also known as ``reinforcement learning'' (see our discussion in Section \ref{sec_bckgrd}).
In the past decade, researchers have paid a significant attention to the confluence of adaptive control, model-free learning and game theory \cite{4445757, 
fudenberg1998theory}. In the domain of CRNs, it is believed that such a trend will lead to a 
promising solution of the various network control/resource allocation problems (e.g., \cite{Akyildiz20062127, liu2010cognitive}). In return, the development of 
the recent network technologies, such as self-organized networks and CRNs, is increasingly demanding more efficient learning mechanisms to be implemented for an adaptive, 
self-organized solution. 

Most of the existing model-free learning methods for network control in CRNs find their origin in the domain of control theory. In the literature, important surveys on these 
model-free learning methods from the perspective of control/game theory include \cite{Kaelbling:1996:RLS:1622737:1622748, 4445757, Buşoniu2010, fudenberg1998theory, 1049606}. 
In the context of network control, existing survey works on the applications of strategy learning usually focus on a certain sub-category of these learning methods. In \cite{lasaulce2011game, 
tembine2012distributed}, comprehensive surveys on distributed learning mechanisms are provided based on the framework of repeated games (see our discussion in Section 
\ref{subsec_game}). In \cite{Yau2012253, 6336689}, the surveys on model-free learning in CRNs place the focus more directly on the Q-learning based methods (see our discussion in 
Section \ref{sec_single_agent}). Apart from the aforementioned works, other survey works on strategy learning in wireless networks usually focus on a specific sub-domain of 
applications such as wireless ad-hoc networks \cite{4496871} and sensor networks \cite{6805162}. 
To assist the readers in obtaining an overview of the development of model-free learning methods and their relationship with the network management problems in CRNs,  
we summarize the aforementioned survey works according to the domains they belong to in Table \ref{table_confluence}.

\subsection{Organization of the Paper}
This paper is devoted to providing a comprehensive survey on the current development of model-free learning in the context of the cognitive wireless networks. In order to 
highlight the difference in the existing level of information incompleteness/locality (from another perspective, the degree of information coupling) for different learning 
mechanisms, we organize the survey on the applications of learning in CRNs into
three major categories: (a) strategy learning based on the single-agent systems, (b) strategy learning based on the loosely coupled multi-agent systems and (c) strategy learning 
in the context of games. 
In Section \ref{sec_bckgrd}, the necessary background and the preliminary concepts of learning in the single-agent system, the distributed, multi-agent systems and games
are provided. In Section \ref{sec_app_crn}-\ref{sec_app_game}, the recent research on the applications of the three major categories of model-free learning mechanisms in CRNs is 
reviewed according to the different system models that the learning mechanisms are based on. In Section \ref{sec_open_issue}, some important open issues for the application of 
model-free learning in CRNs are outlined in order to provide the insight into the future research directions. Finally, we summarize and conclude the paper in Section 
\ref{conclusion}. In Table \ref{tab_abbreviations_1} and Table \ref{tab_abbreviations_2}, we provide an acronym glossary of the terms used in the paper.

\begin{table}[!t]
  \caption{Summary of acronyms for wireless networking terminologies}
  \centering
 \begin{tabular}{ p{130pt}| p{90pt}}
  \hline
  Terminologies & Abbreviations\\
  \hline
  Base station & BS (Section \ref{sec_app_crn}, \ref{sec_app_game})\\
  \hline
  Cognitive radio &  CR (Section \ref{sec_intro}, \ref{sec_app_crn}, \ref{sec_app_dist}, \ref{sec_app_game}) \\
  \hline
  Cognitive radio networks  & CRNs (Section \ref{sec_intro}, \ref{sec_app_crn}, \ref{sec_app_dist}, \ref{sec_app_game})\\
   \hline
  Dynamic channel assignment & DCA (Section \ref{sec_app_crn})\\
  \hline  
  Dynamic spectrum access & DSA (Section \ref{sec_intro}, \ref{sec_app_crn})\\
   \hline
   Key performance indicator & KPI (Section \ref{sec_open_issue})\\
   \hline
  Network operator & NO (Section \ref{sec_app_game})\\
  \hline
  Primary user & PU (Section \ref{sec_app_crn}, \ref{sec_app_dist}, \ref{sec_app_game})\\
   \hline
  Signal-to-interference-plus-noise-ratio & SINR (Section \ref{sec_app_dist}, \ref{sec_app_game})\\
  \hline
  Signal-to-noise-ratio & SNR (Section \ref{sec_app_crn}, \ref{sec_app_game})\\
   \hline 
  Service provider & SP (Section \ref{sec_app_game})\\
  \hline
  Secondary user & SU (Section \ref{sec_app_crn}, \ref{sec_app_dist}, \ref{sec_app_game})\\
  \hline
  Heterogeneous networks & HETNET (Section \ref{sec_app_dist})\\
  \hline
\end{tabular}
  \label{tab_abbreviations_1}
\end{table}

\begin{table}[!t]
  \caption{Summary of acronyms for model-free learning terminologies}
  \centering
 \begin{tabular}{ p{130pt}| p{90pt}}
  \hline
  Terminologies & Abbreviations\\
   \hline
  Actor-critic learning & AC-learning (Section \ref{sec_bckgrd}, \ref{sec_open_issue}) \\
  \hline
  Actor-critic learning automata & ACLA (Section \ref{sec_bckgrd}) \\
  \hline
  Correlated equilibrium & CE (Section \ref{sec_bckgrd}, \ref{sec_app_game}) \\
  \hline
  Correlated-Q learning & CE-Q learning (Section \ref{sec_app_game})\\
  \hline
  Constrained Markov decision process & CMDP (Section \ref{sec_app_crn})\\
  \hline
  COmbined fully DIstributed PAyoff and Strategy-Reinforcement Learning & CODIPAS-RL (Section \ref{sec_bckgrd}, \ref{sec_app_game})\\
  \hline
  Derivative-action gradient play & DAGP (Section \ref{sec_app_game})\\
   \hline
  Dynamic programming & DP (Section \ref{sec_bckgrd}) \\
  \hline
  Distributed reward and value function & DRV function \ref{sec_app_dist}\\
  \hline
  Distributed value function & DVF (Section \ref{sec_app_dist})\\
  \hline
  Experience-weighted attraction learning & EWAL (Section \ref{sec_open_issue})\\
   \hline
  Fictitious play & FP (Section \ref{sec_bckgrd}, \ref{sec_app_game}) \\
  \hline
  Greedy policy searching in the limit of infinite exploration & GLIE (Section \ref{sec_bckgrd})\\
  \hline
  Gradient play & GP (Section \ref{sec_app_crn})\\
 \hline 
  Learning automata & LA (Section \ref{sec_bckgrd}, \ref{sec_app_game})\\
  \hline
  Linear-reard-inaction algorithm & $L_{R-I}$ (Section \ref{sec_bckgrd}, \ref{sec_app_game})\\
  \hline
   Multi-agent Markov decision process & MAMDP (Section \ref{sec_bckgrd}, \ref{sec_app_crn})\\
  \hline
  Multi-agent system & MAS (Section \ref{sec_bckgrd}, \ref{sec_app_dist}, \ref{sec_app_game}, \ref{sec_open_issue})\\
    \hline 
  Markov decision process &  MDP (Section \ref{sec_bckgrd}, \ref{sec_app_crn})\\
  \hline
  Nash equilibrium &  NE (Section \ref{sec_bckgrd}, \ref{sec_app_game}, \ref{sec_open_issue}) \\
  \hline
  Observation-orient-decision-action loop & OODA loop (Section \ref{sec_intro})\\
  \hline
  Partially observable Markov decision process & POMDP (Section \ref{sec_app_crn}, \ref{sec_app_dist})\\
  \hline
  Single-agent Markov decision process & SAMDP (Section \ref{sec_bckgrd}, \ref{sec_app_crn})\\
  \hline
  State-action-reward-state-action & SARSA (Section \ref{sec_bckgrd}, \ref{sec_app_crn})\\
  \hline
  Single-agent system & SAS (Section \ref{sec_bckgrd}, \ref{sec_app_crn})\\
  \hline
  Stochastic games & SGs (Section \ref{sec_bckgrd}, \ref{sec_app_game}) \\
  \hline
  Smoothed/Stochastic fictitious play & SFP (Section \ref{sec_app_game})\\
   \hline
  Simultaneous perturbation stochastic approximation & SPSA (Section \ref{sec_app_game})\\
  \hline
  Temporal difference learning & TD-learning (Section \ref{sec_bckgrd}, \ref{sec_app_crn}, \ref{sec_open_issue})\\
  \hline
  Transfer learning & TL (Section \ref{sec_open_issue})\\
  \hline
\end{tabular}
  \label{tab_abbreviations_2}
\end{table}

\section{Background: Model-Free Learning in the Domains of Distributed Control and Game Theory}
\label{sec_bckgrd}
\begin{table*}[!t]
  \caption{Sequential decision-making models in a nutshell}
  \centering
 \begin{tabular}{ c|  p{45pt} | c |  p{40pt} | c | p{50pt}}
  \hline
  General Model &  Specific Model & Tuple-Based Model Description & Agent-Strategy Coupling & Objective  & Utility Measurement \\
  \hline
  \multirow{3}{50pt}{Multi-agent Markov Decision Process (MDP) / Stochastic Game (SG)} & Single-agent MDP & 
  ${\langle}{\mathcal{S}}, {\mathcal{A}}, r, \Pr(\mathbf{s}'|\mathbf{s},a) {\rangle}$ & N/A & Utility optimization & Accumulated utility\\
  \cline{2-6}
  & Multi-agent MDP & ${\langle}{\mathcal{N},\mathcal{S}}=\times\mathcal{S}_n, {\mathcal{A}=\times\mathcal{A}_n}, \{r_n\}_{n\in\mathcal{N}}, \Pr(\mathbf{s}'|\mathbf{s},
  \mathbf{a}) {\rangle}$& Allowed &  Utility optimization & Accumulated utility\\
  \cline{2-6}
  & Stochastic games & ${\langle}{\mathcal{N},\mathcal{S}}=\times\mathcal{S}_n, {\mathcal{A}=\times\mathcal{A}_n}, \{r_n\}_{n\in\mathcal{N}}, \Pr(\mathbf{s}'|\mathbf{s},
  \mathbf{a}) {\rangle}$& Always &  Reaching equilibria & Accumulated utility\\
  \cline{2-6}
  & Repeated games & ${\langle}\mathcal{N}, \mathcal{A}=\times\mathcal{A}_n, \{r_n\}_{n\in\mathcal{N}}{\rangle}$ & Always &  Reaching equilibria & Accumulated utility\\
  \cline{2-6} 
  & Static games & ${\langle}\mathcal{N}, \mathcal{A}=\times\mathcal{A}_n, \{r_n\}_{n\in\mathcal{N}}{\rangle}$   & Always & Reaching equilibria & Instantaneous utility\\
    \hline  
\end{tabular}
  \label{tab_nutshell}
\end{table*}
Although the applications of model-free learning in wireless networks only became more commonplace in the early 2000s, the fundamental development of the model-free learning 
theory can be traced back much earlier, to the 1980s \cite{1102893, Watkins:1989}. In this section, we provide a necessary introduction of the general-purpose learning 
methods that are developed in the domains of distributed control and game theory. To assist our discussion about learning techniques applied to cognitive wireless 
networks, we categorize the learning methods by the degree of coupling among the decision-making agents with respect to different system models. In what follows, we will 
briefly introduce the general-purpose learning 
algorithms that are built upon the decision-making models of single-agent systems, loosely coupled multi-agent systems and game-based multi-agent systems. Before 
proceeding to more details of the learning mechanisms, we first provide an overview of these decision-making models in Table \ref{tab_nutshell}. The 
notations used in this section are list in Table \ref{notation_sec_bckgrd}.

\begin{table}[t]
  \caption{Summary of the main notations in Section \ref{sec_bckgrd}}
  \centering
 \begin{tabular}{ p{40pt}| p{180pt}}
 \hline
 Symbol & Meaning\\
  \hline
  $t$ & Timing index\\
  \hline
  $a$ & A single action of the decision-making agent in a single-agent system\\
  \hline
  $a_{-n}$ & The joint action of the adversary agents for agent $n$ in a game \\
  \hline
  $\mathcal{A}_n$ &  A finite set of actions for agent $n$ in a multi-agent system\\
  \hline
  $s$  & A single environment state of the agent in a single-agent system\\
   \hline
  $\mathcal{S}_n$  & A finite set of environment states for agent $n$  in a multi-agent system\\
  \hline  
  $u_n(s_n,a_n)$ or $u_n$ & Instantaneous utility function of agent $n$  in a multi-agent system\\
  \hline
  $\Pr(\cdot)$ & State transition probability function \\
   \hline
  $\beta$ & The discount factor for a discounted-reward MDP\\
  \hline
  $\pi(s,a)$ or $\pi$ & The policy mapping function of an agent from a given state to an action\\
  \hline
  $\pi^*$ & An optimal or equilibrium policy\\
  \hline
  $\pi(s,a_{-n})$ or $\pi_{-n}$ & The joint policy of the adversary agents for agent $n$ in a game \\
  \hline 
  $V_{\beta}^{\pi}(s)$ & The state-value function of a discounted-reward MDP from the starting state $s$\\
  \hline
  $Q_{\beta}^{\pi}(s,a)$ & The state-action value function of a discounted-reward MDP from taking action $a$ at the starting state $s$\\
  \hline
  $h^{\pi}(s)$ & The state-value function of an average-reward MDP from the starting state $s$\\
  \hline
  $V^{\pi}(s)$ & The bias utility of an average-reward MDP from taking policy $\pi$ at starting state $s$\\
  \hline
  $\alpha_t$, $\theta_t$ & The learning rates\\
  \hline
  $\tilde{r}$ & The (normalized) value of environment response used by learning automata algorithms\\
  \hline
\end{tabular}
  \label{notation_sec_bckgrd}
\end{table}

\subsection{Single-Agent Strategy Learning}
\label{sec_single_agent}
\begin{figure}[!t]
  \centering
  \includegraphics[width=0.25\textwidth]{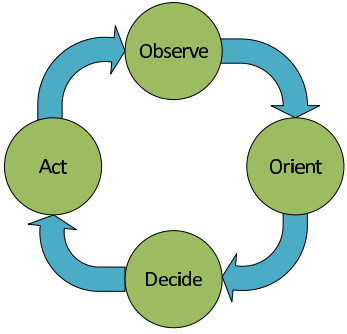}
  \caption{The OODA loop, often known as the cognition cycle \cite{2000mitola_cognitive_radio}.}
  \label{fig1}
\end{figure}

In the context of distributed control and robotics, single-agent learning has been considered as the most fundamental class of the strategy-learning methods. Single-agent 
learning generally assumes that the learning agent has full access to the state information that can be obtained about the system. Frequently, the terminologies ``reinforcement 
learning'' and ``model-free learning'' are (partially) used interchangeably to refer to the decision-making process of a single agent. The agent learns 
to improve its performance by merely observing the state changes in its operational environment and the utility feedback that it received after taking an action. In the recent 
surveys on reinforcement-learning theory and its applications \cite{Yau2012253,6336689}, such a decision-learning process is described by an abstract model, namely, the 
Observe-Orient-Decision-Action (OODA) loop \cite{haykin2012cognitive}. The OODA loop (Figure \ref{fig1}) can be considered as a generalized model of the cognition cycle in the 
context of cognitive wireless networks (Figure \ref{fig_single_cog_cycle}), and it provides a generic description of the information flow in the 
intelligent decision-making process. However, it is the task of the specific reinforcement-learning methods to define the rules of agent behaviors that guide the interaction with 
the to-be-explored environment. Since in most of the practical scenarios, a learning agent needs to deal with environment uncertainty, in the literature, a Markov Decision 
Process (MDP) \cite{ibe2008markov} becomes a prevalent tool for abstracting the model of the agent-environment interaction. Based on the MDP framework, various 
model-free learning methods such as Temporal Difference (TD) learning \cite{Watkins1992} and learning automata \cite{1049606} can be adopted to define the behavior rules of 
an agent.

The standard (single-agent) MDP model is used to describe a stochastic Single-Agent System (SAS). Mathematically, a single-agent MDP is defined as follows:
\begin{Definition}[Single-agent MDP \cite{4445757}]
 \label{def_MDP}
  A single-agent MDP is defined as a 4-tuple: ${\langle}{\mathcal{S}}, {\mathcal{A}}, u, \Pr(\mathbf{s}'|\mathbf{s},a) {\rangle}$, in which 
  \begin{itemize}
   \item $\mathcal{S}=\{s_1,\ldots,s_{|{\mathcal{S}}|}\}$ is a finite set of environment states,
   \item ${\mathcal{A}}=\{a_1,\ldots, a_{|{\mathcal{A}}|}\}$ is a finite set of agent's actions,
   \item $u: \mathcal{S}\!\times\!\mathcal{A}\!\times\!\mathcal{S}\!\rightarrow\!\mathbb{R}$ is the instantaneous utility function,
   \item $\Pr:\mathcal{S}\times\mathcal{A}\times\mathcal{S}\rightarrow [0,1]$ is the state transition probability function, which retains the Markovian property.
  \end{itemize}
\end{Definition}
\noindent In the MDPs, the underlying environment is a stationary stochastic process, and the consequences of the decisions can be probabilistic. The goal of a decision-learning 
agent is to find the proper stationary policy, $\pi\!=\!\Pr(a|s)$ that probabilistically maps state $s$ to action $a$ so that the accumulated long-term utility 
of the agent is optimized. With respect to different applications, the objectives of the MDPs may appear in different forms. In this survey, we will 
mainly consider two types of the infinite-horizon objectives \cite{Kaelbling:1996:RLS:1622737:1622748} as follows:
\begin{itemize}
 \item the discounted-reward MDP with the discount factor $\beta\!\in\![0,1]$:
 \begin{equation}
  \label{eq_dis_mdp}
  V^\pi_{\beta}(s)=E_{\pi}\left(\sum_{t=0}^{\infty}\beta^tu_t(s_t,a)\right),
 \end{equation}
 \item the average-reward MDP:
  \begin{equation}
  \label{eq_avg_mdp}
    h^\pi(s)=\lim_{T\rightarrow\infty}\frac{1}{T}E_{\pi}\left(\sum_{t=0}^{T-1}u_t(s_t,a)\right).
  \end{equation}
\end{itemize}
Both types of MDPs can be represented in the form of the Bellman optimality equation. For the discounted-reward MDP, the Bellman equation can be 
represented either by the state-value function starting from state $s$ under policy $\pi$:
\begin{equation}
 \label{eq_value_fn}
 V^{\pi}_{\beta}({s})=E_{\pi}(u(s,a))+\sum_{s'\in\mathcal{S}}\Pr(s'|s,\pi)V^{\pi}_{\beta}(s'),
\end{equation}
or by the state-action value function (Q-function) that starts from taking action $a$ at state $s$ and follows policy $\pi$ thereafter:
\begin{equation}
 \label{eq_action_value_fn}
 Q^{\pi}_{\beta}({s},a)=u(s,a)+\sum_{s'\in\mathcal{S}}\Pr(s'|s,a)V^{\pi}_{\beta}(s').
\end{equation}

In order to express the average-reward MDP in the form of the Bellman equation, the average adjusted sum of utility (i.e., bias) following policy $\pi$ is introduced as 
follows:
\begin{equation}
 \label{eq_bias}
 V^{\pi}(s)=\lim_{T\rightarrow\infty}E_{\pi}\left(\sum_{t=0}^{T-1}\left(u_t(s_t,a)-h^{\pi}(s)\right)\right),
\end{equation}
with which the average-reward MDP can be expressed by the state-value function\footnote{Due to the space limit, the conditions for the existence of 
a value function in the form of (\ref{eq_bias_value_fn}) is not presented here. The readers are referred to \cite{Kluwer1996} for the details.}:
\begin{equation}
 \label{eq_bias_value_fn}
 V^{\pi}(s)+h^{\pi}(s)=E_{\pi}(u(s,a))+\sum_{s'\in\mathcal{S}}\Pr(s'|s,\pi)V^{\pi}(s').
\end{equation}

With a variety of on-line learning methods that estimate the optimal Q-value or the bias value, a broad spectrum of value-iteration-based learning algorithms have been 
proposed \cite{4445757, Kluwer1996}. Among them, the most widely used model-free learning algorithm is Q-learning \cite{Watkins1992}, which estimates the state-action 
value in (\ref{eq_action_value_fn}) of a discounted MDP based on the time difference of the estimated values for the state-action value function:
\begin{equation}
 \label{eq_q_learning}
 \begin{array}{ll}
 Q_{t+1}(s_t,a_t)\!\leftarrow\!Q_{t}(s_t,a_t)\!+\!\alpha_{t}\bigg(u_t(s_t,a_t)\\
 \qquad\qquad+\beta\max\limits_{a'}Q_t(s_{t+1},a')-Q_{t}(s_t,a_t)\bigg),
 \end{array}
\end{equation}
where $\alpha_t\in(0,1]$ is the learning rate specifying the step that the current state-action value is adjusted toward the TD sample 
$u(s_t,a_t)\!+\!\beta\max_{a'}Q_k(s_{t+1},a')$. Q-learning in (\ref{eq_q_learning}) has been proved to be able to converge to the true optimal value of the state-action 
value function with a stationary deterministic policy, given that $\sum_{t=0}^{\infty}\alpha_t=\infty$, $\sum_{t=0}^{\infty}\alpha^2_t<\infty$ and all actions in all 
states are visited with a non-zero probability \cite{Watkins1992}. The model-free property of Q-learning is reflected in the iterative approximation procedure 
for the Q-values, which does not require knowing the transition map $\Pr(s'|s,a)$ of the MDP in advance.

The counterpart to Q-learning in the average-reward MDP is known as R-learning \cite{Kluwer1996}. In addition to learning the state-action value of the bias expressed in 
(\ref{eq_bias}), R-learning also needs to learn the estimate of the average reward $h^{\pi}$. Therefore, R-learning 
is performed by a two-time scale learning process:
\begin{equation}
 \label{eq_r_learning}
  \begin{array}{ll}
  R_{t\!+\!1}(s_t,a_t)\!\leftarrow\!R_t(s_t,a_t)\!+\!\alpha_t\big(u(s_t,a_t)\!+\!\max\limits_{a'}R_t(s_{t\!+\!1},a')\!\\
  \qquad\qquad\qquad-\!h_t\!-\!R(s_t,a_t)\big),
  \end{array}
\end{equation}
\begin{equation}
\label{eq_r_learning2}
  \begin{array}{ll}
 h_{t+1}\!\leftarrow\!h_{t}+\theta_t\big(u(s_t,a_t)+\max\limits_{a'}R_t(s_{t+1},a')-h_t\\
 \qquad\quad-\max\limits_{a'}R_t(s_{t},a')\big).
   \end{array}
\end{equation}

In contrast to the value-iteration-based learning algorithms given in (\ref{eq_q_learning}), (\ref{eq_r_learning}) and (\ref{eq_r_learning2}), the decision-learning methods based on the 
Learning Automata (LA) allow an agent to directly learn the stationary randomized policy. Instead of updating the action according to the myopic optimal Q-value in 
discounted-reward MDP and bias-value in average-reward MDP, the LA directly 
updates the probabilities of actions based on the utility feedback \cite{1049606}. Let the action probability vector at time instance $t$ be $\pmb{\pi}(t)=(\pi_1(t),\ldots, 
\pi_{|{\mathcal{A}}|}(t))$, where $|{\mathcal{A}}|$ is the size of the action set. Then an LA-based algorithm should be able to achieve the following goal \cite{1049606}:
\begin{equation}
 \label{eq_la_goal}
 \pmb{\pi}^*=\max\limits_{\pi(t)}E{\left[\tilde{r}(t)|\pmb\pi(t),s(t)\right]},
\end{equation}
where $\tilde{r}$ is the value of environment response, and is usually generated based on the instantaneous reward $u_t$ as a normalized value (i.e., $\tilde{r}\in
\{0,1\}$). The general updating rule for LA can be expressed as follows \cite{narendra1989learning}:
\begin{equation}
\label{eq_general_la}
  \left\{
  \begin{array}{ll}
  \pi_i(t\!+\!1)\! =\! \pi_i(t)\!-\!(1\!-\!\tilde{r}(t))f_i(\pi_i(t))\!+\!\tilde{r}(t)g_i(\pi_i(t)), \\
  \qquad\qquad\qquad\qquad\qquad\qquad\qquad\qquad\qquad\forall a(t)\ne a_i,\\
  \pi_i(t\!+\!1)\! = \!\pi_i(t)\!+\!(1\!-\!\tilde{r}(t))\sum_{j\ne i}f_i(\pi_i(t))-\\
  \qquad\qquad\qquad\qquad\tilde{r}(t)\sum_{j\ne i}g_i(\pi_i(t)),\qquad a(t)=a_i,
   \end{array}
     \right.
\end{equation}
where $f$ and $g$ are the penalty and reward functions, respectively. Specifically, different forms of $f$ and $g$ lead to different learning schemes. 
Among them, it has been proved that the linear-reward-inaction (i.e., $L_{R-I}$) algorithm is guaranteed to achieve the $\epsilon$-optimal policies \cite{1104342}. In 
\cite{Kluwer1996}, the automaton-updating procedure based on $L_{R-I}$ is adopted to learn the optimal policy in the ergodic MDPs with average-reward objectives\footnote{For the 
details of $L_{R-I}$, please refer to Section \ref{subsubsec_game_LA}.}. In other 
works such as \cite{ImthiasAhamed20029}, the optimal policy of the discounted-reward MDP is learned by adopting the $L_{R-I}$ algorithm for policy updating and the 
standard Q-learning algorithm in (\ref{eq_q_learning}) for Q-value estimation at the same time.

\begin{figure}[!t]
  \centering
  \includegraphics[width=0.23\textwidth]{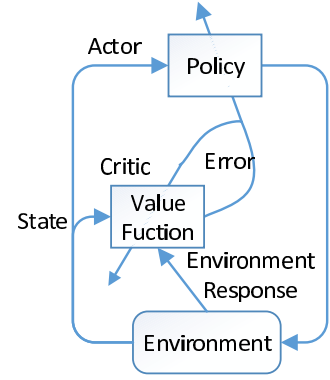}
  \caption{Schematic view of the generalized AC algorithm.}
  \label{fig_general_ac}
\end{figure}

Although the two groups of learning mechanisms, namely, value-iteration-based learning (e.g., TD-based learning such as Q-learning and R-learning) and LA-based learning appear 
distinct from each other, both of them can be considered as special cases in the framework of Actor-Critic (AC) learning \cite{6392457}. In the context of AC learning, 
the concepts of value function and policy are also known as ``critic'' and ``actor'', respectively. Since Q-learning and R-learning only learn a state-action value function and 
there is no explicit function for the policy, the two learning algorithms are also known as the critic-only algorithms. On the contrary, without using any form of a stored value function, LA can be considered 
an actor-only algorithm. Extending from these two special cases, a generalized AC-based mechanism keeps track of both the state-value function and the policy evolution at the 
same time. In this sense, a generalized AC-based mechanism is also known as combined payoff and strategy learning \cite{tembine2012distributed}.
Specifically, if the state-action value of the MDP is learned following the TD-based methods and in the meanwhile the learning agent's policy is updated following 
the LA-based methods, the AC-learning mechanism is also known as Actor-Critic LA (ACLA) \cite{4509588}. A typical rule for jointly updating the estimate of 
the state-value and policy in ACLA can be found in \cite{4509588}. Here it is worth noting that for both critic and actor updating, the learning mechanisms are not limited to the
aforementioned two categories of algorithms. For example, an on-policy learning algorithm, i.e., State-Action-Reward-State-Action (SARSA)\footnote{About the difference between 
Q-learning and SARSA, the readers are referred to \cite{barto1998reinforcement} for more details.}, can be used to replace the Q-learning-based critic-updating mechanism, and 
instead of the LA-like actor-updating mechanism, policy gradient is widely used for actor updating \cite{6392457}. A schematic overview of the generalized AC algorithm is given 
in Figure \ref{fig_general_ac}.

\subsection{Strategy Learning in the Loosely Coupled Multi-Agent System}
\label{subsec_mutli_agent}
A stochastic Multi-Agent System (MAS) can be defined by extending the 4-tuple Single-Agent MDP (SAMDP) (Definition \ref{def_MDP}) into a 5-tuple Multi-Agent MDP (MAMDP): 
${\langle}{\mathcal{N},\mathcal{S}}, 
{\mathcal{A}}, \{u_n\}_{n\in\mathcal{N}}, \Pr(\mathbf{s}'|\mathbf{s},\mathbf{a}) {\rangle}$, in which $\mathcal{N}$ is the set of the decision-making agents, $\mathcal{S}\!=\!
\times\mathcal{S}_n$ is the Cartesian product of the local state spaces of all the agents and $\mathcal{A}\!=\!\times\mathcal{A}_n$ is the Cartesian product of the local action 
spaces of all the agents. When 
considering the learning mechanism in an MAS, it is natural to simply adopt the standard SAS-learning algorithms by assuming that each agent is an independent learner with the 
local utility function $u_n(s_n,a_n)$. In doing so, the activities of the other agents are treated as part of a stationary environment and the learning agents update their 
policy without considering their interactions with the other agents. This approach enjoys popularity especially within the studies in the cooperative decision-making domain 
\cite{Tan:1997:MRL:284860.284934, Panait:2005:CML:1090749.1090753}. Its typical applications can be found in modeling the hunter-prey systems \cite{979961} and team coordination 
\cite{sen1998individual}, just to mention a few. However, it is important to note that multi-agent learning based on SAS learning requires 
the joint learning process to be decomposed into local ones. Thus, individual-agent behaviors are relatively disjoint, and the agents are able to ignore the information raised by 
the interactions with each other. This is also the reason for us to call it a ``loosely coupled multi-agent system''. Otherwise, with concurrent learning, all the 
individual agents need to adapt their policies in the dynamic context of the other learners, in which case the basic assumption of stationary environment for the single-agent 
scenarios will no longer hold. 

Although convergence of SAS-based learning is not guaranteed in most of the practical MAS scenarios, attempts of generalizing the convergence condition for the SAS-based 
learning mechanism can still be found in the literature. By limiting the application scenarios to fully-cooperative MAMDPs (i.e., common-payoff MAMDPs), the convergence property 
of SAS-based learning with Greedy policy searching in the Limit of Infinite Exploration (GLIE) for MAMDPs is discussed in \cite{Claus98thedynamics}:
\begin{Proposition}
 \label{thm_IL_convergence}
 For the multi-agent Q-learning schemes obeying the individual updating rule in (\ref{eq_q_learning}) in a cooperative MAS system, assume that the following conditions are satisfied:
 \begin{itemize}
  \item the learning rate $\alpha_t$ decreases over time such that $\sum_t\alpha_t=\infty$ and $\sum_t\alpha^2_t<\infty$,
  \item each agent samples each of its actions infinitely often,
  \item the probability of agent $i$ choosing action $a\in\mathcal{A}_i$ is nonzero,
  \item the probability of taking a non-optimal action decreases to 0 when $t\rightarrow\infty$ during the exploration stage,
 \end{itemize}
 let $\pi^*_{i}(t)$ be a random variable denoting the probability of action-taking in a (deterministic) equilibrium strategy profile being played at time $t$. Then for SAS-based  
 learning, for any $\xi, \epsilon>0$, there exists $T(\xi, \epsilon)$ such that
 \begin{equation}
  \label{eq_thm_il_conv}
  \Pr(|\pi^*_{i}(t)-1|<\epsilon)>1-\xi, \forall t>T(\xi, \epsilon).
 \end{equation}
\end{Proposition}

Although lacking a formal mathematical proof, Proposition \ref{thm_IL_convergence} has been widely accepted in related studies \cite{Buşoniu2010, KER:8495877}. A more general 
convergence condition for SAS-based learning in MAS scenarios is given by \cite{Littman200155}:
\begin{Proposition}
 \label{thm_IL_convergence_general}
  In an MAS environment, an agent following the updating rule in (\ref{eq_q_learning}) will converge to the optimal response Q-function with probability 1 as long as all the 
  other agents converge in behaviors with probability 1. If the agent follows a GLIE policy and its best response policy is unique, it will also converge in behavior with 
  probability 1.
\end{Proposition}

Propositions \ref{thm_IL_convergence} and \ref{thm_IL_convergence_general} provide theoretical support for the convergence property of a number of SAS-based learning 
algorithms that can be considered a variation of (\ref{eq_q_learning}) (e.g., distributed Q-learning in cooperative MAMDPs \cite{Lauer00analgorithm} and policy hill-climbing 
in two-agent MAMDPs \cite{Bowling2002215}). Again, it is worth pointing out that for most MAS scenarios (e.g., general-sum stochastic games) convergence of SAS-based learning is 
not guaranteed. Furthermore, even when convergence can be reached, it usually takes a significant amount of time for merely determining switching between one pair of actions. 
As a results, most of the practical SAS-based learning mechanisms are limited in the special scenarios such as the fully-cooperative MAS or two-agent MAS. 
In the framework of the independent learning algorithm using standard Q-learning \cite{Claus98thedynamics}, other SAS-based learning algorithms for MAS usually try to eliminate 
the uncertainty caused by the actions of the other agents while still retaining the distributivity of the decision-making process. One typical example can 
be found in \cite{Lauer00analgorithm}, which projects the global Q-table of a deterministic MAMDP (namely, the state transition is deterministic in the MDP) using centralized 
Q-learning with joint action $\mathbf{a}=(a_1,\ldots, a_n)$, 
$Q(s, \mathbf{a})$, to the local Q-table of agent $i$ with only local action information $a_i$, $Q(s, a_i)$. Following the standard Q-learning rule, the projection-based 
independent learning adopts an optimistic assumption that all the other agents will act optimally. However, the learning result of such a distributed algorithm is greedy with 
respect to the centralized Q-table with the joint action. Additionally, its convergence when extended to the scenarios of stochastic MAMDPs is 
not guaranteed since it cannot discern the influence of the behaviors of the other agents from that of the state dynamics. It is important to note that without explicit 
coordination, which is at the cost of losing the distributiveness of the decision-making process, all the independent-learning-based algorithms will suffer for the same 
reason as in the tightly coupled, MAS-based scenarios.

Despite all the limitations of independent learning, one important benefit of adopting the disjoint learning processes in the MAS is that it creates the opportunities of 
experience sharing among individual agents. In 
\cite{979961, Price:2003:ARL:1622434.1622450}, the ``implicit imitation'' mechanism by the observer agents is proposed to incorporate the experience of the expert agents 
in the MAS. Under the framework of distributed, independent MDPs, 
it is frequently assumed that the learning agents are analogous to each other in terms of state space, state transition and action set 
\cite{Price:2003:ARL:1622434.1622450}. Then experience transferring can be implemented by modifying the estimated state-action value of the observer agent based on the 
expertise evaluation of the mentor agents and the weighted combination of their respective Q-values \cite{979961}. When experience transferring is considered beyond the 
framework of model-free learning and the model-based policy-learning mechanism is adopted, 
the observer agent can also implement the experience learning by maintaining the estimation of the mentor's transition map from observation, and incorporating the estimation into
its own value-iteration process \cite{Price:2003:ARL:1622434.1622450}.

\subsection{Multi-Agent Strategy Learning in the Context of Games}
\label{subsec_game}
In most of the practical scenarios, the dynamics of the multi-agent MDP (e.g., the transition probabilities and the local payoff) is determined by the joint policy of all 
the agents. To facilitate distributed policy learning, the multi-agent MDP is usually viewed as a Stochastic Game (SG). Mathematically, an SG shares exactly the same 
5-tuple structure as an MAMDP, ${\langle}{\mathcal{N},\mathcal{S}}, {\mathcal{A}}, \{u_n\}_{n\in\mathcal{N}}, \Pr(\mathbf{s}'|\mathbf{s},\mathbf{a}) {\rangle}$. 
However, the goal of each agent in the SG is to maximize its individual payoff \cite{han2012game}. Based on the definition of SGs, a repeated game can be obtained 
as a 3-tuple, ${\langle}\mathcal{N}, {\mathcal{A}}\!=\!\times\mathcal{A}_n, \{u_n\}_{n\in\mathcal{N}} {\rangle}$, by fixing the environment state as invariant while maintaining
the objective of each player as maximizing its individual discounted/average payoff over the infinite time horizon.
In the repeated game, the system dynamics is reduced to only the mapping between the action and the payoff: $u_n: \mathcal{A}\rightarrow \mathbb{R}$. Further, when the repeated 
game is played only once, it is reduced to a static game. In return, any single shot of an SG or a repeated game is a static game and is known as a single stage
or one-shot game of the original game \cite{mackenzie2006game}.

One important reason for adopting the game theoretic models lies in the requirements that decisions are to be made in a distributed manner with the limited ability of both
information acquisition and action coordination. This may be either due to the overwhelming dimension of the state-action space as the number of agents grows, or due to the 
overhead for information exchange among agents. In the game-based decision-making model, the individual-rationality property of the agents leads to the concept of the best 
response. In an SG, the best response of agent $n$ is defined as the policy $\{\pi_n=\Pr(\mathbf{s},a_n):\mathbf{s}\in\mathcal{S}\}$ such that the long-term payoff under 
local policy $\pi_n$ is not worse than that under any other local policies: $V_n(\pi_n, \pi_{-n})\ge V_n(\pi'_n, \pi_{-n})$, given the joint adversary policy $\pi_{-n}$. Here, 
$\pi_{-n}$ is the joint strategy of the adversary agents except agent $n$ and 
$V_n$ can be either the discounted long-term payoff or the average long-term payoff. If $\forall n\in\mathcal{N}$, the policy is a best response to the joint strategy of the 
other agents, we say that the policy profile $(\pi_1,\ldots,\pi_{|\mathcal{N}|})$ is a Nash Equilibrium (NE) \cite{han2012game}. In the context of games, the goal of policy 
learning now becomes finding the policy updating rules for reaching a specific equilibrium. Apart from the most commonly used solution concept of NEs, a policy learning mechanism 
may resort to other types of equilibria for the convenience such as ensuring convergence or improving performance. In order to facilitate our discussion on different learning 
algorithms, we provide the formal definition of several equilibria in discounted-reward SG $G={\langle}{\mathcal{N},\mathcal{S}}, {\mathcal{A}}, \{u_n\}_{n\in\mathcal{N}}, 
\Pr(\mathbf{s}'|\mathbf{s},\mathbf{a}) {\rangle}$ as follows:
\begin{Definition}[Nash Equilibrium (NE)]
 \label{def_NE}
 In a game $G$, an NE point is a tuple of strategies $(\pi^*_1,\ldots,\pi^*_{|\mathcal{N}|})$ such that $\forall \mathbf{s}\in\mathcal{S}$, $\forall n\in\mathcal{N}$ and $\forall 
 \pi_{n}\in\Pi_n$,
 \begin{equation}
  V_{\beta,n}(\mathbf{s}, \pi^*_1,\!\ldots\!,\!\pi^*_n\!\ldots\!,\!\pi^*_{|\mathcal{N}|})\!\ge\!V_{\beta,n}(\mathbf{s}, \pi^*_1,\!\ldots\!,\!\pi_n\!\ldots\!,\!
  \pi^*_{|\mathcal{N}|}),\nonumber
 \end{equation}
  in which $V_{\beta,n}(\mathbf{s}, \pi^*_1,\ldots,\pi^*_{|\mathcal{N}|})$ is given by (\ref{eq_value_fn}) with a slight abuse of notation.
\end{Definition}
\begin{Definition}[Correlated Equilibrium (CE)]
 \label{def_CE}
 In a game $G$, a CE point is a joint strategy $\pi^*=(\pi^*_n,\pi^*_{-n})$ such that $\forall n\in\mathcal{N}$, $\forall \mathbf{s}\in\mathcal{S}$ and $\forall a_n,
 a'_n\in\mathcal{A}_n$,
  \begin{eqnarray}
  \begin{array}{ll}
  \displaystyle\sum\limits_{a_{-n}\in\mathcal{A}_{-n}}\pi^*(\mathbf{s},a_n, a_{-n})Q^{\pi^*}_{\beta,i}(\mathbf{s},a_n,a_{-n})\ge\\
  \displaystyle\sum\limits_{(a_n,a_{-n})\in\mathcal{A}}\pi^*(\mathbf{s},a_n, a_{-n})Q^{\pi^*}_{\beta,i}(\mathbf{s},a'_n,a_{-n}),\nonumber
  \end{array}
 \end{eqnarray}
 in which $Q^{\pi^*}_{\beta,i}(\mathbf{s},a_n,a_{-n})$ is given by (\ref{eq_action_value_fn}) with a slight abuse of notation and $\pi^*(\mathbf{s},a_n, a_{-n})=
 \pi^*(\mathbf{s}, a_{-n}|a_n)\pi^*(\mathbf{s},a_n)$. 
\end{Definition}
\begin{Definition}[$\epsilon$-Equilibrium]
 \label{def_epsilonE}
 Let $\epsilon\!>\!0$, the profile $\pi^*\!=\!(\pi^*_n,\pi^*_{-n})$ is an $\epsilon$-equilibrium of game $G$ if by following $\pi^*$ no player can improve its payoff by more than 
 $\epsilon$ at any stage. 
 
 Specifically, given the condition of the NE (Definition \ref{def_NE}), $\pi^*\!=\!(\pi^*_n,\pi^*_{-n})$ is an $\epsilon$-NE if $\forall \mathbf{s}\in\mathcal{S}$, $\forall n\in
 \mathcal{N}$ and $\forall\pi_{n}\in\Pi_n$,
 \begin{equation}
  V_{\beta,n}(\mathbf{s},\pi^*_n,\pi^*_{-n})\!\ge\!V_{\beta,n}(\mathbf{s}, \!\pi_n,\pi^*_{-n})-\epsilon.\nonumber
 \end{equation}
 Given the condition of CE (Definition \ref{def_CE}),  $\pi^*\!=\!(\pi^*_n,\pi^*_{-n})$ is an $\epsilon$-CE if $\forall \mathbf{s}\in\mathcal{S}$, $\forall n\in
 \mathcal{N}$ and $\forall a_n, a'_n\in\mathcal{A}_n$,
 \begin{eqnarray}
  \begin{array}{ll}
  \displaystyle\sum\limits_{a_{-n}\in\mathcal{A}_{-n}}\pi^*(\mathbf{s},a_n, a_{-n})Q^{\pi^*}_{\beta,i}(\mathbf{s},a_n,a_{-n})\ge\\
  \displaystyle\sum\limits_{(a_n,a_{-n})\in\mathcal{A}}\pi^*(\mathbf{s},a_n, a_{-n})Q^{\pi^*}_{\beta,i}(\mathbf{s},a'_n,a_{-n})-\epsilon. \nonumber
  \end{array}
 \end{eqnarray}
\end{Definition}

Based on Definitions \ref{def_NE}-\ref{def_epsilonE}, the conditions of equilibria for repeated/static games can be obtained in a similar way. From the perspective of strategy
derivation, a CE can be considered a generalized form of an NE since it does not require the individual player's strategy to be independent with each other. Although the adoption
of a CE is recognized as being able to provide a better performance of an NE, such a performance improvement is usually at the cost of introducing an arbitrator or coordinator 
into the game \cite{han2012game}. From the perspective of convergence reaching, an $\epsilon$-equilibrium can be considered a form of both NE and CE with relaxed condition. For 
learning algorithm design in repeated games, the introduction of $\epsilon$-equilibrium helps develop the learning mechanisms that guarantee the convergence to near-equilibrium 
with a limit-inferior bound. However, it is worth noting that for a general SG, the existence of a stationary $\epsilon$-equilibrium is not guaranteed beyond the case of 
two-player SGs \cite{Vieille20021833}.

According to the Folk theorem \cite{lasaulce2011game}, for every infinite-horizon, $n$-player, discounted repeated/stochastic game with a finite number of actions, 
the existence of a stationary policy $\pmb\pi^*$ as a subgame-perfect NE \cite{han2012game} is guaranteed. By proving the existence of a subgame-perfect NE, the Folk theorem 
implies that when compared with the static one-shot game, policy learning may be able to obtain a better payoff with the new NE in the repeated games. Such a benefit
is also considered a major motivation for the engineers to adopt the game-based learning algorithms in the domain of distributed decision-making. However, 
the implementations of the learning algorithms heavily rely on the game structures and the forms of the equilibria, and may differ significantly. Within 
the past two decades, numerous methods have been proposed for strategy learning in games. In order to facilitate our survey on their applications in cognitive wireless 
networks, we categorize the model-free learning algorithms along the following dimensions\footnote{All the game-based learning methods to be discussed in the following sections 
originate from these algorithms, and in Section \ref{sec_app_crn} more details will be provided for each of them.}:
\begin{itemize}
 \item[1)] Value iteration vs. policy iteration: in SGs, most of the learning algorithms based on the state-action value estimation fall into the category of 
 value-iteration based algorithms. These algorithms include minimax Q-learning \cite{Littman94markovgames}, NSCP-learning \cite{Weinberg:2004:BML:1018410.1018798}
 Nash Q-learning \cite{Hu:2003:NQG:945365.964288},  Nash R-learning \cite{li2007reinforcement} and CE-Q learning \cite{Greenwald03correlatedq-learning}.
 In contrary to value-iteration-based learning, 
 the policy-iteration-based learning algorithms directly update the action-probability vectors of each agent, using either the observation of the adversary agents' 
 action pattern or the payoff received from interaction with the environment. These algorithms include standard Fictitious Play (FP) \cite{fudenberg1998theory}, 
 asynchronous best response \cite{topkis1979equilibrium}, LA-based learning algorithms (e.g., $L_{R-I}$ learning \cite{1104342} and Bush-Mosteller learning 
 \cite{1049610}), gradient-play-based better reply \cite{1406126} and no-regret learning \cite{ECTA:ECTA153}. In the cases when both the strategy and  the local expected 
 payoff are to be learned, the AC-like, multiple-timescale learning algorithms \cite{leslie2003convergent} provide an efficient strategy-learning approach (e.g., stochastic FP 
 \cite{fudenberg1998theory}) for the  agents. Further, when the joint action or the payoff of the adversary agents is not directly observable, conjecture-variation-based
 learning \cite{JeanMarie2006399} works as an alternative way of the aforementioned learning algorithms. 
 In the literature, these joint policy-value-iteration mechanisms for games are also known as the COmbined fully DIstributed PAyoff and Strategy-Reinforcement 
 Learning (CODIPAS-RL) mechanisms \cite{tembine2012distributed}.
 
 \item[2)] NE vs. other equilibria: most of the learning algorithms in 1) such as proposed in \cite{Littman94markovgames, Weinberg:2004:BML:1018410.1018798, Hu:2003:NQG:945365.964288, 
 li2007reinforcement, topkis1979equilibrium, 1104342, 1049610, 1406126} aim at finding the NE of the repeated games/SGs. By contrast, the goal of CE-Q learning 
 \cite{Greenwald03correlatedq-learning} and some no-regret learning algorithms \cite{ECTA:ECTA153} is to learn the CE in the SG and the repeated game, respectively.  By relaxing 
 the condition of an NE from the profile of real actions to the profile of agent beliefs, conjecture-variation-based learning \cite{JeanMarie2006399} converges to the 
 conjecture equilibrium \cite{Wellman:Hu:1998}. In most practical scenarios based on the framework of general repeated games, FP and stochastic FP only guarantee that the 
 $\epsilon$-equilibrium can be reached \cite{fudenberg1998theory}. In the literature, $\epsilon$-equilibrium is sometimes known as the Logit equilibrium when the Logit 
 function\footnote{About the definition of a Logit function, please refer to Section \ref{subsubsec_game_LA}.} is used for strategy updating.
 
 \item[3)] Noncooperative games, cooperative games and team games: technically, these three major categories cover most of the game-based models in the applications of 
 distributed control. Provided that the noncooperative games satisfy certain properties (e.g., being supermodular/submodular \cite{ECTA:ECTA376} or having a unique NE),
 all of the aforementioned learning algorithms in 1) and 2) may ensure to reach one of the equilibria in the game. For cooperative games, which are usually featured 
 by the process of bargaining or coalition formation among agents, the Nash bargaining solution can be learned through FP \cite{tembine2012distributed}. A team game is defined as 
 the game in which the agents share the common payoff function, thus considered as a fully cooperative case of the general 
 SG-based games. Since every team game can be modeled as a potential game \cite{han2012game}, it is possible to apply best-response-based learning \cite{4814554}, stochastic FP 
 \cite{ECTA:ECTA376} or no-regret learning \cite{Jafari01onno-regret} to learn the NE of a repeated team game. In the case of team SGs, each agent can also be associated 
 with one single learning automaton at one game state. Then by applying $L_{R-I}$ learning a pure-strategy NE is guaranteed to be reached \cite{4539483}.
\end{itemize}

\subsection{A Summary of Model-Free Learning Algorithms}
\label{sub_sec_nutshell_learning}
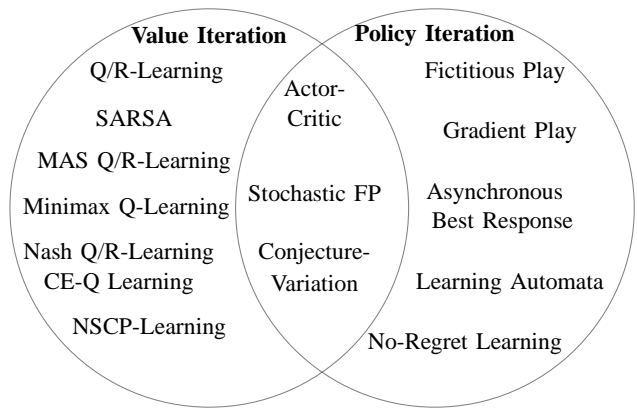
\begin{figure}[!t]
  \centering
  \begin{tikzpicture}
  \tikzstyle{every node}=[font=\small]
  \tikzset{venn circle/.style={draw,circle,minimum width=5.30cm,opacity=0.5}}
  \node [venn circle = white] (A) at (0,0) {};
  \node [venn circle = white] (B) at (0:3cm) {};
  
  \node at (0,2.3) {\bf{Value Iteration}};
  \node at (3,2.3) {\bf{Policy Iteration}};
  
  \node at (-0.70,1.8) {Q/R-Learning};
  \node at (-1.0,1.2) {SARSA};
  \node at (-1.0,0.6) {MAS Q/R-Learning};
  \node at (-1.1,0.0) {Minimax Q-Learning};
  \node at (-1.2,-0.6) {Nash Q/R-Learning};
  \node at (-1.2,-1.0) {CE-Q Learning};
  \node at (-0.8,-1.6) {NSCP-Learning};
  
  \node at (3.8,1.8) {Fictitious Play};
  \node at (4.0,1.0) {Gradient Play};
  \node at (3.8,0.2) {Asynchronous};
  \node at (3.9,-0.2) {Best Response};
  \node at (4.0,-1.0) {Learning Automata};
  \node at (3.4,-1.8) {No-Regret Learning};
  
   \node at (1.4,1.6) {Actor-};
  \node at (1.4,1.2) {Critic};
  \node at (1.4,0.2) {Stochastic FP};
  \node at (1.4,-0.6) {Conjecture-};
  \node at (1.4,-1.0) {Variation};
\end{tikzpicture} 
  \caption{A quick summary of the model-free learning algorithms.}
  \label{fig_nutshell_learning}
\end{figure}

Before proceeding to the next section, we provide a summary of the learning mechanisms that have been introduced in this section in Figure \ref{fig_nutshell_learning}. In Figure 
\ref{fig_nutshell_learning}, the learning mechanisms are categorized according to the experience updating approach (i.e., value iteration or policy iteration) that they apply.
In Table \ref{tab_nutshell_learning}, we further summarize the characteristics of these learning mechanisms in terms of stability property and the system models (SAS, MAS and 
games) that they are built upon. Figure \ref{fig_nutshell_learning} and Table \ref{tab_nutshell_learning} together provide a quick sketch of the algorithms that are to be 
surveyed with respect to their applications in cognitive wireless networks. More details of the characteristics of each learning mechanism will be provided in the following 
sections.
\begin{table}[!t]
  \caption{Brief characteristics of model-free learning mechanisms}
  \centering
 \begin{tabular}{ p{65pt}| p{70pt} | c }
  \hline
  Learning Mechanism &  System Model & Stability Property \\
  \hline
  Q/R-learning  & Single-agent MDP & Optimality learning\\
  \hline  
  SARSA & Single-agent MDP & Optimality learning\\
  \hline
  MAS Q/R-learning &  Multi-agent MDP & Optimality learning\\
  \hline
  Minimax Q-learning & Noncooperative SGs & NE learning\\
  \hline
  Nash Q/R-learning & Noncooperative SGs & NE learning\\
  \hline
  CE Q-learning & Noncooperative SGs & CE learning\\
  \hline 
  NSCP-learning & Noncooperative SGs & NE learning\\
  \hline
  FP & Noncooperative SGs/repeated games & $\epsilon$-Equilibrium learning\\
  \hline 
  Gradient play & Noncooperative repeated games & NE learning\\
  \hline
  Asynchronous Best Response & Noncooperative/Team repeated games & NE learning\\
  \hline
  LA & Noncooperative/Team repeated Games & $\epsilon$-equilibrium learning\\
  \hline
  No-regret learning & Noncooperative/Team repeated games/SGs & NE/CE learning\\
  \hline
  Actor-critic learning &  Single/Multi-agent MDP & Optimality learning\\
  \hline
  Stochastic FP & Noncooperative repeated games & NE learning\\
  \hline
  Conjecture-variation-based learning & Noncooperative SGs/repeated games & $\epsilon$-equilibrium learning\\
  \hline
\end{tabular}
  \label{tab_nutshell_learning}
\end{table}

\section{Applications of Single-Agent-Based Learning in Cognitive Wireless Networks}
\label{sec_app_crn}
Thanks to the property of self-organization, a model-free learner is able to reduce the level of required a-priori knowledge about the network model 
as well as the level of overhead due to explicit information exchange. It is also possible for the learner to adapt quickly
to the changes of the network environment. As a result, model-free learning is particularly suitable for resource management and scheduling problems that demand 
self-exploration and self-organization of the network devices.
Starting from this section, we will provide a comprehensive survey on the applications of model-free learning 
across different protocol layers in cognitive wireless networks following the broad-sense definition of CRNs. In a nutshell, the survey on the applications of model-free 
learning is organized based on the categorization of the learning mechanisms that is provided in Section \ref{sec_bckgrd}. According to the three types of mathematical models for 
decision-making, Sections \ref{sec_app_crn}, \ref{sec_app_dist} and \ref{sec_app_game} are devoted to the applications of learning algorithms based on single-agent systems, 
loosely coupled multi-agent systems and game-based multi-agent systems, respectively. The notations used in this section are summarized in Table 
\ref{notation_subsec_app_single}.

\subsection{Applications of Learning in Single-Agent Systems}
\label{subsec_app_single}

\begin{table}[t]
  \caption{Summary of the main notations in Section \ref{sec_single_agent}}
  \centering
 \begin{tabular}{ p{40pt}| p{180pt}}
 \hline
 Symbol & Meaning\\
  \hline
  $\mathcal{O}$ &  A finite set of observation states in a partially observable Markov decision process\\
  \hline
  $o$  & A single observation state in a partially observable Markov decision process\\
   \hline
  $w(s)$  & A weighting function to map a set of states $s$ to a new state for state abstraction\\
  \hline  
  $c_t(s,a)$ or $c$ & Instantaneous cost function of a constrained MDP\\
  \hline
  $\lambda$ & Lagrange multiplier \\
  \hline
\end{tabular}
  \label{notation_subsec_app_single}
\end{table}
The early attempts in applying learning algorithms to wireless networking problems appeared even before the concept of cognitive radios was proposed. 
Generally, the a-priori knowledge of the environment evolution dynamics (e.g., the transition probabilities of the MDPs) is not required by the MDP-based, value-iteration 
learning schemes. Thus, the schemes are widely applied to the problems in the time-varying dynamics of the wireless environment that cannot be perfectly sensed.
These problems include dynamic packet routing 
\cite{Boyan94packetrouting}, Dynamic Channel Assignment (DCA) \cite{790549, 809089} and joint radio resource management for multi-rate transmission control in WCDMA networks 
\cite{1262128}, just to mention a few. The strategy-learning schemes in these studies are featured by a single/centralized agent, and are usually based on the standard Q-learning
algorithm given in (\ref{eq_q_learning}). In early studies, the learning schemes are built upon the simplified system models. Thus, the issues such as the convergence 
conditions of the learning schemes are still not the focus of the discussion. As a result, the existence of Markovian property is simply assumed in most of these works 
\cite{790549, 809089, 
1262128}. Also, in order to reduce the complexity of the system model, the original MDPs modeling the network dynamics are usually transformed into new MDPs with reduced 
state-action space using state abstraction \cite{Li06towardsa} or Q-table projection methods. However, the equivalence between the original MDPs and the re-transformed MDPs is 
generally not guaranteed (see the example of \cite{1262128}). In most of these works (e.g., \cite{Boyan94packetrouting, 790549}), the learning rules are designed in a heuristic 
manner. Sometimes the standard Q-learning schemes are modified by introducing the neural networks in order to represent the table of the state-action values and approximate 
the Q-value-updating function \cite{Boyan94packetrouting, 809089, 1262128}. With these simplifications, the convergence to an optimal strategy of the learning schemes in these 
studies is also not guaranteed.

Among different approaches for simplifying the MDP-based model of the network-control process, state abstraction \cite{Li06towardsa} becomes a necessary way of trading 
off optimality for the efficiency of the single-agent-based learning mechanisms. The necessity of state-action-space reduction lies in the need for computational tractability
of the learning schemes in the case of state-action-space explosion. This is especially necessary when a single agent is learning the strategy from a large 
set of candidate actions in a system with a huge number of states. In the context of networking problems, state abstraction maps an original network-control
model based on one MDP into a new MDP with a smaller state-action set. Mathematically, state abstraction in MDPs can be defined as follows:
\begin{Definition}[State abstraction \cite{Li06towardsa}]
 For two MDPs $M={\langle}{\mathcal{S}}, {\mathcal{A}}, u, \Pr({s}'|{s},a) {\rangle}$ and $\overline{M}={\langle}\overline{\mathcal{S}}, {\mathcal{A}}, \overline{u}, 
 \Pr(\overline{{s}}'|{{s}},a) {\rangle}$, $\phi:\mathcal{S}\rightarrow\overline{\mathcal{S}}$ is such a mapping that  $\{\phi^{-1}(\overline{{s}})|\overline{{s}}\in
 \overline{\mathcal{S}}\}$ partitions the state space $\mathcal{S}$. Define a weighting function $w: \mathcal{S} \rightarrow [0,1]$, where $\forall \overline{{s}}\in\overline{
 \mathcal{S}}, \sum_{{s}\in\phi^{-1}(\overline{{s}})}w(s)=1$. $\overline{M}$ is an abstracted  MDP of $M$, if the following conditions are satisfied:
 \begin{equation}
  \label{eq_abstraction_reward}
  \overline{u}(\overline{{s}},{a})=\sum_{{s}\in\phi^{-1}(\overline{{s}})}w(s)u(s,a),
 \end{equation}
 and
 \begin{equation}
  \label{eq_abstraction_transition}
  \overline{\Pr}(\overline{{s}}'|\overline{{s}},{a})=\sum_{{s'}\in\phi^{-1}(\overline{{s}}')}\sum_{{s}\in\phi^{-1}(\overline{{s}})}w(s)\Pr(s'|s,a).
 \end{equation}
\end{Definition}

However, the state-abstraction method generally requires that the state transition in the new MDP with reduced complexity to be well-defined. Namely, the 
linear-combination-based mapping in (\ref{eq_abstraction_reward}) and (\ref{eq_abstraction_transition}) needs to be established and the condition $\sum_{\overline{{s}}'}
\overline{\Pr}(\overline{{s}}'|\overline{{s}},{a})=1$ needs to be satisfied. Since with model-free learning, the transition models are generally not known, it will be practically 
impossible to obtain an accurate model of the reduced MDP. In order to address such an issue, approximate abstraction is proposed in \cite{1311428, 1494473}. In \cite{1311428, 
1494473}, an on-policy reinforcement learning method, SARSA, is applied to the DCA problem in a multi-cell, multi-channel network with the consideration of handoffs. In the 
considered cellular network, $N$ cells provide $M$ channels to mobile stations, thus forming an $N\!\times\!(M\!+\!1\!)\times\!M$ state-action set. The arbitrary 
state-aggregation method proposed in \cite{1311428, 1494473} aggregates the rarely encountered states by reducing the size of the channel state space to a fraction of the total 
number of the channels. The state variable representing the number of currently allocated 
channels is also excluded, which leads to a 98\% reduction from the original state-action space. A more complicated state-action-space abstraction method can be found in 
\cite{1262128}. It adopts the feature extraction method and maps the original state vector based on four dimensions, namely, the mean and variance of the interference from 
the existing connections, the transmission type and the required transmission rate, into a vector of the resultant interference profile. The feature extraction method is further 
adopted in stochastic-game-based modeling for strategy learning in CRNs \cite{4814773, 4581648}. In \cite{4814773, 4581648}, the central spectrum moderator allocates the 
transmission opportunities to the CRs through an iterative, second-price auction (see \cite{han2012game} for the definition of auctions), whose dynamics is jointly determined by 
the Signal-to-Noise-Ratio (SNR) of the channels and 
the buffer states of all the CRs. In \cite{4814773, 4581648}, multi-stage bidding is adopted. Since for each CR, the value of tax to be paid for
using the channels are based on the inconvenience it causes to the other CRs, the individual CRs use their local tax announced by the central spectrum moderator to classify the 
channel-buffer states that the other (adversary) CRs are in. Therefore, individual CRs only need to exchange the pricing information with the central spectrum moderator,
and no extra information exchange between the CRs is required. In these works, the feature extraction method does not only achieve 
the goal of state abstraction, but also help avoid the explicit information exchange between individual CRs.
\begin{figure}[!t]
    \centering
    \includegraphics[width=0.47\textwidth]{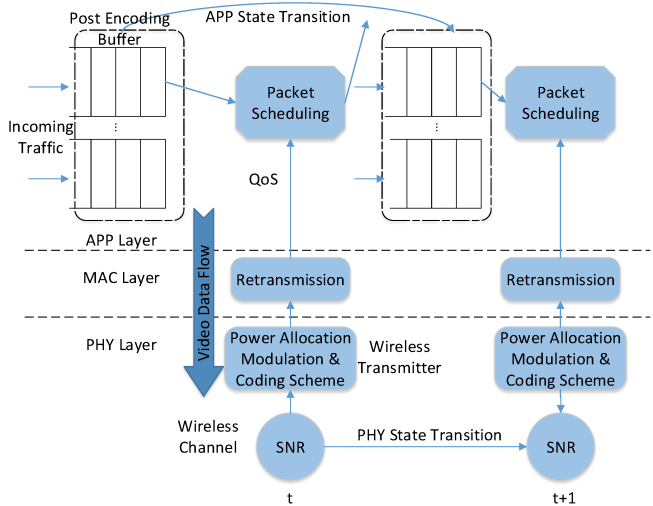}
    \caption{Real-time video streaming process (adapted from \cite{5433062}).}
    \label{fig_layered_MDP_System}
\end{figure}

\begin{figure}[!t]
\centering
  \includegraphics[width=0.49\textwidth]{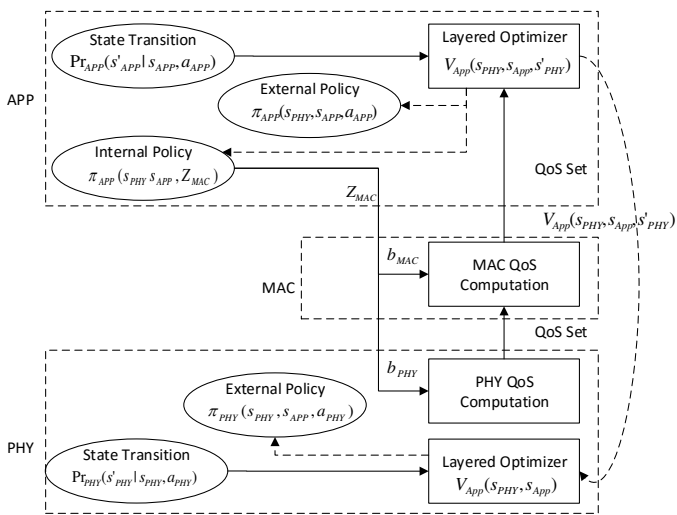}
 \caption{The operation and message exchange in the layered MDP (adapted from \cite{5433062}).}
    \label{fig_layered_MDP}
\end{figure}

With the development of MDP-based modeling in different protocol layers of the wireless networks (see examples in MAC layer \cite{4410464}, link layer \cite{4600185} 
and application layer \cite{1354562}), the SAS-based learning mechanisms in the cognitive wireless networks also gain more capabilities in addressing the radio resource 
management problems. In \cite{5433062}, the problem of real-time video transmission over a single-hop, slow-varying flat fading channel is formulated as a systematic 
layered MDP (see Figure \ref{fig_layered_MDP_System} and Figure \ref{fig_layered_MDP} for a schematic view of the system and the corresponding layered MDP model). With the proposed 
problem formulation, the discrete system state is composed of three components, i.e., the SNR as the channel state in the PHY layer, the transmission opportunity as the state of 
the MAC layer and the amount of both the incoming traffic and the buffered packets as the state of the application layer (see Figure \ref{fig_layered_MDP}). The evolution 
of the joint state $(s_{APP}, s_{PHY})$ is modeled as a Markov chain controlled by the joint action $(a_{APP}, a_{MAC}, a_{PHY})$, in which $a_{MAC}$ is composed of two internal
actions $b_{PHY}$ and $b_{MAC}$. The 
joint action is determined by the power allocation, the channel resource payment made to the spectrum moderator and the packet scheduling algorithm. The cross-layer management of 
packet transmission is formulated as a layered MDP. This is because for the Bellman optimality equation of the state value, the Dynamic Programming (DP) based 
expression can be decomposed into a two-loop DP-based optimization. In the two-loop optimization, it is assumed that both layers have access to the global state in each time 
slot. The inner loop (i.e., the application-layer optimization) only needs to know the joint MAC-application action and the reported state value of the PHY layer for policy 
updating, while the outer loop (i.e., the PHY-layer optimization) only needs to know the PHY-layer action information and the reported state value from the application layer for 
policy updating. The layered Q-learning \cite{5299114} can be applied to learn the optimal strategy for transmission, with the standard Q-value updating rule in
(\ref{eq_q_learning}) modified in each layer by incorporating the estimated Q-value from the other layer into the estimation of the local Q-values.

Apart from lacking the a-priori knowledge about the statistics of the underlying Markov process, the decision-making entity in the network may frequently face the constraints 
on the available resources. To tackle these constrained radio resource allocation/scheduling problems, the unconstrained MDP models are extended to the 
Constrained MDPs (CMDPs), based on which, modified reinforcement learning algorithms are also proposed \cite{Fei:2006:EQP:1147607.1147616, 6120364, 6585729, 5683502, 
Sun:2011:CMV:1996558.1996579, 4357390}. Mathematically, a CMDP is defined by expanding the 4-tuple MDP model (Definition \ref{def_MDP}) to be a 5-tuple, 
${\langle}{\mathcal{S}}, {\mathcal{A}}, u, c, \Pr(\mathbf{s}'|\mathbf{s},a) {\rangle}$, with the additional cost/constraint element $c$ \cite{altman1999constrained}. Taking the 
average-reward CMDP as an example, a generic CMDP optimization problem can be stated as follows:
\begin{equation}
 \label{eq_cmdp}
 \begin{array}{ll}
 \max\limits_{\pi} &h^{\pi}(s)\!=\!\lim\limits_{T\rightarrow\infty}\sup\displaystyle\frac{1}{T}E_{\pi}\left\{\sum\limits_{t\!=\!0}^{T-1}u_t(s_t,a_t)\right\},\\
 \textrm{s.t.} &C^{\pi}(s)\!=\!\lim\limits_{T\rightarrow\infty}\sup\displaystyle\frac{1}{T}E_{\pi}\left\{\sum\limits_{t\!=\!0}^{T-1}c_t(s_t,a_t)\right\}\!\le\!C_{\max}.
 \end{array}
\end{equation}
According to Theorem 12.7 of \cite{altman1999constrained}, we have the following theorem for the average-reward CMDP:
\begin{theorem}
 \label{dual_CMDP}
 If the underlying Markov chain of the CMDP, ${\langle}{\mathcal{S}}, {\mathcal{A}}, u, c, \Pr(\mathbf{s}'|\mathbf{s},a)
{\rangle}$, is unichain and the sequence of the immediate cost $c_t$ is bounded below and satisfies the following growth condition:
\begin{itemize}
 \item[] for $c\!:\!\mathcal{K}\!\rightarrow\!\mathbb{R}$ there exists a sequence of increasing compact subsets $\mathcal{K}_i$ of $\mathcal{K}$ such that $\cup_i\mathcal{K}_i\!=\!
 \mathcal{K}$ and $\lim_{i\rightarrow \infty}\inf\{c(\kappa);\kappa\notin\mathcal{K}_i \}=\infty$,
\end{itemize}
then there exists an optimal Lagrange multiplier $\lambda^*$ such that the optimal solution of the CMDP is equivalent to the optimal solution of the unconstrained MDP,
${\langle}{\mathcal{S}}, {\mathcal{A}}, g=u-\lambda^*c, \Pr(\mathbf{s}'|\mathbf{s},a){\rangle}$.
\end{theorem}
According to Theorem \ref{dual_CMDP}, the non-structured learning schemes for the unconstrained MDP based on the Lagrangian dual function can be developed for solving the 
resource management/scheduling problems in the form of both R-learning \cite{4357390, 5683502, 6120364} and Q-learning \cite{Pietrabissa201189, 
Sun:2011:CMV:1996558.1996579, Fei:2006:EQP:1147607.1147616}, depending on the form of the reward/cost of the CMDP. Apart from the primal-dual equivalence based solution,
it is also possible to develop constrained learning algorithms by exploiting the structure of the specific problems. The special structure is featured by
the convexity of the objective and constraint functions in the original CMDP,
or the modularity of the objective or the constraint functions \cite{Sun:2011:CMV:1996558.1996579, 4156378}. When certain structural property of the network control problems is 
satisfied (specifically, when both the instantaneous payoff and the constraint cost are multi-modular), the constrained structured-learning algorithm can be applied in the 
form of primal projection or submodular parameterization \cite{4156378}.
\begin{table*}[!t]
  \caption{Applications of SAS-based learning schemes in cognitive wireless networks: a summary}
  \centering
 \begin{tabular}{ c|  p{65pt} | p{40pt} |  p{35pt} | p{70pt} | p{85pt}| p{75pt}  }
  \hline
  Network Type &  Application & Problem Formulation & Reference & Learning Scheme & Learning Scheme Variation  & Convergence \\
  \hline
  \multirow{4}{40pt} {Cellular} & Dynamic channel allocation & MDP \cite{790549, 809089, 1311428, 1494473}, semi-MDP \cite{1673076} & \cite{790549, 809089, 1311428, 1494473, 
  1673076} & Q-learning \cite{790549, 809089, 1673076}, SARSA \cite{1311428, 1494473} & Neural Network \cite{809089}, State abstraction \cite{1311428, 1494473}, N/A 
  \cite{1673076} & N/A \\  
  \cline{2-7}
  & Multirate transmission control & MDP & \cite{1262128} & Q-learning & Neural network with feature extraction \cite{1262128} & N/A\\
  \cline{2-7}
  & Call admission control & CMDP &\cite{Fei:2006:EQP:1147607.1147616, Pietrabissa201189} & Q-learning \cite{Fei:2006:EQP:1147607.1147616, Pietrabissa201189} 
  & State abstraction \cite{Fei:2006:EQP:1147607.1147616, Pietrabissa201189} & N/A \\
  \cline{2-7}
  & Joint admission-bandwidth control & CMDP & \cite{1354620, 4357390} & Q-learning \cite{1354620}, Neural network \cite{4357390} & N/A & N/A\\
  \hline
  \multirow{3}{40pt} {Single link} & Cross-layer resource allocation & Layered-MDP & \cite{5433062, 5299114} & Layered Q-learning \cite{5433062} & 
  Virtual experience tuples \cite{5299114} & N/A\\
  \cline{2-7}
  & Scheduling-admission control & CMDP & \cite{6585729} & Stochastic sub-gradient & N/A & Deterministic optimal policy \cite{6585729} \\
  \cline{2-7}
  & V-BLAST power-rate control & CMDP & \cite{4156378} & Q-learning & Constrained structured Q-learning & Randomized optimal policy\\
  \hline 
  \multirow{1}{40pt} {MANETs}  & QoS routing & POMDP & \cite{4224867} & Actor-critic learning & N/A & N/A\\
  \hline  
  \multirow{3}{40pt} {CRNs} & Dynamic spectrum access & MDP \cite{4658293}, CMDP \cite{6120364, 5683502}, POMDP \cite{6226310}& \cite{6120364, 5683502, 6226310, 4658293} & 
  Actor-critic learning \cite{4658293}, R-learning \cite{6120364, 5683502}, policy gradient \cite{6226310} 
  & N/A \cite{6120364, 6226310, 4658293}, Arbitrary state reduction \cite{5683502} & Deterministic optimal policy \cite{6120364}, N/A \cite{4658293, 5683502}, Local optimum 
  policy \cite{6226310}\\
  \hline
  \multirow{2}{40pt} {HETNETs} & Vertical handoff & CMDP & \cite{Sun:2011:CMV:1996558.1996579} & Q-learning & N/A
  & Optimal randomized policy\\
  \cline{2-7}
  & Admission control & MDP & \cite{4815247} & Q-learning & Q-learning based on neural-fuzzy-inference network & N/A \\
  \hline  
\end{tabular}
  \label{tab_sas_learning}
\end{table*}

In addition to not knowing the environment evolution dynamics and being limited by the resource constraints, the learning agents in a wireless network may also lack the ability 
of complete state-information acquisition. This can be a common issue in scenarios such as DSA networks, in which the secondary devices lack the capability of performing 
full-spectrum sensing due to the limited number of antennas \cite{4155374}. The common approach to handle such a problem is to model the radio resource management problem 
as a Partially Observable Markov Decision Process (POMDP). Extending from Definition \ref{def_MDP}, an unconstrained POMDP can be defined as a 6-tuple, ${\langle}{\mathcal{S}}, 
\mathcal{O}, {\mathcal{A}}, u, \Pr({s}'|{s},a), \Pr(o|{s},a) {\rangle}$, in which $\mathcal{O}$ is the set of observations $o$, and $\Pr(o|{s},a)$ denotes the 
mapping probability between the system states and the observations. Instead of directly observing the state information of $s$, the learning agent can only obtain the network
observation $o$. In the POMDPs, the random process associated with the observation is no longer a Markov process. A standard model-based solution to the POMDP is to convert 
the recorded state observations into belief states, and obtain a new unconstrained MDP with a continuous state space of the belief states. However, when the state-transition 
and the state-observation mapping is unknown, the TD-based learning schemes cannot be directly used for learning the optimal strategies of the POMDPs. Instead, other
learning algorithms such as actor-critic learning \cite{Konda:1999:ALA:335624.335633} and policy-gradient-based 
learning \cite{baxter2000reinforcement} are applied. In \cite{4224867}, a delay-constrained least-cost routing problem in MANETs is modeled as a POMDP, the belief state of which 
captures the link-delay uncertainty due to the imprecise link state information. The belief-policy mapping is considered as a parametric function, the policy parameter of which 
is learned through a standard actor-critic learning method. In \cite{6226310}, to solve the DSA problem in a CRN, the channel access process of the Secondary Users (SUs) is 
first modeled as a constrained POMDP. In the constrained POMDP, a reward function is used to collect the instantaneous reward of the SUs, while a cost function reflects the 
instantaneous cost of the Primary Users (PUs) due to the channel interference from the SUs. The partial observation in the problem comes from the imperfect spectrum
sensing of the SUs over the primary channel state. After converting the original constrained POMDP into an unconstrained POMDP with the help of the Lagrange multiplier, the 
learning algorithm based on policy gradient \cite{baxter2000reinforcement} is applied for finding a local optimal policy.

To summarize this section, we categorize in Table \ref{tab_sas_learning} the aforementioned works (and some more) on SAS learning according to the networking applications 
that they focus on. As shown by Table \ref{tab_sas_learning}, the SAS-based learning algorithms are powerful in addressing a number of radio resource allocation problems, as long 
as they can be formulated as a single-link-centric one. However, it is worth noting that although the theoretical support for the convergence of the SAS-learning 
schemes has been well studied, such an issue still needs to be addressed under practical circumstances.

\section{Applications of Learning Based on Loosely Coupled Multi-Agent Systems}
\label{sec_app_dist}
The multi-agent learning scheme naturally leads to the framework of distributed decision making, thus the possibility of self-organization without a dedicated central 
coordinator. Therefore, it is considered especially appropriate for the network management problems in the CRNs, device-to-device (D2D) networks, heterogeneous networks 
(HETNETs) and ad-hoc networks, as long as the networks consist of multiple independent decision-making entities. However, although the framework of distributed decision making 
naturally leads to the consideration of adopting the multi-agent decision learning scheme for network control, it is worth noting that for most cases it may be difficult to 
directly adopt the learning mechanisms based on the loosely coupled MAS by simply ignoring the interactions between the network entities and treat each of them as an independent 
learner. Due to the existence of device interaction, it is necessary to carefully investigate into both the advantage and the limitation of formulating a distributed network 
control problem as a loosely coupled MAS. Furthermore, when adopting the model of learning in the loosely coupled MAS, it is still necessary to check to what level the 
information exchange between the learning agents is needed, and in what ways it can help improving the performance of the network.

The new notations used in this section are summarized by Table \ref{notation_sec_app_dist}.
\begin{table}[t]
  \caption{Summary of the new notations in Section \ref{sec_app_dist}}
  \centering
 \begin{tabular}{ p{40pt}| p{180pt}}
 \hline
 Symbol & Meaning\\
  \hline
  $\gamma_r^{i,F}$ & The received SINR for femto/pico link $i$ over resource block $r$\\
  \hline
  $P_r^{iF}$ & The transmit power of femto/pico BS\\
  \hline
  $g_{ii,r}^{FF}$ & The link gain between the femto/pico BS and its user\\
  \hline
  $g_{ii,r}^{MF}$ &  The link gain between the macro BS and the femto/pico user\\
  \hline
  $\sigma^2$  & noise power\\
  \hline  
  $I[x]$ or $I(x,y)$ & The indicator function\\
  \hline
  $w_i(j)$ or $w'_i(j)$ & The weight assigned by agent $i$ for its neighbor $j$'s instantaneous reward or estimated state value\\
  \hline
  $Y$ & The social reward of a group of agents\\
  \hline
  $y$ & The private reward that an individual agent chooses\\
  \hline
\end{tabular}
  \label{notation_sec_app_dist}
\end{table}

\subsection{Applications of Distributed Learning Based on the Model of Loosely Coupled Multi-Agent Systems}
\label{subsec_app_dist}
\begin{figure}[!t]
    \centering
    \includegraphics[width=0.36\textwidth]{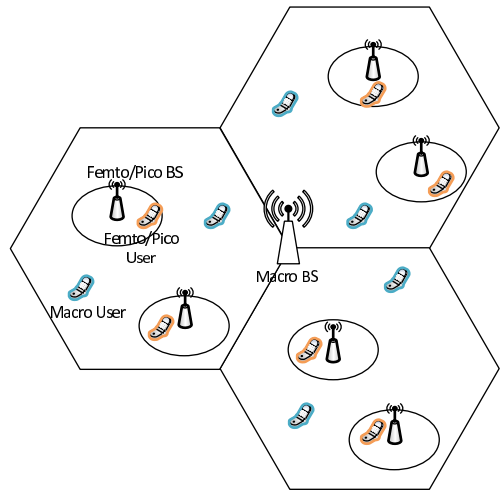}
    \caption{Structure of a HETNET with both inter-cell and cross-layer interference. A HETNET is featured by the hierarchy in the network
    structure, which is comprised by the high-power, high-capacity, wide-range macrocells and the low-power, low-capacity, small-range femtocells/picocells 
    \cite{5876496}.}
    \label{fig_femto_cells}
\end{figure}
For distributed learning in wireless networks, it is usually difficult to definitely classify between a non-game-based, multi-agent decision learning scheme and an SG-based 
learning scheme. The reason for this lies in the inherited nature of strategy coupling in most of the practical networking problem setups. One typical example is illustrated in 
\cite{5493950, 6503987}, which consider that $L$ macrocells and $N$ femtocells/picocells operate over the same frequency band (see Figure \ref{fig_femto_cells}) in a 
HETNET. In order to develop a self-organized power allocation scheme for the downlink transmission in the HETNET, the Shannon capacity of a link 
is considered as the individual utility of a cell, which is a function of the Signal-to-Interference-plus-Noise-Ratio (SINR) of the transmitting link in that cell. Take the 
femtocells/picocells as an example, when both the intra-cell interference and the cross-tier interference are considered, for femotocell/picocell link $i$, the SINR at the 
receiver is determined as follows:
\begin{equation}
 \label{eq_generic_SINR}
 \gamma_{r}^{i,F}=\frac{P_r^{i,F}g_{ii,r}^{FF}}{\sum_{j=1}^{L}P_r^{j,M}g_{ji,r}^{MF}+\sum_{k=1,k\ne i}^{N}P_r^{i,F}g_{ki,r}^{FF}+\sigma^2},
\end{equation}
where $P_r^{i,F}$ is the transmit power of femto/pico Base Station (BS) $i$ over the resource block $r$, $g_{ii,r}^{FF}$ is the link gain between the femto/pico BS and its 
user, $g_{ji,r}^{MF}$ is the link gain between macro BS $j$ and the femto/pico user, $g_{ki,r}^{FF}$ is the link gain from another femto/pico BS $k$ to the user of femto/pico 
BS $i$, and $\sigma^2$ is the noise power. 

Apparently, the capacity of femto/pico link $i$ is determined not only by the transmit power of femto/pico BS $i$, $P_r^{i,F}$, but 
also by the inter-cell interference $\sum_{k=1,k\ne i}^{N}P_r^{i,F}g_{ki,r}^{FF}$ and the cross-tier interference $\sum_{j=1}^{L}P_r^{j,M}g_{ji,r}^{MF}$. Therefore, the private
utility of femto/pico link $i$ is also a function of the strategy of the other femto/pico BS $k$ ($k=1,\ldots, n, k\ne i$) and all the macro BSs $j$ ($j=1,\ldots,L$). The goal 
of the local cells for maximizing the individual utilities conflicts with each other, and it is difficult to decompose the strategy coupling between the cells. As a result, many 
works formulate the same problem as a noncooperative repeated game \cite{6542770, 6990372}. However, it is still possible to tackle such a power control problem by treating the 
strategies of the other BSs as part of the environment dynamics. For example, in \cite{5493950} the system state from the perspective of a femto/pico link is designed as a 
binary one:
\begin{equation}
  \label{eq_state_thresholding}
  I_{t}^{i,r}=\left\{
  \begin{array}{ll}
    1,\textrm{ if } \gamma_{r,t}^{i,F}<\gamma_{\textrm{Th}},\\
    0, \textrm{ otherwise},
  \end{array}\right.
\end{equation}
which is based on hard thresholding (compared with the permitted SINR given as $\gamma_{\textrm{Th}}$) of the macrocell user with the interference from the femto/pico links. The 
similar network-state formulation can be found in the 
other works such as \cite{6503987}. By adopting a standard Q-learning scheme based on the assumption of independent state-value evolution, it is assumed in 
\cite{5493950, 6503987} that the dynamics of the aggregated interference to the macrocell user is a stationary Markov process. Consequently all the strategies of the other
femto/pico users are treated as stationary ones hence part of the wireless environment. In most of the cases, such a formulation/solution with the distributed MDPs and 
independent Q-learning algorithm may not guarantee the convergence to any equilibrium. However, empirical studies show that when using the distributed Q-learning scheme, 
convergence can still be achieved given a sufficiently large number of iterations \cite{5493950, 5415567, 5700414}, and the distributed Q-learning algorithm is also able to 
achieve a better performance compared with the non-adaptive algorithms \cite{5415567, 5700414, 6503987}. Although not mathematically proved, one possible explanation for such a
result may lie in Proposition \ref{thm_IL_convergence_general}, since one independent Q-learning agent is always able to converge as long as the other agents happen to converge 
in behavior.

\begin{figure}[!t]
\centering
  \includegraphics[width=0.42\textwidth]{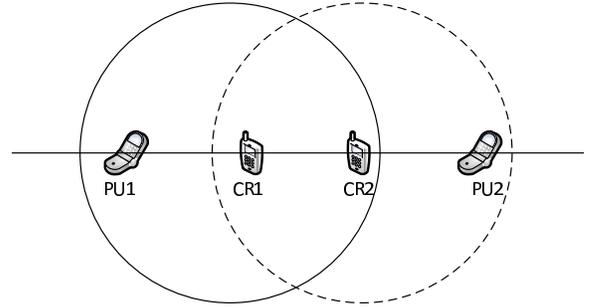}
 \caption{Illustration of the interference map for the two-SU-two-PU DSA network \cite{4531946}.}
    \label{fig_two_user_pomdp}
\end{figure}

Generally, for the network management problems with strategy coupling, directly adopting the distributed learning schemes in the loosely coupled MAS (e.g., multi-agent learning 
in the form of distributed, independent Q-learning) can be considered 
as an approach that trades off the certainty of algorithm convergence for the simplicity of system analysis and learning-rule design. Except for heterogeneous networks, 
applications that follow such a design pattern can be found in the problem formulation such as distributed DSA with the SU collisions 
\cite{Wu:2010:SMC:1838194.1838199, 5426758, Biggelaar2012}, power allocation in the overlay, cognitive wireless mesh network \cite{5670390} and dynamic spectrum management in 4G 
cellular networks \cite{6912482}. Although with many studies that adopt such a design pattern for the learning schemes, it is important to reiterate that 
overlooking strategy coupling may result in poor performance of each learning agent. In \cite{4531946}, a problem of DSA management with 2 SUs over 2 primary
channels (see Figure \ref{fig_two_user_pomdp}) is used to exemplify how the lack of coordination between individual agents may impact the agent performance. In \cite{4531946}, 
the availability of a primary channel
is modeled as a two-state discrete Markov chain. The SUs try to access the idle primary channel while avoiding the collision with the other SU. The adaptation 
of the channel-access strategies is formulated as a POMDP, in which the observation of an SU includes 3 states: busy, collision and success. Based on the assumption that the 
presence of the other SU can be ignored, a model-based single-user approach for strategy updating is proposed. When compared with the cooperative approach, which allows the SUs 
to exchange their belief state vectors of the POMDP, the performance of the single-user-based approach is shown to be significantly inferior. Moreover, the simulation results in
\cite{4531946} show that the performance of the single-agent-based approach is even worse than that of the deterministic channel-assignment scheme, which indicates that in the 
situation of strategy coupling, allowing some degree of cooperation will be essential.

In order to balance between the simplicity of the learning mechanism (namely, the distributiveness of strategy learning) and the optimality of the learning algorithm, careful 
modeling is needed with respect to different network scenarios. In \cite{6175021}, a set of decision-learning mechanisms based on distributed Q-learning is adopted for a 
scalable DSA mechanism in an overlay CRN. The goal of the learning mechanism design is to obtain the near-optimal strategies without the explicit coordination among the SUs.
It is shown in \cite{6175021} that by properly designing the private/local objective functions of the individual SUs, the needs of both agent coordination and distributed 
decision-learning can be fulfilled. In \cite{6175021}, the SUs are assumed to share the temporarily free band roughly equally. It means that the reward of an individual SU with 
DSA, $u_i(t)$, is approximately equal to the average of the social reward of all the SUs that attempt to use the same primary band (denoted by $Y(t)$):
\begin{equation}
 \label{eq_local_dsa_reward}
 u_i(t)\!=\!\frac{1}{|\mathcal{N}_i(t)|\!+\!1}Y(t)\!=\!\frac{1}{|\mathcal{N}_i(t)|\!+\!1}\sum_{i=1}^{|\mathcal{N}_i(t)|\!+\!1}u_i(\mathcal{N}_i(t)),
\end{equation}
where $\mathcal{N}_i(t)$ is the set of SUs that interfere with SU $i$ over the same band at time $t$. The PU activity is also modeled as a two-state Markov chain. 
In \cite{6175021}, two guidelines are proposed for designing the private/individual objective function of each SU:
\begin{itemize}
 \item [1)] \emph{alignedness}, which reflects agent coordination, and the full alignedness requires the SUs not working against each other when maximizing their own private 
 objectives;
 \item [2)] \emph{sensitivity}, which reflects the efficiency of the individual learning processes and requires the SUs to be able to discern the impact of their own action 
 changes so as to learn about the better local strategies fast enough.
\end{itemize}

In \cite{6175021}, the measurable indices of ``factoredness'' and ``learnability'' are introduced to measure alignedness and sensitivity of the private objective function, 
respectively. Denoting the selected private objective function as $y_i$ ($y_i$ may not be the same as $u_i$) and the joint deterministic strategy by the SUs over the same band 
as $\pmb\pi=(\pi_i,\pi_{-i})$, the degree of factoredness and learnability can be expressed as in (\ref{eq_factoredness}) and (\ref{eq_learnability}), respectively:
\begin{gather}
 \label{eq_factoredness}
 F_{y_i}\!=\!\frac{\sum_{\pi_{i}}\sum_{\pi_{-i}}I[\left(y_i(\pmb\pi)\!-\!y_i(\pmb\pi')\right)\left(Y(\pmb\pi)\!-\!Y(\pmb\pi')\right)]}{\sum_{\pi_{i}}\sum_{\pi_{-i}}1},\\
 \label{eq_learnability}
 L_{i,y_i}(\pmb\pi)=\frac{E_{\pi'}[|y_i(\pmb\pi)-y_i(\pi_{-i},\pi'_{i})|]}{E_{\pi'_{-i}}[|y_i(\pmb\pi)-y_i(\pi_{-i},\pi'_{i})|]},
\end{gather}
where $I[x]$ is the indicator function, $I[x]=1$ if $x>0$ and $I[x]=0$ otherwise, and $F_{y_i}$ ($0\!\le\!F_{y_i}\!\le\!1$) measures the consistence between the local objective 
and the social payoff. The higher the degree of factoredness (i.e., the value of $F_{y_i}$) is, the more likely a change of the local action by SU $i$ 
will have the same impact on both its private reward and the global reward. $L_{i,y_i}$ measures the sensitivity of local reward to the local action changes. According to 
(\ref{eq_learnability}), the higher the sensitivity (i.e., the learnability), the more the dependence of $y_i(\pmb\pi)$ on the local actions of SU $i$. By employing the property
described by
(\ref{eq_local_dsa_reward}), namely, the private reward $y_i(t)=u_i(t)$ being proportional to the global reward $Y(t)$, it is shown in \cite{6175021} 
that a good objective function can be obtained by removing from $Y(t)$ the effects of all SUs other than SU $i$. A general form of such a local objective function can be 
expressed as follows:
\begin{equation}
 \label{eq_good_local_payoff}
 D_i(\pmb\pi)=Y(\pmb\pi)-Y(\pi_{-i}).
\end{equation}
Since $u_i(t)$ is a function of both $Y(t)$ and the cardinality of the interfering-SU set $\mathcal{N}_i(t)$, all that SU $i$ needs to obtain the value of $D_i$ is to estimate 
$|\mathcal{N}_i(t)|$ given the information that SU $i$ observes locally. It is shown that with the proposed objective function (\ref{eq_good_local_payoff}), the 
distributed learning scheme achieves better spectrum efficiency than those learning with both private reward and global reward. From the game theoretic perspective, spectrum 
access with the individual reward as in (\ref{eq_local_dsa_reward}) can be interpreted as a cardinal potential game \cite{han2012game}, in which (\ref{eq_good_local_payoff}) is 
in the exact form of a potential 
function. In this sense, the design of the objective function in \cite{6175021} can be considered as a special case of global-reward-based learning, and may not 
be easily extended to a general radio resource management problem such as \cite{5493950,6503987}. Although the two indices in (\ref{eq_factoredness}) and (\ref{eq_learnability})
provide important guidelines on individual utility function design for distributed learning, it is still needed to find appropriate approaches other than that given by 
(\ref{eq_local_dsa_reward}) for the networking applications which cannot be modeled as a potential game.

\begin{figure}[!t]
\centering
  \includegraphics[width=0.35\textwidth]{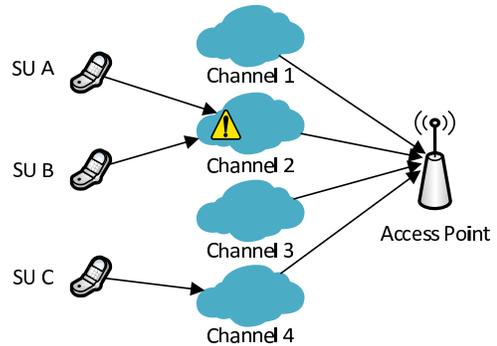}
 \caption{Channel access competition and conflict in an Aloha-like multi-user-multi-channel CR system \cite{Li:2010:MQA:1863662.1928710}.}
    \label{fig_aloha_collision}
\end{figure}

Instead of designing a different objective function, the learning scheme itself can also be tailored to meet the requirement of radio resource management. One example of learning
scheme design in the strategy-coupling scenario is provided by \cite{Li:2010:MQA:1863662.1928710}, which studies an Aloha-like spectrum access scheme without any negotiation in a 
multi-user, multi-channel CRN (Figure \ref{fig_aloha_collision}). In \cite{Li:2010:MQA:1863662.1928710}, $N$ primary channels are modeled as $N$ independent, two-state Markov 
chains, while the SUs are assumed to have no mutual communication and need to learn the collision-avoidance strategies online. Instead of adopting the standard state-value 
evolution model given in (\ref{eq_action_value_fn}) and the TD-based strategy-learning mechanism given in (\ref{eq_q_learning}), the expected one-time reward is adopted as 
(\ref{eq_one_time_q_value}):
\begin{equation}
 \label{eq_one_time_q_value}
 Q_{ij}^{\mathbf{s}}=E[u_i|a_i(t)=j,\mathbf{s}(t)=\mathbf{s}],\\
\end{equation}
and a learning mechanism without considering the future reward is designed as (\ref{eq_myopic_q_learning}):
\begin{equation}
 \label{eq_myopic_q_learning}
 Q_{ij}^{\mathbf{s}}(t\!+\!1)\!=\!(1\!-\!\alpha_{ij}(t))Q_{ij}^\mathbf{s}(t)\!+\!\alpha_{ij}(t)u_i(t)I(a_i(t),j).
\end{equation}
In (\ref{eq_one_time_q_value}) and (\ref{eq_myopic_q_learning}), $a_i(t)=j$ represents the action of SU $i$ to select channel $j$ for transmission, $\mathbf{s}$ is the vector of 
the channel states, $\alpha_{ij}(t)$ is the learning step, $u_i(t)$ is the instantaneous reward of SU $i$ and $I(x,y)$ is the indicator function (i.e., $I(x,y)=0$ if $x\ne y$ and 
$I(x,y)=1$ if $x=y$). Although (\ref{eq_myopic_q_learning})
appears in a similar form to distributed Q-learning, it is derived based on the analysis of the channel contention as an SG. It is shown in 
\cite{Li:2010:MQA:1863662.1928710} that with the Boltzmann distribution-based strategy exploration, the learning scheme in (\ref{eq_myopic_q_learning}) is equivalent to the
Robbins-Monro iteration \cite{Robbins1951} and converges asymptotically to a stationary point (i.e., an NE) with probability one.

Generally, the aforementioned multi-agent learning schemes can be divided into two categories, namely, distributed learning based on the assumption of purely independent state-value 
evolution (e.g., \cite{5493950, 5415567, 5700414, 6503987, Wu:2010:SMC:1838194.1838199, 5426758, Biggelaar2012,5670390,6912482}) and distributed learning based on the 
structural property of the specific resource management problems (e.g., \cite{6175021, Li:2010:MQA:1863662.1928710}). Although both of them do not require explicit information 
exchange among network devices, sometimes introducing a certain level of information exchange (at the cost of more overhead) can help improve the network performance. 
In the literature, the learning schemes with explicit information exchange is usually referred to as learning based on Distributed Value Function (DVF).  With DVF, local
devices are required to share their state-value/reward functions with the neighbors. Instead of learning the Q-value based on the individual reward or local state values, 
individual decision making aims at the maximization of both the local and the neighbors' weighted sum of rewards/state-values. By modifying (\ref{eq_q_learning}), a typical 
learning mechanism with DVF can be expressed as
\begin{equation}
 \label{eq_q_DVF}
 \begin{array}{ll}
 Q_i^{t+1}(s_i,a_i)\!\leftarrow\!(1-\alpha_{t})Q_i^t(s_i,a_i)+\\
 \qquad\alpha_{t}\left(u_i^t(s_i,a_i)+\beta\sum\limits_{j\in\mathcal{N}(i)}w_i(j)V_j(s_j)\right),
 \end{array}
\end{equation}
in which $\mathcal{N}(i)$ is the set of device $i$'s neighbors (including $i$) and $w_i(j)$ is the weight that determines the contribution of device $j$'s state-value to 
device $i$'s estimation of $V_i$. 
\begin{table*}[!t]
  \caption{Applications of independent-learner learning schemes in cognitive wireless networks: a summary}
  \centering
 \begin{tabular}{ l|  p{90pt} |  p{35pt} | p{80pt}  | p{110pt} | p{55pt}}
  \hline
  Network Type &  Application & Reference & Strategy Coupling Assumptions & Learning Scheme & Convergence \\
  \hline
    \multirow{4}{40pt} {CRNs} & Aggregated interference control & \cite{5415567} & None & Independent Q-learning for MDP and neural network for 
    POMDP \cite{5415567}  & N/A \\  
  \cline{2-6}
  & Joint spectrum and power management &\cite{Wu:2010:SMC:1838194.1838199, 5670390} & None \cite{Wu:2010:SMC:1838194.1838199}, coupling as a noncooperative game \cite{5670390} 
  & Independent Q-learning \cite{Wu:2010:SMC:1838194.1838199, 5670390} & N/A\\
  \cline{2-6}
  & Dynamic spectrum access &\cite{5426758,6912482,6175021,Li:2010:MQA:1863662.1928710} & None \cite{5426758, 6912482}, coupling with fully connected topology \cite{6175021},
  Coupling as a noncooperative game \cite{Li:2010:MQA:1863662.1928710} & Independent Q-learning \cite{5426758}, Win-or-Learn-Fast (WoLF) \cite{6912482}, 
  Independent learning (unspecified) \cite{6175021}, Modified independent Q-learning without considering the future states \cite{Li:2010:MQA:1863662.1928710} & N/A 
  \cite{5426758, 6912482}, near optimal strategy \cite{6175021} or NE \cite{Li:2010:MQA:1863662.1928710}\\
  \cline{2-6}
  & Joint sensing-time and power allocation & \cite{Biggelaar2012} & None & Independent Q-learning & N/A\\
  \hline
  \multirow{2}{40pt} {HETNETs} & Femto-user power allocation & \cite{5493950} & None & Independent Q-learning & N/A\\
  \cline{2-6}
  & Inter-cell interference coordination & \cite{6503987} & None & Independent Q-learning & N/A \\
  \hline 
  \multirow{1}{40pt} {Sensor networks}  & Coverage and energy consumption management & \cite{4496881} & Coordinated decision-making & DVF-based learning & N/A\\
  \hline  
  \multirow{1}{40pt} {Cooperative networks} & Power and relaying probability management & \cite{5512673} & Coordinated decision-making & Q-learning based on distributed reward 
  and value function, & Local optimal point\\
  \hline
  \multirow{1}{40pt} {Cellular networks}  & Power allocation and experience sharing & \cite{Wang:2011:ECS:2093256.2093311} & Coordinated decision-making & DVF-based learning & 
  N/A \\
  \hline
\end{tabular}
  \label{tab_mas_learning}
\end{table*}

The applications of the DVF-based learning mechanism in  wireless networks can be found in 
\cite{4496881, 5512673, Wang:2011:ECS:2093256.2093311}. In \cite{4496881}, DVF-based learning is used in an ad-hoc sensor network to coordinate the sensing and hibernation 
operation as the state of the grid-point coverage changes. To encourage the sensor node with a larger coverage area to perform the sensing operation, the individual reward is 
designed as a function of the number of the covered grid points.
It is shown that DVF-based learning outperforms the independent learner-based learning algorithm, especially under the condition of high sensor node densities. In \cite{5512673}, 
a learning algorithm based on the exchange of both the instantaneous reward and the estimated local state-value is proposed for the joint power control and relay selection in 
a distributed cooperative network. The proposed learning scheme is featured by weighting over both the instantaneous reward and the estimated local state-value that are shared by
the neighbor nodes, and thus is called learning with the Distributed Reward and Value (DRV) function. By extending (\ref{eq_q_DVF}), the rule of learning with DRV can be 
expressed as follows:
\begin{eqnarray}
 \label{eq_q_DRV}
 \begin{array}{ll}
 Q_i^{t+1}(s_i,a_i)\!\leftarrow\!(1-\alpha_{t})Q_i^t(s_i,a_i)+\\
 \alpha_{t}\left(\sum\limits_{j\in\mathcal{N}(i)}w'_i(j)u_j^t(s_j,a_j)+\beta\sum\limits_{j\in\mathcal{N}(i)}w_i(j)V_j(s_j)\right),
 \end{array}
\end{eqnarray}
in which $w'_i(j)$ and $w_i(j)$ are the weight of node $i$ given to its neighbor $j$'s instantaneous reward and estimated state value, respectively. With the learning scheme given 
in (\ref{eq_q_DRV}), each node in the network maintains a vector of both the channel/buffer state of its direct link and the channel/buffer state of its cooperative link. It is 
shown in \cite{5512673} that learning based on sharing both the instantaneous rewards and the local state values can achieve a better power efficiency than that using only 
the local reward or the local state value information. In \cite{Wang:2011:ECS:2093256.2093311}, the DVF-based learning scheme is adopted in a real-time multimedia cellular network to 
adapt the power allocation of interfering links. In addition to coordinating the individual links, the Q-value updating mechanism (\ref{eq_q_DVF}) is also used to 
improve the convergence of the newly adopted links in the network.

In Table \ref{tab_mas_learning}, we categorize the works discussed in this subsection according to their respective applications. For applications of multi-agent 
independent-learning schemes in wireless networks, convergence of learning remains an open issue in most of the existing studies. Compared with the SAS-based learning 
algorithms, adopting independent learning schemes requires more attention for any specific networking optimization problem.

\subsection{Experience Sharing Based on Distributed Learning}
\label{subsec_app_docition}
Apart from improving the expected network performance with shared information in the form of structured reward/state-value functions (e.g., using the social reward and the DVF/DRV 
functions), another consideration in MAS-based learning is whether information sharing can also help the individual learning agents to speed up their learning processes.
To answer this question, it is necessary to investigate into the homogeneity of the distributed learning processes so that we can check whether one learning process may be able 
to benefit from the ``shared experience'' offered by another learning process, and furthermore, in what form such a ``shared experience'' would be.

\begin{figure}[!t]
  \centering
  \includegraphics[width=0.48\textwidth]{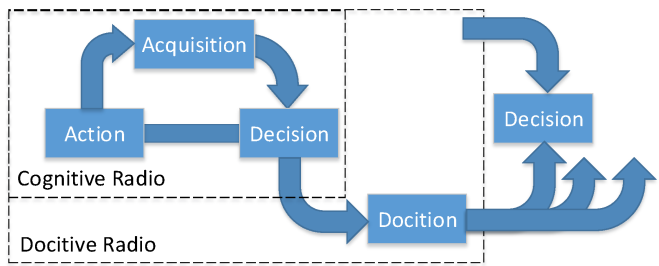}
  \caption{Docitive cycle which extends the cognitive cycle by cooperative teaching (adapted from \cite{Giupponi:2010:CDT:1864807.1864871}).}
  \label{fig_docition}
\end{figure}

We call a group of distributed learning processes homogeneous when the distributed learning agents apply the same learning method with an evolution determined by exactly the same stochastic 
process. In the framework of homogeneous learning processes, it is possible for individual agents to share their private experience (e.g., strategies, estimated Q-values) 
with the other agent in order to accelerate the learning process and improve the performance. Recently, the possibility of applying the teacher-pupil paradigm in human cognition 
to solve the wireless networking problems has been discussed in a series of studies \cite{5547921, 5577685, Giupponi:2010:CDT:1864807.1864871, 5683552}.
In these pioneering studies, the paradigm of ``docitive network'' was proposed based on the extension of distributed cognitive networks (Figure \ref{fig_docition}). In the 
framework of docitive networks, ``docition'' (teaching) is performed by a more experienced network agent to accelerate the learning process of the other agents. Depending 
on the degree of docition among the wireless devices, the teaching-learning process can be distinguished into 3 categories \cite{5547921}:
\begin{itemize}
 \item \emph{Startup docition}: each wireless node learns independently. When a new node joins the network, instead of learning from zero experience, it learns the 
 policies from docitive nodes which have already acquired a certain level of expertise on strategy selection.
 \item \emph{Adaptive docition}: the nodes exchange information about the performance of their learning processes. The docitive nodes share policies and the learning nodes learn 
 from  the expert neighbors which have the best performance. 
 \item \emph{Perfect docition}: each node in the network is able to observe the joint action and all individual rewards. Based on the observation, every docitive node models its 
 interaction with the rest of the network as a complete centralized MDP separately and selects its individual actions.
\end{itemize}

The basic prerequisite for implementing docition in any networking problems is that the individual learning processes can be modeled as parallel, homogeneous MDPs, through 
which imitating the strategies of the docitive nodes by the learning nodes will not influence the policies of the docitive nodes. However, empirical studies have 
shown that relaxing such a constraint in the situation of a noncooperative game-like scenario may also help improve the performance of the learning nodes \cite{5577685, 5683552, 
6662729}. In \cite{5577685}, the distributed downlink power allocation problem in an IEEE 802.22 WRAN (underlay to the TV-Broadcasting bandwidth) is studied. An aggregated 
interference model from the SUs to the PU is considered. The channel state experienced by the individual SUs is defined by a binary state according to hard thresholding on the aggregated 
interference, which is similar to (\ref{eq_state_thresholding}). Each secondary BS ignores the impact of the other BSs on the channel state and adopts a standard independent 
Q-learning scheme to learn its own power selection strategy. The docition process is based on exchanging the Q-tables among the neighbor secondary BSs. In this case, the learning
nodes perform either the startup docition or the adaptive docition periodically by adopting the Q-tables of the expert nodes with the best performance. The simulations in 
\cite{5577685} show that the docitive paradigm significantly speeds up the learning process with respect to the case of independent learners. A similar approach is adopted in 
\cite{5683552, 6662729}, which study the power allocation problem in self-organized heterogeneous networks with femotocells. In these studies, a cross-tier interference model is 
adopted in a manner similar to (\ref{eq_generic_SINR}), while strategy coupling among the femto links is also ignored by individual learners. Again, here docition is performed 
through exchanging the Q-tables among the neighbor nodes. In \cite{5683552}, the similarity metric to measure of the correlation between the femto BS strategy and the aggregated 
interference to the macrocell is introduced as a user-defined gradient. The proposed metric measures the similarity of the policies between two neighbor 
nodes. With the similarity metric, the learning 
nodes can not only adopt the Q-tables from the neighbor nodes with the best performance, but also take into account the degree of the similarity between their own action-state 
correlation and their neighbors'.

While it is relatively easy to implement docition in the framework of independent Q-learning based on the model of parallel, homogeneous MDPs, it generally remains an open issue 
to estimate the similarity of the policies between two neighbor learners when the learning processes are heterogeneous. Especially, in the scenario of strategy coupling and 
interest conflict, imitating the strategies or the Q-tables of the adversary neighbor node with the best performance may result in strategy oscillation. Such a situation can be
illustrated by revisiting the power allocation problem defined by (\ref{eq_generic_SINR}). In the simplified situation of mutual interference with only two femto BSs, 
increasing the transmit power of one BS will result in the performance deterioration for the other BS, because the interference to the other BS is also increased. Consider 
the case that the BS with the smaller transmit power decides to adopt the strategy of its rival BS by increasing its transmit power. If independent Q-learning is used by both BSs to 
learn their power selection strategies, the other BS will soon discover that it will benefit from increasing its current transmit power too. This creates an ``arm race'' 
situation in which 
each BS begins to increase its transmit power in turn until both the BSs reach their maximum power level, which is a typical situation of the prisoner's dilemma in
noncooperative games. Such an unwanted situation can be avoided if both BSs treat the power allocation process as a noncooperative game and adopt the learning methods in games 
such as Fictitious Play (FP) and best response without any docition procedure\footnote{Studies adopting the same mutual interference model as in (\ref{eq_generic_SINR}) within 
the framework of repeated games can be found in \cite{983324}. In \cite{983324}, the best response without docition ensures the convergence to the Pareto dominant equilibria.}.
As a result, in works such as
\cite{6215470} the docitive paradigm and the game-based learning paradigm are considered two controversial frameworks for strategy learning. However, it is worth noting that 
with emerging techniques such as transfer learning \cite{Taylor:2009:TLR:1577069.1755839} and experience-weighted attraction learning \cite{ECTA:ECTA054}, incorporating the 
teaching process in the game-based framework of learning is no longer impossible. For this part, we will leave the discussion of more details to Section \ref{sec_open_issue}.

\section{Applications of Game-Based Learning in Cognitive Wireless Networks}
\label{sec_app_game}
Generally, there are the limitations of the distributed learning mechanisms (e.g., the algorithms reviewed in \ref{sec_app_dist}) that post the necessity of introducing the game-based 
learning mechanisms in CRNs. By modeling distributed network control problems as games, it is possible to better address the problems raised by device interactions in the networks.
Also, it is possible to design learning schemes that theoretically guarantee the convergence of the individual strategies to a fixed point or equilibrium, while such convergence
is usually not guaranteed by the distributed learning mechanisms. In this section, we consider the repeated games as the special cases of SGs and introduce the applications of 
learning algorithms based on repeated games and SGs separately.  We will organize the learning algorithms based on the three  game property dimensions discussed in Section 
\ref{subsec_game}. Our major focus will be (a) the rules in each learning scheme; (b) the conditions and properties of the games with which a specific learning scheme may 
converge; and (c) the degree of information exchange required by each learning scheme to achieve convergence.
The new notations used by this section can be found in Table \ref{notation_sec_app_game}.

\begin{table}[t]
  \caption{Summary of the main notations in Section \ref{sec_app_game}}
  \centering
 \begin{tabular}{ p{50pt}| p{170pt}}
 \hline
 Symbol & Meaning\\
  \hline
  $\kappa_i^{t}(a_{-i})$ & The statistic of player $i$ for its opponent's actions\\
  \hline
  $\theta_i^{t}(a_{-i})$  & The estimated probability of player $i$ for its opponent to play action $a_{-i}$\\
  \hline
  $\textrm{BR}(\cdot)$ & The set of best-response actions\\
  \hline
  $\nu$  & The learning parameter for perturbation in SFP \\
  \hline  
  $\alpha_t$ & The learning factor in SFP\\
  \hline
  $q_i^t(a)$ & The estimated frequency of local actions\\
  \hline
  $\epsilon_t$ & The time-varying step size in GP\\
  \hline
  $\mu$, $\mu_t$ & The learning parameters in GP, LA and no-external-regret learning\\
  \hline
  $p_i$ & The probability of accessing a channel in a random medium access game\\
  \hline
  $R_m$ & The transmit rate over channel $m$\\
  \hline
  $R_i(\pi_i,a_i|a_{-i})$ & The regret of player $i$ for playing strategy $\pi$ instead of playing $a_i$\\
  \hline
  $R^t_i(a^t_i,a'_i)$ & The regret for agent $i$ not playing $a'_i$ at time $t$\\
  \hline
  $c_i^t(\mathbf{s},a_{-i})$ & The conjecture of opponent policy $\pi_{-i}(\mathbf{s})$ by agent $i$ at time $t$\\
  \hline
  $\overline{\pi}_i^t(\mathbf{s},a_i)$ & The reference point for conjecture learning\\
  \hline
\end{tabular}
  \label{notation_sec_app_game}
\end{table}
\subsection{Applications of Learning in the Context of Repeated Games}
\label{subsec_app_game}
Repeated games play an important role in problem formulation for distributed network control. When the network evolution is not subject to a stochastic environment, most of 
the network control problems that requires considering the interactions among distributed devices can be formulated as a repeated game instead of an MAMDP. In contrast to the 
MDP-based learning mechanisms that heavily depend on value iteration, policy iteration now plays an important role in deriving the learning rules for repeated games. In the 
context of repeated games, model-free learning emphasizes more on the situation of information locality. This is because in many practical scenarios, the information of the local utilities,
actions or strategies of one network device may not be available to the other devices due to either the concern of privacy or the lack of enough resources for information 
exchange. In this subsection, we will organize our survey on the applications of learning in repeated games according to the prototypical learning schemes that they are 
based on. These prototypical learning schemes include (i) fictitious play, (ii) gradient play, (iii) learning automata and (iv) no-regret learning.

\subsubsection{Fictitious Play and Stochastic Fictitious Play}
\label{subsubsec_game_FP}
The basic prerequisite of the standard FP is that the agents are willing to reveal their (discrete) action information to the others after each round of play, so they can 
track the frequency of action selection by the other agents \cite{fudenberg1998theory}. With FP, agent $i$ assesses the distribution of its opponent's actions at 
round $t$ as follows:
\begin{equation}
 \label{eq_fp_frequency}
 \kappa^t_i(a_{-i})=\kappa^{t-1}_i(a_{-i})+I(a^{t-1}_{-i},a_{-i}).
\end{equation}
Agent $i$ estimates the probability for the opponent agents to play the 
joint action $a_{-i}$ at round $t$ as:
\begin{equation}
 \label{eq_fp_probability}
 \theta_i^t(a_{-i})=\frac{\kappa_i^t(a_{-i})}{\sum_{a'_{-i}\in\mathcal{A}_{-i}}\kappa_i^t(a'_{-i})}.
\end{equation}
In this sense, FP is sometimes considered as a model-based learning mechanism since with (\ref{eq_fp_probability}) it tries to build the model of the opponents' joint policy from 
accumulated experience.
However, compared with other model-based, non-learning mechanisms such as dynamic programming for MDPs, FP does not need any a-priori knowledge of the system or other players.
Based on (\ref{eq_fp_probability}), FP is defined as any rule that assigns the best response to agent $i$ given its current estimation of the opponent policy $\theta_i^t(a_{-i})$. 
Usually, such an operation is represented by $a_i^t(a_{-i})\in\textrm{BR}_i(\theta_i^t(a_{-i}))$, where the operator $\textrm{BR}(\cdot)$ derives the best-response 
action set. Typically, $\textrm{BR}_i$ can be derived by maximizing the estimated expected payoff of agent $i$: $\textrm{BR}_i(\theta_i^t(a_{-i}))=\arg\max_{a\in\mathcal{A}_i}
E[u_i(a,\theta^t_i(a_{-i}))]$. The convergence property of FP in a general repeated game is given by Theorem \ref{th_fp_convergence} \cite{fudenberg1998theory}.
\begin{theorem}[Convergence of FP]
 \label{th_fp_convergence}
  1) Strict NE\footnote{This is equivalent to the condition when the best-response payoffs in the NE are strictly greater than the other possible payoffs for all the 
  agents.} are the absorbing state for the process of fictitious play. 2) Any pure-strategy steady state of fictitious play must be an NE.
\end{theorem}

Theorem \ref{th_fp_convergence} gives the sufficient condition for FP to converge to an NE. Thereby, the convergence of FP-based learning is guaranteed in any repeated games that 
possess at least one pure-strategy NE. According to Theorem \ref{th_fp_convergence}, a typical way of checking the convergence condition for FP in a game is to check if the game 
possesses certain properties (such as being potential or S-modular \cite{han2012game}) that guarantee the existence of a pure-strategy NE.

As long as the learning agents are able to observe the actions of the rival agents or afford the overhead for action information exchange, FP can be employed as the basic 
solution for many resource management games in wireless networks. In \cite{4518957}, an FP-based multi-agent learning algorithm is employed by the secondary nodes in an 
ad-hoc DSA network to learn the strategies for forwarding delay-sensitive packets. In \cite{4518957} the condition of channel availability is characterized by the matrix of
spectrum opportunity, and the condition of channel contention is characterized by the interference matrix from both the PUs and the rival SUs.
With the learning scheme proposed in \cite{4518957}, each SU needs to collect the information 
about the spectrum opportunity matrix locally, and establish its local interference matrix according to the action information collected from its neighbors. Then, every SU tracks the 
frequency of action selection by its neighbors according to a modified version of (\ref{eq_fp_frequency}) with a discount factor $\kappa^{t-1}_i$. Each SU also needs to 
determine a subset of feasible actions that do not interfere with higher priority traffic. This is done through estimating the expected interference based on the 
policy estimation model in (\ref{eq_fp_probability}). The local deterministic best response is calculated based on minimizing the expected effective transmission time over the 
candidate links. 

\begin{figure}[!t]
  \centering
  \includegraphics[width=0.45\textwidth]{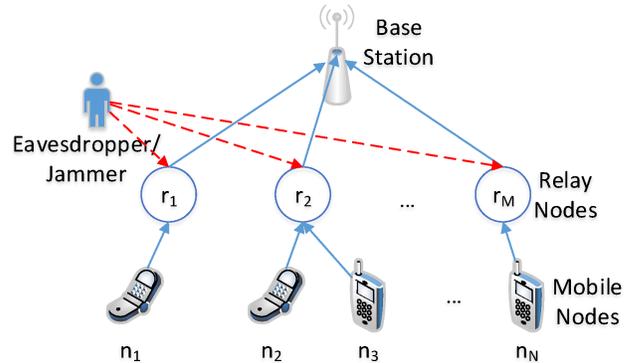}
  \caption{A network consisting of $M$ one-hop relays and $N$ wireless users that is subject to eavesdropping/jamming from one active malicious node \cite{6127463}.}
  \label{fig_eavesdropping}
\end{figure}

Another example of FP can be found in \cite{6127463}, which applies FP to obtain a defense mechanism against eavesdropping and jamming attacks in the uplink of a cellular network
consisting of multiple relays (Figure \ref{fig_eavesdropping}). 
In the defense-attack game, the normal/malicious nodes are assumed to be able to observe the actions of other nodes, so they can use the models in (\ref{eq_fp_frequency}) and 
(\ref{eq_fp_probability}) to estimate the other nodes' policies. Instead of directly obtaining a deterministic strategy based on the local best response, each normal node 
updates its mixed strategy at time slot $t$ as follows \cite{1406126}:
\begin{equation}
 \label{eq_modified_mixed_probability}
 \pi_i^t(m)=\pi_i^{t-1}(m)+\frac{1}{t}(I(a^t_i,m)-\pi_i^{t-1}(m)), 
\end{equation}
in which $m$ is the index of the candidate relays. The malicious node adopts a similar policy-updating rule based on its own action set for attacking. The actions of each node
at round $t$ are selected from the best response based on the expected private utility with the locally estimated policy vector $(\pi^t_i,\theta^t_i)$. The same learning rule as 
in \cite{6127463} can be found in \cite{5462017}, which uses the local policy updating rule in (\ref{eq_modified_mixed_probability}) to learn the strategy in a continuous 
strategy space for power allocation. In \cite{5462017}, such a learning scheme is referred to as the \emph{best response dynamics} of the power allocation game, and is proved to 
be able to converge to the $\epsilon$-equilibria. Such a learning rule is also adopted in \cite{5426293}, which formulates a hierarchical network formation game for nodes 
in a multi-hop wireless network to select relays. In \cite{5426293} the relay selection game is decomposed into multi-layers and solved using a backward induction method 
from the sink to the source. The learning scheme defined by (\ref{eq_fp_frequency})-(\ref{eq_modified_mixed_probability}) is applied to each layer-game and the mixed strategies 
are obtained from the local best responses.

With the standard FP, local actions are updated based on the best responses, which are generally of pure strategies. As pointed out by \cite{fudenberg1998theory}, one drawback of
such an FP-based learning scheme lies in the discontinuity of agent behaviors, for a small change in the opponent-policy estimation may result in an abrupt local-behavior change.
Due to this, a Smoothed-FP (SFP) procedure was proposed through searching the best response with a modified local objective function that is perturbed by a 
differentiable, strictly concave function. Assume that the best response is obtained through maximizing a payoff function $u_i(\pi_i,\pi_{-i})$. Then the operation for obtaining 
the smoothed best response $\textrm{BR}(\cdot)$ can be used to replace the original best response $\arg\max{u_i(\pi_i,\pi_{-i})}$:
\begin{equation}
 \label{eq_stoch_br_obj}
 \textrm{BR}(\pi_{-i})=\arg\max_{\pi_i}\left\{u_i(\pi_i,\pi_{-i})+\nu\eta_i(\pi_i)\right\},
\end{equation}
in which the perturbation function $\eta_i$ is typically given as the entropy function of $\pi_i$:
\begin{equation}
 \label{eq_stoch_br_per}
 \eta_i(\pi_i)=\sum_{a_i\in\mathcal{A}_i}-\pi_i(a_i)\log\pi_i(a_i).
\end{equation}
Problem (\ref{eq_stoch_br_obj}) with (\ref{eq_stoch_br_per}) can be explicitly solved as:
\begin{equation}
 \label{eq_stoch_br}
 \textrm{BR}(\pi_{-i})=\frac{\exp(({1}/{\nu})u_i(a_i,\pi_{-i}))}{\sum_{a\in\mathcal{A}_i}(\exp({1}/{\nu})u_i(a,\pi_{-i}))},
\end{equation}
in which $\nu$ is the weight of the perturbation term that controls the strategy exploration rate. It has been proved that for any average-reward repeated game, we can always 
find the $\nu$ that makes the payoff of agent $n$ under $\textrm{BR}(\pi_{-n})$ to be sufficiently near the real best-response payoff (Proposition 4.5 of 
\cite{fudenberg1998theory}). The SFP-based learning scheme is also known as the stochastic FP. Unlike standard FP, in SFP it is not necessary to observe the opponents' actions or 
even know the structure of the local utility functions. Instead, the expected payoff ${u}_i^t(a_i,\pi_{-i})$ in (\ref{eq_stoch_br}) is estimated based on local information 
as follows:
\begin{equation}
 \label{eq_stoch_estimate_u}
 \tilde{u}_i^t(a_i)\!=\!\frac{1}{\kappa_{i}^{t\!-\!1}}I(a_i^t,a_i)\left({u}_i^t(a_i)\!-\!\tilde{u}_i^{t\!-\!1}(a_i)\right)\!+\!\tilde{u}_i^{t\!-\!1}(a_i),
\end{equation}
in which $\kappa_{n}^{t}$ and $I(a_n^t,a_n)$ follow the same definitions as in (\ref{eq_fp_frequency}), and $\tilde{u}_n^t(a_n)$ is the estimate of the expected utility 
${u}_i^t(a_i,\pi_{-i})$. The local mixed policy is usually updated in the following form:
\begin{equation}
 \label{eq_stoch_estimate_policy}
 \pi_i^t(m)=\pi_i^{t-1}(m)+\alpha_t\left(\textrm{BR}(\tilde{u}_i^t(a_i))-\pi_i^{t-1}\right), 
\end{equation}
in which $\textrm{BR}(\tilde{u}_i^t(a_i))$ is calculated based on (\ref{eq_stoch_br}) with the payoff estimated by (\ref{eq_stoch_estimate_u}) and $\alpha_t$ is a learning 
factor.

It is worth noting that with both value iteration in (\ref{eq_stoch_estimate_u}) and policy iteration in (\ref{eq_stoch_estimate_policy}), SFP is usually considered as a typical
form of CODIPAS-RL methods (see the example in \cite{6117774}).
Generally, the convergence conditions of SFP are based on the analysis of Lyapunov stability of the corresponding perturbed best response dynamic \cite{ECTA:ECTA376}.
A summary of these conditions for different types of games is given as follows\footnote{About the definitions of ESS, rest point and supermodular game, please refer to 
\cite{han2012game} for more details.}:
\begin{theorem}[Convergence of FP \cite{ECTA:ECTA376}]
 \label{thm_SFP_convergence}
 Consider SFP defined by (\ref{eq_stoch_br_obj})-(\ref{eq_stoch_estimate_policy}) starting from an arbitrary strategy in game $G$.
 \begin{itemize}
  \item [(i)] If $G$ is a two-player symmetric game with an interior Evolutionary Stable Strategy (ESS) or a two-player zero-sum game, then SFP converges with probability one 
  to an NE.
  \item [(ii)] If $G$ is an $N$-player potential game, then SFP converges in a subset of the rest points of the perturbed best response dynamic. If all the rest points of 
  the perturbed best response dynamic are hyperbolic and two-order continuous, SFP converges to an NE with probability one.
  \item [(iii)] If $G$ is an $N$-player supermodular game, then SFP converges almost surely to a rest point of the perturbed best response dynamic. In particular, if
  the rest point is unique, then SFP converges to the NE with probability one.
 \end{itemize}
\end{theorem}

With the property of requiring no information exchange, SFP is considered an important tool in self-organized learning for resource allocation games. In \cite{4278411}, 
SFP is applied to the power control game in wireless ad-hoc networks. According to Theorem
\ref{thm_SFP_convergence}, SFP is guaranteed to converge to a stationary point (with a non-zero probability to an NE) for a supermodular/potential game. In order 
to take advantage of such a property, a supermodular utility function is designed for each node in \cite{4278411}, and the convergence with SFP is thus guaranteed. However, since
the utility function in \cite{4278411} is monotonically decreasing, the learning scheme will finally converge to the unique NE of that game, which corresponds to all users 
transmitting with zero power. This problem of utility function design is addressed in \cite{6990372} by studying the power allocation problem in a small-cell network through a 
non-trivial Stackelberg game \cite{han2012game}. This game design is intended to balance the femtocell power efficiency and interference control in the macrocell. The 
supermodularity property is retained for the femto link utility, and the SFP-based scheme give in (\ref{eq_stoch_br}-\ref{eq_stoch_estimate_policy}) is applied to the follower game 
among the femtocells. The same learning mechanism is adopted in \cite{6542770}, which considers the power allocation in the femtocells as a common-payoff game (thus a potential 
game). With the assumption of the common-payoff game, it is proved in \cite{6542770} that the $\epsilon$-equilibrium is guaranteed to be reached in the potential game by the 
SFP-based learning algorithm.

\subsubsection{Gradient Play}
\label{subsubsec_game_GP}
Compared with FP, Gradient Play (GP) adjusts the strategy of one agent based on the gradient ascent dynamics instead of directly jumping to the best response based on the 
empirical frequencies of the opponent agents' action selection. Therefore, GP can be viewed as a ``better response'' algorithm. Mathematically, following the learning scheme of the 
standard GP, each agent in the repeated game updates its strategy on selecting $a_i$ according to \cite{1406126}:
\begin{equation}
 \label{eq_GP}
 \pi^{t+1}_i(a_{i})=\left[q^{t}_i(a_{i})+\epsilon_t(\nabla_{\pi_i}u_i(\pi_i,\theta^t_{i}(a_{-i})))\right]^{{\Pi}_i},
\end{equation}
where $\epsilon_t$ is the time-varying step size, $[\cdot]^{\Pi_i}$ defines the projection onto the strategy space $\Pi_i$ of agent $i$, $\theta^t_{i}(a_{-i})$ is the 
estimated opponent-action frequency, which can be derived following (\ref{eq_fp_probability}), and $q^{t}_i(a_{i})$ is the estimated local-action frequency, which can be derived 
in the same manner as (\ref{eq_modified_mixed_probability}):
\begin{equation}
 \label{eq_dagp_q}
 q_i^{t+1}(a_i)=q_i^t(a_i)+\frac{1}{t+1}(I(a_i^t,a_i)-q_i^t(a_i)),
\end{equation}
where action $a_i^t$ is generated as random outcomes of the evolving strategies $q_i^t$. Following (\ref{eq_GP}) and (\ref{eq_dagp_q}), the strategy of 
each agent is a 
(projected) combination of its own empirical action frequency and a gradient step based on the estimated opponents' action frequency. According to \cite{1406126, 1430262}, 
GP in continuous games is guaranteed to converge within a distance of order of $\epsilon_t$ of the NE of the game, if the NE is a strict one. However, GP cannot 
converge to a completely-mixed NE of the game (see Lemma 4.1 of \cite{1406126}). Due to such a limitation on convergence condition, the basic form GP in (\ref{eq_GP}) is rarely used 
directly in the solution to networking problems.

As an improvement to the basic form GP, Derivative-Action GP (DAGP) is developed in 
\cite{1406126}. By introducing parameter $v^t_i(a_i)$ to approximate the first-order derivative of $q_i$, the updating mechanism of DAGP is defined as follows \cite{1430262}:
\begin{equation}
 \label{eq_dagp_r}
 v_i^{t+1}(a_i)=v_i^t(a_i)+\frac{\mu_t}{t+1}(q_i^t(a_i)-v_i^t(a_i)),
\end{equation}
\begin{equation}
 \label{eq_DAGP}
 \begin{array}{ll}
 \pi^{t+1}_i(a_{i})=\big[q^{t}_i(a_{i})+\epsilon_t(\nabla_{\pi_i}u_i(\pi_i,\theta^t_{i}(a_{-i})))+\\
 \qquad\qquad\quad\quad\mu_t(q_i^t(a_i)-v_i^t(a_i))\big]^{\Pi_i},
 \end{array}
\end{equation}
where $q_i^t$ is updated following (\ref{eq_dagp_q}), $[\cdot]^{\Pi_i}$, $\epsilon_t$ and $\theta^t_{i}$ are obtained in the same way as in (\ref{eq_GP}), and $\mu_t$ 
is a large factor satisfying $\mu_t>0$. According to \cite{1406126, 1430262}, for large $\mu_t>0$, if $\epsilon$ satisfies certain conditions (see Theorem 4.2 of 
\cite{1406126} and Theorem 3.1 and Theorem 3.3 in \cite{1430262} for more details), the strategy $\pi_i^t$ is asymptotically locally stable and converges to the NE with a 
non-zero probability.

GP and DAGP not only require the agents to be able to track the frequency of both the local actions and the opponent actions, but also require that the structure of local utility 
functions is known to each agent. Compared with FP and SFP, the most important feature of the GP-based learning algorithms is that the updating mechanism can be easily extended 
to the cases of continuous games. In \cite{4604737}, standard GP is applied to the continuous, random medium access game, in which a set of wireless nodes learn 
to play the random access strategies $p_i$ ($0\le p_i\le 1$) after observing the vector of channel contention signal $\mathbf{q}_i$. Instead of 
directly adapting to the contention signal $\mathbf{q}_i$, each wireless node introduces a price function $C_i(\mathbf{q}_i)$ to adjust its local net payoff with the original 
utility function $U_i(p_i)$ as $u_i(\mathbf{p})=U_i(p_i)-p_iC_i(\mathbf{q}_i)$. In \cite{4604737}, the random access game is proved to have a unique nontrivial NE (namely,
$\nabla_{p_i}u(p^*_i, p^*_{-i})=0$ at the NE $(p^*_i, p^*_{-i})$), and that the standard GP converges geometrically to the nontrivial NE if a certain condition is satisfied with 
the step size $\epsilon^t$ in (\ref{eq_GP}). The application of standard GP can also be found in the power control game of a multi-cell CDMA network with dynamic handoffs 
between cells \cite{Alpcan:2004:HSM:1032151.1032156}. After introducing a pricing mechanism with the cost function based on the local power consumption, the game formulation in 
\cite{Alpcan:2004:HSM:1032151.1032156} adopts a payoff function that is twice continuously differentiable, non-decreasing and strictly convex. It is proved in 
\cite{Alpcan:2004:HSM:1032151.1032156} that standard GP is able to exponentially converge to the smallest convex set which contains all the possible NE of the power control
game, if the spreading factor of the CDMA system satisfies certain conditions\footnote{``Exponential convergence'' is used to 
describe the property of learning when asymptotically converging to the convex set. If $\Vert\pi_i^t-\pi^*_i\Vert=O(\mu^t)$ for some $\mu<1$, we say that the learning process
achieves exponential convergence.}.

One typical example of applying the DAGP-based learning to networking problems can be found in \cite{4290040}, which formulates the interference coordination problem in a 
multi-link MIMO system as a noncooperative game. In the game, the covariance matrix of the signal of each link is considered as the local strategy and is drawn from a common, 
continuous strategy space. The matrix form of (\ref{eq_DAGP}) is adopted and guaranteed to converge to a unique NE of the game, if the covariance matrix of the total interference
and noise at the receiver of each link satisfies a certain condition.

\subsubsection{Learning Automata}
\label{subsubsec_game_LA}
As introduced in Section \ref{sec_single_agent}, LA is featured by the process of action selection based on policy iteration using only local information. For non-game-based
wireless networking
problems, (distributed) LA has been shown to be efficient in the scenarios which can be formulated to be of single state and controlled by a single active decision-making 
entity at one time instance. Successful applications of LA in these scenarios can be found in the works such as multipath on-demand multicast routing in CRNs \cite{7265222} and 
multicast routing in mobile ad-hoc networks \cite{AkbariTorkestani2010721}.
When it comes to the more complicated framework of network control games, most of the LA-based learning schemes are employed to obtain NE policies. As a special case of the 
general LA updating rule 
(\ref{eq_general_la}), $L_{R-I}$ learning has been widely applied to network control problems due to its simplicity and convergence property. By abusing the notations in 
(\ref{eq_general_la}), the rules of $L_{R-I}$ learning can be expressed as (\ref{eq_la_lri}):
\begin{eqnarray}
 \label{eq_la_lri}
 \pi_i^{t+1}(a_i)=\left\{
 \begin{array}{ll}
   \pi_i^t(a^t_i)+\mu\tilde{r}_i^t(1-\pi_i^t(a^t_i)), \quad \textrm{ if } a_i^t=a_i,\\
   \pi_i^t(a_i)-\mu\tilde{r}_i^t\pi_i^t(a_i),\qquad\quad\;\; \textrm{ if } a_i^t\ne a_i,
 \end{array}\right.
\end{eqnarray}
where $\mu$ ($0<\mu<1$) is a learning parameter. The convergence property to the NE for the learning mechanism in (\ref{eq_la_lri}) in a general noncooperative game 
has been proved in \cite{293490}:
\begin{theorem}
 \label{thm_la_ne}
 In a repeated game $G\!=\!{\langle}\mathcal{N}, {\mathcal{A}}\!=\!\times\!\mathcal{A}_n,\{0\le\tilde{r}_n\le 1\}_{n\in\mathcal{N}}{\rangle}$, with each agent employing $L_{R-I}$ 
 learning, the following statements are true if $\mu$ in (\ref{eq_la_lri}) is sufficiently small:
 \begin{itemize}
  \item all stationary points that are not NE are unstable, and
  \item all strict NE in pure strategies are asymptotically stable.
 \end{itemize}
\end{theorem}

However, no uniform expression is provided in the literature to obtain the normalized environment response function $\tilde{r}_i^t$ in (\ref{eq_la_lri}). For 
example, in \cite{6151775}, standard $L_{R-I}$ learning is adopted to manage the opportunistic spectrum access by $N$ SUs over $M$ primary channels with a fixed 
transmit rate $R_m$ on channel $m$. In this case, the normalized random reward $\tilde{r}_i^t$ is obtained as follows:
\begin{equation}
 \label{ea_normalized_reward}
 \tilde{r}_m^t={u}_m^t/(\max_{n} R_n),
\end{equation}
where $u_m^t$ is the instantaneous reward of SU $m$ after considering the PU activities and the channel contention with its rival nodes. The opportunistic spectrum access game is
further modeled as an exact potential game. Therefore, at least one pure-strategy NE exists for the game \cite{han2012game}. According to Theorem \ref{thm_la_ne}, $L_{R-I}$ 
learning ensures the convergence to the pure-strategy NE in the opportunistic spectrum access game. Apart from \cite{6151775}, the standard $L_{R-I}$ 
learning scheme can be found as a frequent solution to the problems whenever the convergence property of Theorem \ref{thm_la_ne} is satisfied and the existence of a pure-strategy 
NE can be proved. The applications of the standard $L_{R-I}$ learning scheme range 
from relay-selection in the cooperative network \cite{6877699} to the CSMA-based DSA management \cite{6942209} and the MIMO-based DSA management \cite{5502112} in the CRNs. 

In contrast to the aforementioned works, the variation of the standard $L_{R-I}$ learning mechanism using a different strategy-updating rule can also be found in the studies such as 
\cite{4479503}. In \cite{4479503}, a discrete power control problem in a CDMA-like cellular network with mutual interference is modeled as a repeated noncooperative game. In 
the power control game, each node only knows its local payoff measured as the power efficiency. The modified linear-reward-inaction updating rule in \cite{4479503} is defined as 
follows:
\begin{eqnarray}
 \label{eq_la_lri_variation1}
 \pi_i^{t+1}(a_i)=\left\{
 \begin{array}{ll}
   \pi_i^t(a^t_i)-\mu\tilde{r}_i^t\pi_i^t(a^t_i), \qquad\quad\ \textrm{ if } a_i^t\ne a_i,\\
   \pi_i^t(a_i)+\mu\tilde{r}_i^t\sum\limits_{a\ne a_i}\pi_i^t(a),\quad\;\; \textrm{ if } a_i^t=a_i.
 \end{array}\right.
\end{eqnarray}
Let $u_i^t$ denote the utility of node $i$ by choosing a discrete power level $a^t_i$ for transmission at time $t$. Then, the normalized utility feedback $\tilde{r}_i^t$ is 
obtained as follows:
\begin{equation}
 \label{ea_normalized_reward_variation1}
 \tilde{r}_i^t=\frac{u_i^t-\min_{i}\{u_i\}}{\max_i\{u_i\}-\min_{i}\{u_i\}}.
\end{equation}
The major difference between (\ref{eq_la_lri_variation1}) and (\ref{eq_la_lri}) lies in the way of updating the probability of choosing an action when the action results in a new 
reward. Under this learning algorithm, the evolution of the power selection becomes a Markov process. 
Following the same approach of proving the convergence property based on Ordinary Differential Equation (ODE) analysis and Lyapunov's stability theorem 
as in \cite{293490}, it is proved in \cite{4479503} that the LA-based learning scheme in (\ref{eq_la_lri_variation1}) will only converge to the mixed-strategy NE of the 
considered power control game if the learning step $\mu$ is sufficiently small.

In addition to $L_{R-I}$ learning, other learning schemes based on the general LA updating rule in (\ref{eq_general_la}) are also employed for resource allocation 
in the CRNs. In \cite{5072224}, an LA mechanism based on the softmax (Logit) function is applied to learn the $\epsilon$-optimal solution to the 
traffic allocation problem in a multi-hop cognitive wireless mesh network. With the proposed LA mechanism, node $i$'s local action to select link $k$ for transmitting at
the $n$-th possible rate is determined by the softmax function:
\begin{equation}
 \label{ea_la_softmax}
 \pi_{i,k}^n=\frac{\exp({w_{i,k}^n})}{\sum_{m=0}^{N}\exp({w_{i,k}^m})},
\end{equation}
where $N$ denotes the number of possible transmit rates and the intermediate parameter $w_{i,k}^n$ is updated according to the following LA rules:
\begin{eqnarray}
 \label{eq_la_lri_variation2}
 w^n_{i,k}(t\!+\!1)\!=\!\left\{
 \begin{array}{ll}
   w^n_{i,k}(t)+\alpha_t\Xi(t)(1-\frac{\exp({w_{i,k}^n})}{\sum_{s=0}^{N}\exp({w_{i,k}^m})})\\
   +\sqrt{\alpha_t}\xi_{i,k}^n(t), \qquad\quad\quad\textrm{ for } n=j;\\
   w^n_{i,k}(t)+\sqrt{\alpha_t}\xi_{i,k}^n(t), \quad\textrm{for } n\ne j.
 \end{array}\right.
\end{eqnarray}
In (\ref{eq_la_lri_variation2}), $\alpha_t$ ($0\!<\!\alpha_t\!<\!1$) is the learning rate and $\xi_{i,k}^q(t)$ is obtained from a set of i.i.d. random variables with zero mean. 
$\Xi(t)$ is the normalized utility feedback that is provided by the gateway node. In order to ensure the convergence of the learning algorithm 
in (\ref{eq_la_lri_variation2}), the traffic engineering game is modeled as a team game with the identical payoff (hence a potential game). Thus the SUs need to share the 
information on the global, normalized utility feedback $\Xi(t)$ for updating the value of $w^q_{i,k}(t)$. In \cite{5072224}, the value of $\Xi(t)$ is obtained from arbitrarily 
scaling the sum of the local payoff functions down to the range of $[0,1]$. By allowing information exchange and constructing an $N$-person potential game, it is proved in
\cite{5072224} that for sufficiently small values of $\alpha_t$ and the variance of $\xi_{i,k}^q(t)$, the LA mechanism in (\ref{eq_la_lri_variation2}) is guaranteed to achieve 
the $\epsilon$-optimal solution to the traffic engineering problem.

\begin{figure}[!t]
  \centering
  \includegraphics[width=0.42\textwidth]{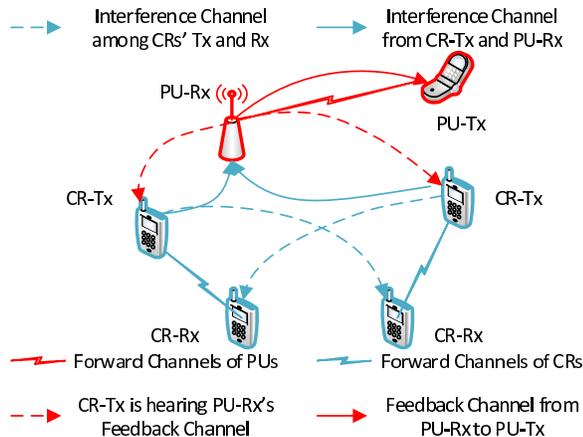}
  \caption{A toy example of power allocation in the multi-user CRN with limited ability of acquiring the strategy information from other CRs \cite{6117762}.}
  \label{fig_cr_power_allocation}
\end{figure}
In \cite{6117762}, Bush-Mosteller LA \cite{1049610} is adopted for learning the NE of the repeated power control game in a CRN with the set of power constraints on the aggregated
interference experienced by each PU (Figure \ref{fig_cr_power_allocation}). Bush-Mosteller learning, also known as the linear reward-penalty LA, can be viewed as a general form 
of $L_{R-I}$ learning \cite{poznyak1997learning}. In \cite{6117762}, the CRN is assumed to be composed of $N$ SUs and $M$ PUs. The wireless channels are assumed to be stationary, 
and the SUs are able to monitor each PU's feedback indicating the sum of interference to each PU receiver. It is also assumed that no SU can observe the strategies of the other 
SUs (see Figure \ref{fig_cr_power_allocation}). Let $U_k(\pi_k,\pi_{-k})$ be the expected utility of SU link $k$ and $W_l(\pi_k,\pi_{-k})$ be the corresponding
expected interference at PU $l$, the constrained game is transformed into an unconstrained game with the help of the Lagrange multipliers. The Lagrange function of
SU $k$ is defined with a regularization term ${\delta}/{2}\left(\Vert{\pi}_k\Vert^2-\Vert{\pmb\lambda}_k\Vert\right)$ as follows:
\begin{eqnarray}
 \label{ea_la_lagrange}
  \begin{array}{ll}
    L^{\delta}_k(\pi_k,\pi_{-k}, {\pmb\lambda}_k)=\!U_k(\pi_k,\pi_{-k})\!-\!\sum_{l\!=\!1}^{M}\lambda_l(W_l(\pi_k,\pi_{-k})\\
    \qquad\qquad\qquad\qquad-\overline{W}_l)-\displaystyle\frac{\delta}{2}\left(\Vert{\pi}_k\Vert^2-\Vert{\pmb\lambda}_k\Vert^2\right),
  \end{array}
\end{eqnarray}
where $\lambda_l$ is the Lagrange multiplier for the constraint from PU $l$, ${\pmb\lambda}_k$ is the vector of $\lambda_l$ and $\overline{W}_l$ is the maximum level of the 
interference to PU $l$. It is shown in \cite{6117762} that finding the equilibrium point of the original constrained power control game is asymptotically equivalent to 
determining the equilibrium point of the unconstrained game with the regularized function given in (\ref{ea_la_lagrange}). The following learning scheme, based on linear
reward-penalty LA, is adopted to update the local policies:
\begin{equation}
 \label{ea_la_Bush_Mosteller}
    \pi_k^{t\!+\!1}\!=\!\pi_k^t\!+\!{\alpha}_k^t[\mathbf{e}_{N_k}(P_k^t)\!-\!\pi_k^t\!+\!\frac{\tilde{r}_k^t(\mathbf{e}^{N_k}\!-\!N_k\mathbf{e}_{N_k}(P_k^t))}{N_k-1}],
\end{equation}
where $P_k^t$ is the power level that SU $k$ chooses at iteration $t$. $\mathbf{e}_{N_k}(P_k^t)$ and $\mathbf{e}^{N_k}$ are defined as follows:
\begin{gather}
 \label{ea_la_unit_vector1}
    \mathbf{e}_{N_k}(P_k^t)=(\underbrace{0,\ldots, 0, 1}_{i_k}, 0, \ldots, 0)^T,\\
 \label{ea_la_unit_vector2}
    \mathbf{e}^{N_k}=(1,\ldots, 1)^T.
\end{gather}
The normalized utility feedback $\tilde{r}_k^t$ is obtained based on the Lagrangian with the expected utility and interference being replaced by the instantaneous payoff and 
interference in (\ref{ea_la_lagrange}). With a user-defined normalization procedure, the value of $\tilde{r}_k^t$ is scaled within the interval $[0,1]$\footnote{For the detailed 
derivation of $\tilde{r}_k^t$, please refer to (31) and (32) in \cite{1049610}.}. The time-varying correction (adaptation) factors ${\alpha}_k^t$ also belong to the unit segment. 
Meanwhile, the Lagrange multiplier is updated as:
\begin{gather}
 \label{ea_la_update_lagrange1}
    \lambda_l^{t+1}=[\lambda_l^{t}-\alpha_{\lambda}^t\psi_l^t]^{\lambda_{l+1}^+}_0,\\
 \label{ea_la_update_lagrange2}
    \psi_l^t=\delta^t\lambda_l^t-\eta_l^t+C_l,
\end{gather}
where $\eta_l$ is the instantaneous sum of interference at PU $l$ and $\delta^t$ is the regularization factor in (\ref{ea_la_lagrange}), and $[\cdot]^{\lambda_{l+1}^+}_0$ is 
a projection operator. The learning
scheme defined by (\ref{ea_la_Bush_Mosteller})-(\ref{ea_la_update_lagrange2}) ensures the convergence to the NE, provided that the sequences $\{\eta^t_l\}$ and $\{\delta^t\}$ 
satisfy certain properties (see Assumptions A1-A3 in \cite{6117762}), and the power control game is diagonal concave \cite{1049610}. Compared with $L_{R-I}$ learning, 
Bush-Mosteller LA requires stricter condition for converging to the NE. This is a major reason for impeding Bush-Mosteller learning from being widely applied to the wireless 
resource allocation problems. Due to the requirement for the game to be diagonal concave, and because the original SINR-based utility does not naturally possess the property of 
diagonal concavity, the authors of \cite{6117762} use an arbitrarily designed utility function to replace the real expected mutual-interference-based local 
utility in order to derive the proper payoff function for the constructed power control game.

\subsubsection{No-Regret Learning}
\label{subsubsec_game_NR}
Usually, the terminology ``no-regret learning'' is used to refer to any learning algorithm that exhibits the property of no-regret when compared with the set of some designated 
strategies \cite{Greenwald03ageneral,ECTA:ECTA153}. Formally, for an infinitely repeated game $G={\langle}\mathcal{N}, {\mathcal{A}}\!=\!\times\mathcal{A}_n, \{u_n\}_{n\in\mathcal{N}} 
{\rangle}$, and given the adversary (deterministic) strategy $a_{-i}$, the regret of agent $i$ for playing strategy $\pi_i$ instead of choosing strategy $a_i$ can be defined as 
the difference in its payoff obtained from playing these strategies:
\begin{equation}
 \label{eq_general_regret}
 R_i(\pi_i,a_i|a_{-i})=u_i(a_i,a_{-i})-u_i(\pi_i,a_{-i}).
\end{equation}
Let $\phi(\cdot)$ denote a modification mapping $\pi'_i=\phi(\pi_i)$, where $\pi'_i(a)=\sum_{b:\phi(b)=a}\pi_i(b)$ ($a,b\in\mathcal{A}_i$). Then, for a sequence of adversary 
strategies $\{a^t_{-i}\}$, we can define a general no-regret learning algorithm (also known as $\phi$-no-regret learning) for agent $i$ as follows \cite{Greenwald03ageneral}:
\begin{Definition}[$\phi$-no-regret learning]
 \label{def_phi_no_regret}
 For a finite subset $\Phi$ of memoryless mapping $\phi$, a learning algorithm that generates $\pi^t_i$
 is said to exhibit $\phi$-no-regret if the regret of that learning algorithm,
 \begin{equation}
 \label{eq_modifled_general_regret}
 R_{i,\phi}(\pi^t_i,\phi(\pi^t_i)|a^t_{-i})=u_i(\phi(\pi^t_i),a^t_{-i})-u_i(\pi^t_i,a^t_{-i}),
\end{equation}
 satisfies the following condition: 
 \begin{equation}
 \label{eq_codition_no_regret}
  D_i=\lim_{T\rightarrow 0}\sup\frac{1}{T}\sum_{t=1}^TR_{i,\phi}(\pi^t_i,\phi(\pi^t_i)|a^t_{-i})=0.
 \end{equation}
\end{Definition}

There are two well-studied categories of the $\phi$-no-regret properties: no-external-regret and no-internal-regret \cite{Greenwald03ageneral}. The no-external-regret property
is to minimize the regret with respect to any comparison class of algorithms that lead to deterministic strategies. In other words, for no-external-regret learning, the mapping 
$\phi(\cdot)$ satisfies $\phi(\pi_i)\!=\!a$ ($a\!\in\!{\mathcal{A}_i}$). The no-internal-regret property is also known as 
no-swap-regret since the property of internal regret swaps the current online strategies as follows:
\begin{eqnarray}
 \label{eq_swap_regret_mapping}
 \phi_{a,b}(\pi_i(c))=\left\{
 \begin{array}{ll}
  \pi_i(c), \quad\quad\quad\;\;\;\textrm{if } c\ne a,b,\\
  0, \quad\quad\quad\quad\quad\textrm{ if } c=a,\\
  \pi_i(a)+\pi_i(b),\textrm{ if } c=b.
 \end{array}\right.
\end{eqnarray}
One well-known example for applying no-external-regret learning to the wireless networking problems is \cite{Nie:2006:ACA:1238642:1238644}, which uses the random 
weighted majority (i.e., Hedge) algorithm \cite{nisan2007algorithmic} for learning the NE strategies in a channel allocation game in a CRN. With a careful utility design, the 
channel-allocation game is proved to be an exact potential game. Let $u^t_i(a_i)$ denote the cumulated instantaneous payoff received by SU $i$ given the sequence of the adversary 
strategy $\{a^t_{-i}\}$, the mixed policy of SU $i$ is updated as follows:
\begin{equation}
 \label{ea_rwm}
    \pi_i^{t+1}(a_i)=\frac{(1+\mu)^{u^t_i(a_i)}}{\sum_{a'_i\in\mathcal{A}_i}(1+\mu)^{u^t_i(a'_i)}},
\end{equation}
where $\mu>0$. It is well-known that the learning scheme in (\ref{ea_rwm}) has a regret bound as $D_i^T\le\mu/2$ \cite{Jafari01onno-regret}. Compared with the widely 
applied best-response-based learning schemes for potential games, which also ensure the convergence to the NE, the random weighted majority algorithm (\ref{ea_rwm}) does not 
need any information sharing between SUs. 

The construction of a no-external-regret learning mechanism can be further illustrated by the example of \cite{5502580}, where the problem of collaborative sensing 
with malicious nodes in an $N$-channel CRN is studied. In the considered CRN, SU $j$ is supposed to collaborate with a set of its neighbor SUs $\mathcal{N}_j$ and to choose
whether to aggregate one of their sensing reports into its local channel-state prediction. At time $t$, a mixed policy ${\pi}_t^j=[\pi^j_{1,t}, \ldots, 
\pi^{j}_{|\mathcal{N}_j|,t}]$ is adopted to choose the reports from the SUs in $\mathcal{N}_j$. With the goal of minimizing the long-term expected loss due to false decision 
by choosing the sequence ${\pi}_t$ instead of the pure-strategy best response (internal regret), we have
\begin{equation}
 \label{eq_loss_sensing}
 \min_{\{{\pi}_t^j\}_{t=1}^T}\sum_{t=1}^{T}\left(\overline{l}^j(\pi_t^j,s^t)-{l}^j(j',s^t)\right), \forall j'\in\mathcal{N}_j,
\end{equation}
where ${l}^j(j',s^t)$ is the instantaneous loss due to adopting the report by SU $j'$, and $\overline{l}^j(\pi_t^j,s^t)$ is the average loss with policy $\pi_t^j$ at 
channel state $s^t$: $\overline{l}^j(\pi_t^j,s^t)=\sum_{j'\in\mathcal{N}_j\cup\{j\}}\pi_{j',t}^j{l}^j(j',s^t)$. In \cite{5502580}, such a decision process is modeled 
as a two-player constant-sum game. In the game, SU $j$ plays against nature\footnote{The definition of nature in an extensive form game can be found in \cite{han2012game}.}, 
which plays as an adversary player and chooses state $s$ aiming at causing the worst cost to SU 
$j$. The strategy-updating mechanism is designed upon the softmax function (\ref{ea_la_softmax}) with the accumulated instantaneous loss 
$\sum_{\tau=1}^tl^j({j'}^{\tau},s^{\tau})$ being the argument of the logarithmic function $\exp(\cdot)$. It is shown in \cite{5502580} that no-regret learning based on the 
softmax function converges to the NE, which is equivalent to the minimax value of the game.

Another category of no-regret learning algorithms that are widely applied in the context of network control aims at minimizing the internal regret and learning the CE in 
repeated games \cite{ECTA:ECTA153}. For a general repeated game $G={\langle}\mathcal{N}, {\mathcal{A}}\!=\!\times\mathcal{A}_n, \{u_n\}_{n\in\mathcal{N}} {\rangle}$, the 
estimated average loss for agent $i$ to play action $a_i^t$ instead of playing $a'_i$ at time $t$ is given by: 
\begin{equation}
 \label{eq_regret_loss}
 D_i^t(a^t_i, a'_{i})=\frac{1}{t}\sum_{\tau\le t}\left(u_i^t(a'_i,a^t_{-i})-u_i^t(a^t_i,a^t_{-i})\right). 
\end{equation}
Based on (\ref{eq_regret_loss}), the regret of agent $i$ for not playing $a'_i$ is
\begin{equation}
 \label{eq_regret}
 R_i^t(a^t_i, a'_{i})=\max\left\{D_i^t(a^t_i, a'_{i}),0\right\}. 
\end{equation}
With (\ref{eq_regret}), the mixed policy of agent $i$ is updated by
\begin{eqnarray}
  \label{eq_learning_no_regret}
 \pi_i^{t+1}(a)=\left\{
 \begin{array}{ll}
  \displaystyle\frac{1}{\mu}R_i^t(a^t_i, a), \qquad\qquad\qquad\forall a\ne a^t_i,\\
  1-\displaystyle\sum\limits_{a'\ne a^t_i}\pi_i^{t+1}(a'), \quad\qquad a=a^t_i,
 \end{array}\right.
\end{eqnarray}
where $\mu$ is a sufficiently large constant to ensure that $\pi_i$ ($i\in\mathcal{N}$) is a well-defined probability.

Like the random weighted majority algorithm, the learning scheme defined by (\ref{eq_regret_loss})-(\ref{eq_learning_no_regret}) to learn the CE does not need the agents to 
exchange the action/utility information. The no-internal-regret learning scheme ensures the asymptotic convergence to the set of the CE, according to Theorem 
\ref{thm_nr_convergence} \cite{ECTA:ECTA153}:
\begin{theorem}
 \label{thm_nr_convergence}
 If every agent plays according to the learning scheme defined by (\ref{eq_regret_loss})-(\ref{eq_learning_no_regret}), the empirical distribution of the joint 
 action selection:
 \begin{equation}
  z_T(\mathbf{a})=\frac{1}{T}\vert t\le T: \mathbf{a}^t=\mathbf{a}\vert
 \end{equation}
  converges almost surely to the set of CE of the game $G$ as $T\rightarrow\infty$.
\end{theorem}

The applications of the learning scheme given by (\ref{eq_regret_loss})-(\ref{eq_learning_no_regret}) to network control problems can be found in \cite{4224253, 5462143, 6076652, 
2012Zheng}. As one of the earliest works that employ no-regret learning in the network control problem, it is aimed at obtaining the CE in a dynamic spectrum access game
with an overlay CR network in \cite{4224253}. No-regret learning is used for the SUs to address the problem of channel contention. It is shown that the performance at the CE obtained 
through learning is almost as good as the optimal equilibrium in the set of CE. In \cite{5462143}, a joint power-channel selection problem is studied in an underlay CRN with a 
free band and a set of price-charging PU channels. The no-regret learning algorithm (\ref{eq_regret_loss})-(\ref{eq_learning_no_regret}) is aggregated with an auction game, which 
considers the SINR to the PU or the allocation power as an item for auction. The joint power-channel selection game is played in two 
levels. In the lower-level subgame, the SUs perform the SINR/power bidding game with a fixed set of PU-channel selection. In the higher-level subgame, the SUs adopt the no-regret 
learning algorithm (\ref{eq_regret_loss})-(\ref{eq_learning_no_regret}) to obtain the CE in the channel-selection game. In \cite{6076652}, the learning scheme of
(\ref{eq_regret_loss})-(\ref{eq_learning_no_regret}) is adopted to obtain the CE strategies in a spectrum sensing game among heterogeneous SUs in an overlay CRN. In the game,
each SU chooses either to cooperatively sense the PU channel that it is assigned to with some power consumption (i.e., with some cost), or to directly access the channel 
as a free rider (i.e., without any cost) based on the sensing reports by the neighbor SUs. With the proposed no-regret learning scheme, the strategies are obtained based on 
minimizing the total regret of the neighborhood set of an SU rather
than the individual regret. It is shown in \cite{6076652} that the learning scheme with the neighborhood regret can significantly outperform the learning algorithm based on the 
local regret. This is also considered as the main reason that motivates local SUs to share their local action and payoff information for neighborhood learning. In \cite{2012Zheng},
the scheme in (\ref{eq_regret_loss})-(\ref{eq_learning_no_regret}) is applied to learn the CE of the subcarrier allocation strategies in a multi-cell OFDMA network. Again, 
each link in the subcarrier allocation game does not need to know the private strategies and utilities of the other links. 

The no-internal-regret learning scheme (\ref{eq_regret_loss})-(\ref{eq_learning_no_regret}) only requires that the structure of the local payoff function is known to each agent. 
Compared with the NE-driven learning methods such as FP and best-response learning, no-internal-regret learning could achieve a better social performance (i.e., in terms of sum 
of the players' rewards). Since the set of CE is a convex polytope with all the NE lying on one of its sections \cite{Robert2004}, it is possible for the 
no-internal-regret learning 
algorithm to reach a CE that is not in the polygon of the NE, thus resulting in a better performance than any NE. Although the learning rule of
(\ref{eq_regret_loss})-(\ref{eq_learning_no_regret}) does not guarantee convergence to the social optimal CE, a number of empirical studies (e.g., no-regret learning in the 
cognitive congestion control games \cite{4784355, 6766288}) show that the no-regret learning scheme can significantly outperform best-response learning and FP \cite{4784355}.
Moreover,
its convergent strategy can be considered as a good approximation of the global optimal solution \cite{6766288}. As a result, many studies consider the no-internal-regret 
learning scheme as an approach to implicitly enforce cooperation within the framework of general-sum noncooperative games.

\subsection{Applications of Learning in the Context of Stochastic Games (SGs)}
\label{subsec_game_SG}
SGs generalizes both the repeated games and the MDPs by allowing the payoff of the players at each round of the game to be dependent on the state variable, whose evolution is 
influenced by the joint actions of the players. Compared with the models based on  repeated games, SGs are considered a more practical tool for modeling the agent interaction in a 
stochastic wireless environment, especially when the elements of the wireless environment (e.g., channel states, buffer states and collision states) evolve stochastically and 
are influenced by the transmission strategies of the wireless agents. In the context of SGs, the model-free learning schemes are referred to the value/policy-iteration algorithms
(e.g., the algorithms summarized in \cite{Bowling00ananalysis}) that do not require any a-priori knowledge about the state transition of the wireless system. We note that such a 
property makes model-free learning especially appropriate for finding the solution to the equilibria of the SGs in the context of wireless networks. This is because in most of 
the practical scenarios it is difficult to obtain all the details of the system dynamics due to the complexity of the network. In what follows, we organize our survey on learning 
in SGs according to the approaches used for experience updating (i.e., value-iteration-based learning vs. non-value-iteration-based learning).

\subsubsection{Value-Iteration-Based Learning}
\label{subsub_value_iter}
In contrast to those model-based solutions which use linear programming to obtain the NE (see the example of a constrained power control SG \cite{4410976}),  
value-iteration-based learning algorithms generally need to construct a series of intermediate ``matrix games'' from the original SGs. Consider a general 
discounted-reward SG, $G={\langle}{\mathcal{N},\mathcal{S}}, {\mathcal{A}}, \{u_n\}_{n\in\mathcal{N}}, \Pr(\mathbf{s}'|\mathbf{s},\mathbf{a}) {\rangle}$, a matrix game is defined 
based on the current estimation of the state value of the SG, which is derived in a similar way as (\ref{eq_action_value_fn}):
\begin{Definition}[Matrix game \cite{li2007reinforcement}]
 \label{def_stage_game}
 An $n$-player matrix game (also known as stage game) in an SG is defined as a tuple $G(\mathbf{s})={\langle}\mathcal{N},\mathcal{A}_1,\ldots,\mathcal{A}_{|\mathcal{N}|},
 \hat{Q}^t_{\beta,1},\ldots, \hat{Q}^t_{\beta,|\mathcal{N}|}{\rangle}$, in which $\hat{Q}^t_{\beta,i}$ ($1\le i\le|\mathcal{N}|$) is given by:
 \begin{eqnarray}
  \label{eq_estimated_q}
  \hat{Q}^t_{\beta,i}(\mathbf{s},\mathbf{a})\!=\!u(\mathbf{s},\mathbf{a})\!+\!\beta\sum_{\mathbf{s}'\in\mathcal{S}}\Pr(\mathbf{s}'|\mathbf{s},\pi)V_{\beta,i}^{\pi}
  (\mathbf{s}'|\mathbf{s},\pi_i,\pi_{-i}).
 \end{eqnarray}
\end{Definition}
We note that in (\ref{eq_estimated_q}), $V_{\beta,i}^{\pi}(\mathbf{s}'|\mathbf{s},\pi_i,\pi_{-i})=E_{\pi}\{\hat{Q}^t_{\beta,i}(\mathbf{s},\mathbf{a})\}$. Under policy $\pi$, 
transition probability $\Pr(\mathbf{s}'|\mathbf{s},\pi)$ can be expressed as follows:
\begin{equation}
 \label{eq_expected_trans_prob}
 \begin{array}{ll}
 \Pr(\mathbf{s}'|\mathbf{s},\pi)\! = \!\displaystyle\sum\limits_{a_1\in\mathcal{A}_1}\ldots\sum\limits_{a_{|\mathcal{N}|\in\mathcal{A}_{|\mathcal{N}|}}}\bigg(
 \Pr(\mathbf{s}'|\mathbf{s},a_1,\ldots, a_{|\mathcal{N}|})\\
 \qquad\qquad\qquad\times\pi_{1}(\mathbf{s},a_1)\cdots\pi_{|\mathcal{N}|}(\mathbf{s},a_{|\mathcal{N}|})\bigg).
 \end{array}
\end{equation}
According to Definition \ref{def_stage_game}, a general form of strategy searching based on value iteration can be implemented as in Algorithm \ref{alg_value_iter}
\cite{Bowling00ananalysis}.
\begin{algorithm}[!tb]
 \begin{algorithmic}
 \REQUIRE  
 Initialize $V^t_{\beta,i}, \forall 1\le i\le|\mathcal{N}|$ arbitrarily.
 \WHILE{convergence criterion is not met}
  \STATE (a) For state $\mathbf{s}$ at round $t$, update the estimated value of $\hat{Q}^t_{i}(\mathbf{s},\mathbf{a})$ of the matrix game.
  \STATE (b) For state $\mathbf{s}$, update the expected state value of $V_{\beta,i}^{t}(\mathbf{s})$ after computing the (mixed) equilibrium strategy $(\pi_i(\mathbf{s}),
  \pi_{-i}(\mathbf{s}))$:
  \begin{equation}
  \label{eq_estimate_v}
  V_{\beta,i}^{t}(\mathbf{s}_t)\leftarrow \textrm{Eval}_{\pi}(\hat{Q}_{\beta,i}(\mathbf{s},\mathbf{a})).
  \end{equation}
 \ENDWHILE
 \end{algorithmic}
  \caption{Value-iteration-based learning algorithm.}
  \label{alg_value_iter}
\end{algorithm}
In (\ref{eq_estimate_v}) of Algorithm \ref{alg_value_iter}, operator $\textrm{Eval}_{\pi}(\cdot)$ computes (estimates) the expected payoff in the NE of the matrix game. The 
equivalence between the NE of the matrix game and the NE of the discounted SG is given by Theorem \ref{theo_SG_matrix_equivalence}.
\begin{theorem}[\!\!\cite{li2007reinforcement}]
 \label{theo_SG_matrix_equivalence}
 The following are equivalent:
 \begin{itemize}
  \item $\pi^*$ is an equilibrium point in the discounted SG, $G$, with equilibrium payoffs $(V_{\beta,1}(\pi^*),\ldots, V_{\beta,|\mathcal{N}|}(\pi^*))$.
  \item For each $\mathbf{s}\in\mathcal{S}$, strategy $\pi^*(\mathbf{s})$ constitutes an equilibrium point in static matrix game $G(\mathbf{s})$ with equilibrium 
  payoffs $(\textrm{Eval}_{\pi^*}(\hat{Q}_{\beta,1}(\mathbf{s},\mathbf{a})),\ldots, \textrm{Eval}_{\pi^*}(\hat{Q}_{\beta,|\mathcal{N}|}(\mathbf{s},\mathbf{a})))$. The value of 
  $\hat{Q}_{\beta,i}(\mathbf{s},\mathbf{a})$ is given by Definition \ref{def_stage_game}.
 \end{itemize}
\end{theorem}

According to Theorem \ref{theo_SG_matrix_equivalence}, Algorithm \ref{alg_value_iter} can be considered a combination of a matrix-game solver and a value-iteration-based state
value learner. It works as the general form of a set of model-free strategy-learning algorithms, 
which differ from each other only in the way of defining operator $\textrm{Eval}_{\pi}(\cdot)$. In \cite{Littman94markovgames}, operator $\textrm{Eval}_{\pi}(\cdot)$ in value 
iteration is implemented by a minimax optimization process, and the Q-value of each learning agent is updated through a standard single-agent Q-learning process. Such a learning 
scheme is known as minimax-Q learning. Specifically, the learning mechanism can be expressed by
\begin{equation}
 \label{eq_minimax_q_learing}
 \begin{array}{ll}  
 Q^{t+1}_{\beta,i}(\mathbf{s}_t, a^t_{i}, a^t_{-i})\leftarrow (1-\alpha_t)Q^{t}_{\beta,i}(\mathbf{s}_t, a^t_{i}, a^t_{-i})+\\
 \qquad\qquad\alpha_t\left(u_i(\mathbf{s}_t, a^t_{i}, a^t_{-i})+\beta V^t_{\beta,i}(\mathbf{s}_{t+1})\right),
 \end{array}
\end{equation}
\begin{equation} 
 \label{eq_minimax_matrix_v}
 V_{\beta,i}^{t}(\mathbf{s}_t)=\max_{\pi(\mathbf{s}_t, a_i)}\min_{a_{-i}}\sum_{\mathbf{a}\in\mathcal{A}}Q^t_{\beta,i}
 (\mathbf{s}_t, a_{i}, a_{-i})\pi(\mathbf{s}_t, {a}_i),
\end{equation}
\begin{equation}
 \label{eq_minimax_pi}
 \pi^{t}(\mathbf{s},a_i)=\arg\max_{\pi(\mathbf{s},a_i)}\min_{a_{-i}}\sum_{a_i}Q^t_{\beta,i}(\mathbf{s},a_i,a_{-i})\pi(\mathbf{s},a_i).
\end{equation}
The solution to (\ref{eq_minimax_pi}) is usually obtained through linear programming, which requires that the matrix game of the SG is of complete information. It is worth 
noting that (\ref{eq_minimax_matrix_v}) is an approximation of the exact state value, $V_{\beta,i}^{t}(\mathbf{s}_t)=\max_{\pi(\mathbf{s}_t, a_i)}\min_{\pi(\mathbf{s}_t,
a_{-i})}\sum_{\mathbf{a}\in\mathcal{A}}Q^t_{\beta,i}(\mathbf{s}_t, \mathbf{a})\pi(\mathbf{s}_t, \mathbf{a})$, which cannot be obtained directly since the local strategies are 
usually private information. Due to the approximation, the updating mechanism in (\ref{eq_minimax_q_learing})-(\ref{eq_minimax_pi}), although proved to be effective by empirical 
studies \cite{Littman94markovgames}, does not provide a strict condition for convergence to the NE.

\begin{figure}[!t]
  \centering
  \includegraphics[width=0.48\textwidth]{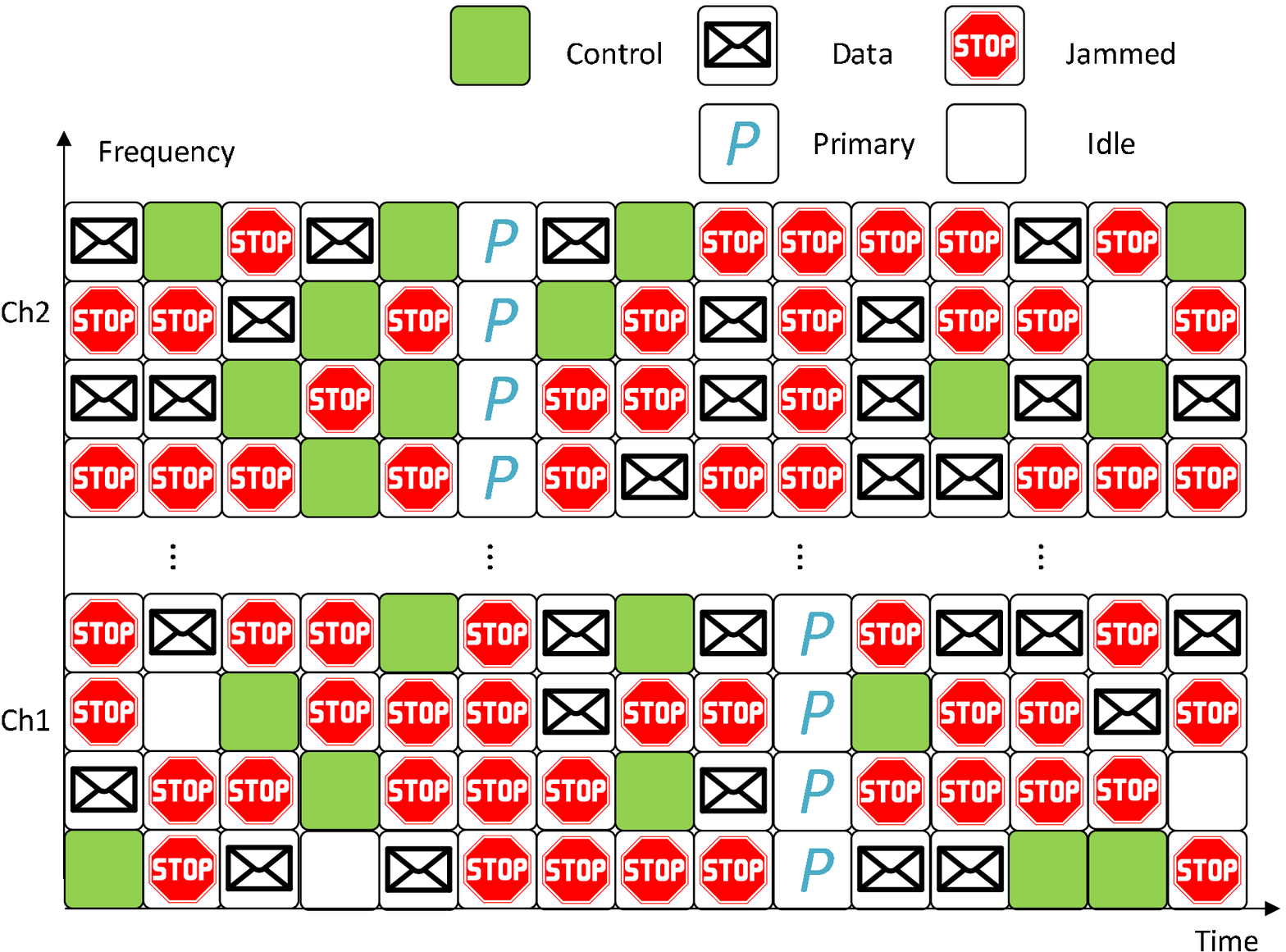}
  \caption{A snapshot of the anti-jamming defense process in a multi-channel CRN (adapted from \cite{5738229}).}
  \label{fig_minimax_q_attack}
\end{figure}
Minimax-Q learning is usually adopted to solve the problems which can be described as a constant-sum (also known as strictly competitive) game. One typical category of its 
applications in wireless networks is strategy-learning in attack-defense problems, since such problems can usually be modeled as a two-player, zero-sum game 
with the group of normal nodes and the group of malicious nodes treated as two super players. In \cite{5738229}, a two-player zero-sum SG is adopted to model the anti-jamming 
process of a group of SUs in the CRN (Figure \ref{fig_minimax_q_attack}). Due to the random activities of the PUs, the channel-availability states viewed by the SUs are modeled 
as a group of independent, two-state Markov chains. In addition, for each channel, the channel quality measured by the local SNR is modeled as a finite state Markov chain. In 
\cite{5738229}, the devices in the CRN are divided into two groups: the normal SUs and jamming nodes. Both the normal SUs and the attackers access the PU channels in a slotted 
manner. At each time slot, the normal SUs will select a subset of channels for transmission while the attackers will select a subset of channels for jamming. The group of 
channels that are 
selected for transmission are further subdivided into control channels and data channels. For a normal SU, the non-zero gain of a channel can only be achieved when the channel is 
used for data transmission and at least one control channel selected by the normal SU is not jammed by the attackers. The goal of the normal SUs is to maximize the local channel 
utility. Based on the formulation of the two-player zero-sum SG, the standard minimax-Q-learning algorithm is applied for the normal SUs to find the equilibrium strategies in 
the stochastic attack-defense game. Convergence of the learning algorithm has been shown by empirical studies. Also, the numerical simulations show that minimax-Q learning 
outperforms both the myopic strategy, which does not consider the future payoff, and the fixed strategy, which uniformly selects the channels regardless of the attacker's 
strategy. 

The application of minimax-Q learning in a similar scenario can be found in \cite{6682689}, which formulates the competition for open access spectrum in a tactical 
wireless network as a competitive mobile network game. The study in \cite{6682689} extends the attack-defense model in \cite{5738229} by dividing the competitive mobile network 
into two sub-networks: the ally network and the enemy network. Each network is composed of both communicating nodes and jamming nodes. The goal of the two networks is to 
achieve the maximum spectrum utility while jamming the opponent transmission as much as possible. The channel-availability state is jointly determined by the transmission-jamming
actions of the two networks as a controlled Markov chain. Channel access in the competitive network is modeled as a two-player, zero-sum game, and standard minimax-Q learning is 
adopted for both the ally and the enemy network to learn their equilibrium strategies. Apart from \cite{6682689}, other applications of minimax-Q learning can be found in 
\cite{6523804, 6554865}, which basically adopt the same framework of the two-player, zero-sum SG as in \cite{5738229, 6682689} to obtain the anti-jamming scheme. In 
\cite{6523804}, minimax-Q learning in the SG is employed in a typical DSA network without considering the impact of jamming the control channels. In \cite{6554865}, the two-player
SG model is extended to the scenarios of stochastic routing in a MANET, and the attack-proof strategy is obtained through minimax-Q learning.

For networking problems that need to be described as an $n$-player general-sum SG, a more general learning scheme can be implemented by replacing the minimax operator for 
$\textrm{Eval}_{\pi}(\cdot)$ with the operator that leads to the payoff of the NE in the general game. For the discounted-reward general-sum SGs, such a learning scheme is known
as Nash Q-learning \cite{Hu:2003:NQG:945365.964288}. Nash Q-learning adopts the same Q-value updating scheme (\ref{eq_minimax_q_learing}) as in the minimax-Q learning algorithm, 
and requires that the value of $V_{\beta,i}^{t}(\mathbf{s}_t)$ is obtained based on the matrix game NE of the SG. According to Theorem \ref{theo_SG_matrix_equivalence},
as long as the NE of each matrix game obtained from the SG in stage $\mathbf{s}_t$ is used in (\ref{eq_minimax_q_learing}) to compute the value of 
$V_{\beta,i}^{t}(\mathbf{s}_t)$, the learning process converges to the NE of the SG. For Nash Q-learning, operator $\textrm{Eval}_{\pi}(\cdot)$ can be expressed by:
\begin{equation} 
 \label{eq_nash_q_matrix_v}
 V_{\beta,i}^{t}(\mathbf{s}_t)=\displaystyle\sum\limits_{a_1\in\mathcal{A}_1}\ldots\sum\limits_{a_{|\mathcal{N}|\in\mathcal{A}_{|\mathcal{N}|}}}\prod_{i=1}^{|\mathcal{N}|}\pi^*_i(\mathbf{s}_t,a_i)Q^t_{\beta,i}(\mathbf{s}_t,\mathbf{a}).
\end{equation}
In (\ref{eq_nash_q_matrix_v}), $\pi^*_i(\mathbf{s})$ is the NE strategy of the matrix game at stage $t$ when the payoff matrix of agent $i$ is 
$Q^t_{\beta,i}(\mathbf{s},\mathbf{a})$.

Theorem \ref{theo_SG_matrix_equivalence} also holds when the SG is based on average reward.
The counterpart to Nash Q-learning in an average-reward SG is known as Nash R-learning \cite{li2007reinforcement}. Nash R-learning adopts the R-learning-based scheme for 
state-action updating as in (\ref{eq_r_learning}) and (\ref{eq_r_learning2}), which can be summarized by the following equations:
\begin{equation}
 \label{eq_nash_r_learing}
 \begin{array}{ll}
 R^{t\!+\!1}_{i}(\mathbf{s}_t, \mathbf{a}_t)\!\leftarrow \!R^{t}_{i}(\mathbf{s}_t, \mathbf{a}_t)+\\
 \alpha_t\big(u_i(\mathbf{s}_t, \mathbf{a}_t)\!+\!V^t_{i}(\mathbf{s}_{t\!+\!1})\!-\!R^t_i(\mathbf{s}_t, \mathbf{a}_t)\!-\!h^t_i(\mathbf{s}_t, 
 \mathbf{a}_t)\big),
 \end{array}
\end{equation}
\begin{equation}
 \label{eq_nash_r_average}
 h^{t+1}_i(\mathbf{s}_t, \mathbf{a}_t)=h^t_i(\mathbf{s}_t,\mathbf{a}_t)+\theta_tV^t_{i}(\mathbf{s}_{t+1}),
\end{equation}
where $V^t_{i}(\mathbf{s})$ is the equilibrium payoff of the stage game and is computed following (\ref{eq_nash_q_matrix_v}). 

When the goal of the learning process is to find the CE of the discounted-reward SG instead of the NE, Correlated-Q (CE-Q) Learning can be implemented based on the updating 
mechanism in (\ref{eq_minimax_q_learing})-(\ref{eq_minimax_pi}) with the state value $V_{\beta,i}^{t}(\mathbf{s}_t)$ estimated at the CE strategies 
\cite{Greenwald03correlatedq-learning}. The equivalence between the CE of the original SG and the CE of the matrix game in each state still holds. Based on Definition 
\ref{def_CE} and Theorem \ref{theo_SG_matrix_equivalence}, we have
\begin{theorem}[CE in the SG \cite{Greenwald03correlatedq-learning}]
 \label{theo_CE}
 For a discounted-reward SG $G$, a stationary policy $\pi$ is a correlated equilibrium if $\forall{i}\in{\mathcal{N}},\forall\mathbf{s}\in\mathcal{S}$, $\forall\mathbf{a}\in
 \mathcal{A}$ with $\pi_i(a_i)>0$, for all $a'_i\in\mathcal{A}_i(\mathbf{s})$
 \begin{equation}
  \label{eq_CE}
  \begin{array}{ll}
  \sum\limits_{a_{-i}\in\mathcal{A}_{-i}}\pi(\mathbf{s}, a_{-i}|a_i)Q_{\beta,i}(\mathbf{s}, (a_{-i},a_i))\ge\\
  \sum\limits_{a_{-i}\in\mathcal{A}_{-i}}\pi(\mathbf{s}, a_{-i}|a_i)Q_{\beta,i}(\mathbf{s}, (a_{-i},a'_i)),
  \end{array}
 \end{equation}
  which defines the CE of the matrix game in $\mathbf{s}$ as $\pi(\mathbf{s})$.
\end{theorem}

For both the NE based Q-learning (Nash-Q and Nash-R) and the CE-based Q-learning (CE-Q), it is not specified how the equilibrium strategies $\pi^*_i(\mathbf{s})$ for each matrix 
game is obtained during the learning process. Since it is necessary for the game to be of complete information in order to immediately obtain the NE/CE of the matrix game, it 
is required that the learning agents should keep track of the entire Q-table from all the other agents at state $\mathbf{s}$ in order to compute the exact stage-game equilibrium. 
In practice, exchanging such information will result in a large transmission overhead, which is usually unaffordable in a wireless network. As a result, most of the 
existing studies apply heuristic methods to approximate the matrix game equilibrium. One example of payoff approximation at the NE of the matrix game can be found in 
\cite{Fu:2013:SGW:2502350.2502357}, which decouples the wireless network into a group of Service Providers (SPs) and a single entity called Network Operators (NOs) for network 
virtualization. Each SP is 
responsible for reallocating the available spectrum resources to a group of end users, while the NO is responsible for allocating the time-varying spectrum resources to the SPs.
Here, resource allocation through the interface between the NO and the SPs at each time slot is treated as an auction game with the NO acting as the auctioneer and the SPs acting 
as the bidders. The auction is performed following the Vickrey-Clarke-Groves (VCG) mechanism \cite{han2012game}. The entire auction process in the stochastic environment is modeled as a 
discounted general-sum SG, in which the channel state and the traffic state are assumed to be Markovian and the SP action is the selection of value functions through choosing the
transmit rate. In \cite{Fu:2013:SGW:2502350.2502357}, the matrix games of the original SG is referred to as the ``current games''. Also, to avoid directly computing the value of 
$V_{\beta,i}(\mathbf{s})$ in (\ref{eq_nash_q_matrix_v}), a conjecture price which approximates the unit-rate price (strategy) of the NO in the future is introduced. A Q-value 
updating scheme which is analogous to the SAS-based Q-learning scheme is proposed, and the value of the conjecture price is updated using the subgradient method.

For networking problems which do not possess the single-server-distributed-agents property as stochastic auction games, the equilibrium strategies can be learned by implementing 
an appropriate amount of local information exchange. In \cite{7012044}, the problem of traffic offloading in a stochastic heterogeneous cellular network is first formulated
as a centralized discrete-time MDP and then as an SG. In the SG, a group of macrocell BSs try to offload their downlink traffic to their corresponding group of small-cell BSs, 
which operate in the open access mode and share the same band with the macro BSs. Before the learning mechanism is implemented, the authors in \cite{7012044} employ a standard state abstraction 
procedure based on linear state-value combination (see our discussion in Section \ref{subsec_app_single}). The Q-values (i.e., the payoff of matrix games) are updated with the 
gradient-ascending method based on the gradient of the new Q-values after state abstraction. The matrix game in a given state $\mathbf{s}$ is modeled as a ``virtual game'' with 
common payoff by allowing the macro BSs to share their instantaneous spectrum utility with each other. Also, the action of each BS is updated using $\epsilon$-exploration instead
of directly computing the mixed strategy of the matrix game. It is proved in \cite{7012044} that convergence (which may not be the NE) is guaranteed with probability one.

A different approach to approximate the matrix game equilibrium with only local information in the SG can be found in \cite{JCM0410790802, 6554824}, which employ the 
learning methods for the repeated games to learn the matrix game equilibrium strategies and then use these intermediate strategies to approximate the state value 
$V^{\pi^*}_{\beta,i}(\mathbf{s})$  of the original SG. In \cite{JCM0410790802}, the interference mitigation problem with a finite action set of discrete powers for both the PUs 
and the SUs in a CRN is modeled as a discounted-reward SG. In \cite{6554824}, the cross-layer resource allocation problem for layered video transmission in a CRN is modeled as a 
discounted-reward SG. In both works, the goal of strategy learning is to find the CE of their respective SG. Both works treat the matrix game at state $\mathbf{s}\in\mathcal{S}$ 
as a repeated game and adopt the no-internal-regret learning method defined by(\ref{eq_regret_loss})-(\ref{eq_learning_no_regret}) to approximate the CE strategy 
$\pi^*_i(\mathbf{s})$ at state $\mathbf{s}$. Let $\tilde{\pi}_i(\mathbf{s})$ define the intermediate strategy that is obtained with (\ref{eq_learning_no_regret}). Since with the 
no-internal-regret learning scheme, no action/payoff information exchange is needed, the strategy estimation in the SG is solely based on local information. The same method as 
in (\ref{eq_minimax_q_learing}) is adopted for Q-value updating, for which state value $V^{\pi^*}_{\beta,i}(\mathbf{s})$ under the CE strategy can be estimated as the expected 
payoff of the matrix game:
\begin{equation}
 \label{eq_SG_NR_estimation}
 V^{t}_{\beta,i}(\mathbf{s}_t)=\sum_{{a}_i\in\mathcal{A}_i}\tilde{\pi}^t(\mathbf{s}_t, \mathbf{a})Q^t_{\beta,i}(\mathbf{s}_t, \mathbf{a}).
\end{equation}
To further reduce the information-exchange overhead, the values of $\tilde{\pi}^t(\mathbf{s}_t, \mathbf{a})$ and $Q^t_{\beta,i}(\mathbf{s}_t, \mathbf{a})$ can be replaced by 
the conditional local strategy (given the adversary actions) and the Q-table based on the local state-action pairs \cite{6554824}, repectively. Such a two-fold, approximate 
learning scheme does not require the information exchange between 
wireless devices. However, compared with the original learning scheme in Algorithm \ref{alg_value_iter}, such a learning algorithm may suffer from using the non-CE policies 
in the matrix game and from the inaccurate estimation of $V^{t}_{\beta,i}(\mathbf{s}_t)$. Although empirical studies show that convergence can be achieved by the two-fold learning 
scheme, no theoretical support is available to guarantee the convergence to the CE.

\subsubsection{Conjecture-Based Learning}
Consider the problem of unguaranteed convergence due to the inaccurate estimation of the equilibrium strategies in the matrix games with two-fold learning, the concept of 
``conjecture'' \cite{tembine2012distributed} about one player's opponent policies is introduced in several recent studies \cite{5452952, 6265055, 6779690}. In an SG, the 
conjecture of agent $i$ can be defined as any belief function $c_i: \mathcal{S}\times\mathcal{A}_i\rightarrow\mathcal{C}$, in which $\mathcal{C}$ is the space of agent $i$'s 
conjectures (e.g., about the opponents' policies and states). In the case of policy conjecture, we can define $c_i^t(\mathbf{s}, a_{-i})$ as the conjecture of opponent policy 
$\pi_{-i}(\mathbf{s})$ by agent $i$ at time $t$. With only local information, the most widely accepted conjecture updating mechanism is
\begin{equation}
 \label{eq_conjecture}
 c_i^{t+1}(\mathbf{s}, a_{-i})={c}^t_i(\mathbf{s}, a_{-i})+\omega_i^{\mathbf{s}}(\overline{\pi}(\mathbf{s},a_i)-\pi_i^t(\mathbf{s},a_i)),
\end{equation}
where $\overline{\pi}_i^t(\mathbf{s},a_i)$ is the so-called reference point and is assumed to be of common knowledge to all the players. With (\ref{eq_conjecture}), the conjecture is used 
by local agent $i$ to maximize its individual payoff in the condition of not knowing what the strategies of the other players are, or what their payoff functions are.
(\ref{eq_conjecture}) is obtained based upon the assumption that the other players will be able to observe player $i$'s deviation from the reference point 
$\pi_i^t(\mathbf{s},a_i)$, and in response to such a deviation, they will deviate from their own reference point by a quantity that is 
proportional to this deviation \cite{tembine2012distributed}. With conjecture $c_i(\mathbf{s}, a_{-i})$, the conjecture equilibrium can be defined as follows (extended from the 
definition in \cite{5452952}):
\begin{Definition}[Conjecture equilibrium]
 \label{def_conjecture_equilibrium}
 In the stochastic game $G$, a configuration of conjectures $\mathbf{c}$ and a joint policy $\pi^*$ constitute a conjecture equilibrium if $\forall i\in\mathcal{N}$
 \begin{align}
  \label{eq_con_equ1}
   &c^*_i(\mathbf{s}, \pi^*_i)=c_i(\mathbf{s}, \pi^*),\\
   \label{eq_con_equ2}
   &\pi^*_i=\arg\max_{\pi_i}Q_i(\mathbf{s}, \pi_i,  c^*_i(\mathbf{s}, \pi_i)).
 \end{align}
\end{Definition}

\begin{figure}[!t]
  \centering
  \includegraphics[width=0.40\textwidth]{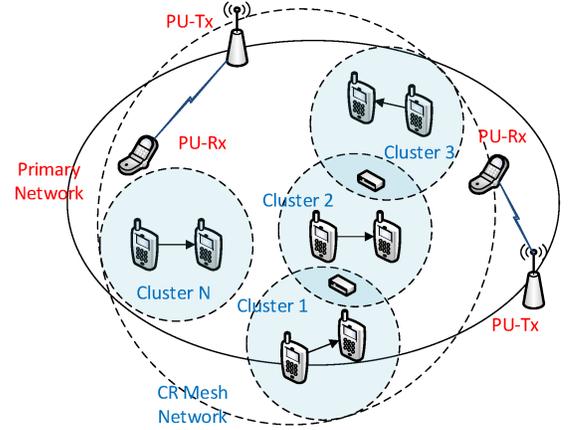}
  \caption{Structure of underlay CR mesh network (adapted from \cite{6265055}).}
  \label{fig_cr_mesh_network}
\end{figure}
We take \cite{6265055} as an example to explain the details of employing conjecture to learn in SGs. In \cite{6265055}, the power allocation problem in an underlay CR mesh 
network (Figure \ref{fig_cr_mesh_network}) is studied. The multi-node power allocation process is modeled as an SG, in which the local binary state 
of a secondary link is determined by the SINR level of its receiver. The local payoff is measured by the power efficiency. Compared with the standard matrix-game-based 
strategy-learning mechanism in (\ref{eq_estimate_v})-(\ref{eq_minimax_q_learing}), the authors in \cite{6265055} constructs the Q-table with only local states and actions. Here, 
the policy conjecture is introduced to approximately learn the matrix game equilibrium strategy and the Q-value of the SG. Based on the conjecture-updating scheme in 
(\ref{eq_conjecture}), the Q-value updating mechanism is defined as follows:
\begin{equation}
 \label{eq_conjecture_Q}
 \begin{array}{ll}
 Q^{t+1}_{\beta,i}(s_i,a_i)=(1-\alpha^t)Q^{t}_{\beta,i}(s_i,a_i)+\\
 \alpha^t\left(\sum\limits_{a_{-i}\!\in\!\mathcal{A}_{-i}}c_i^{t}({s}_i, a_{-i})u_i(s_i,{a}_i,a_{-i})\!+\!\beta\max\limits_{b_i\in\mathcal{A}_i}Q^t_{\beta,i}(s'_i,b_i)\right).
 \end{array}
\end{equation}
The local policy $\pi_i$ is updated using the Logit function (\ref{ea_la_softmax}). It is proved in \cite{6265055} that the second term on the right-hand side of 
(\ref{eq_conjecture_Q}) is a contraction mapping operator and the learning scheme converges with sufficiently large number of iterations.

\subsubsection{Other Learning Algorithms in SGs}
For algorithms that do not work in the framework of hierarchical learning that is separated into learning in the matrix games and the original SG, we simply refer them to the 
category of the ``other learning algorithms''. In these algorithms, the Q-learning-based value-iteration scheme for the payoff of the matrix game may not necessarily be applied, 
or the computation of the state value of the SG may not be needed. Due to the complexity of a general SG, most of the existing learning methods in this category cannot be 
represented by a single  prototypical algorithm.

We note that for an SG, the property of the MDP generally requires that the state value of the game be computed following the Bellman optimality equation (in the general form 
as (\ref{eq_value_fn})), whenever a stationary 
policy is to be obtained. Extending from the value-iteration-based algorithm, we can construct a general learning scheme, which is composed of two learning loops:
an inner loop that uses an appropriate scheme to approximate the SG equilibrium strategies $\pi^*$ and an outer loop that employs an appropriate method to estimate the 
state value $V_{\beta,i}(\mathbf{s})$ of each player. Within this general framework, the construction of matrix games is not necessary. We can generalize the two-layer 
learning process in SG $G={\langle}
{\mathcal{N},\mathcal{S}}, {\mathcal{A}}, \{u_n\}_{n\in\mathcal{N}}, \Pr(\mathbf{s}'|\mathbf{s},\mathbf{a}) {\rangle}$ as Algorithm \ref{alg_general_iter}.
\begin{algorithm}[!tb]
 \begin{algorithmic}
 \REQUIRE  
 Initialize $V^t_{\beta,i}$ and $\pi^t_i$, $\forall 1\le i\le|\mathcal{N}|$.
 \WHILE{convergence criterion is not met}
  \STATE Outer loop: $V^{t+1}_{\beta,i}\leftarrow \textrm{UpdateStateValue}(u^t_i, V^t_{\beta,i}, \pi^t_i, \pi^t_{-i})$
   \STATE Inner loop: $\pi^{t+1}_i\leftarrow \textrm{UpdateStrategy}(V^t_{\beta,i}, \pi^t_i, \pi^t_{-i})$.
 \ENDWHILE
 \end{algorithmic}
  \caption{Two-layer learning mechanism in the SG.}
  \label{alg_general_iter}
\end{algorithm}

One widely-used two-layer approach for strategy learning in wireless SGs is to adopt FP-based policy updating as the inner-loop learning scheme.  Such an approach of policy 
evolution can be found rooted in the model-based learning algorithms (namely, with known state-transition maps) \cite{Ganzfried:2009:CEM:1661445.1661469}. Since the standard 
FP-based algorithm with (\ref{eq_fp_frequency}) and (\ref{eq_fp_probability}) requires that each wireless node to track the opponent actions, extending FP-based learning 
from repeated games to the SG is considered a challenge due to the explosion of state-action dimensionality. In \cite{Liu2013817}, such a challenge is resolved by regulating the 
SG into a sequential game, in which only one wireless node is allowed to update its action in each round. In\cite{Liu2013817}, the problem of joint channel selection 
and power allocation for the SUs in an overlay DSA network is studied. With the assumption of a sequential game, each SU adopts a standard SAS-based Q-learning scheme as in 
(\ref{eq_q_learning}) for updating the Q-table based on the local state-action pairs. To further reduce the state-action space, Q-learning is only applied to the 
strategy-learning for channel selection. The power adaptation is performed only after the channels are selected by the SUs. The FP-based mixed-strategy-updating scheme in 
\cite{Liu2013817} can be considered as a variation of the best-response-based strategy learning schemes described in (\ref{eq_modified_mixed_probability}). 

It is also necessary to consider a different approach to update the state value for FP-based learning when the players in the SGs update their strategies simultaneously, because 
the state value of the MDP cannot be easily estimated by only tracking the opponents' actions. For those works that directly estimate the state value without using the 
TD-learning-based methods, it is also necessary to track the frequency of state transition in order to estimate the state transition probabilities. Examples of learning the state
transition can be found in \cite{4814773, 4581648}. In \cite{4814773}, secondary wireless stations compete with each other for network resources to transmit delay-sensitive in a 
stochastic CRN. In \cite{4581648}, a similar problem is specified in an overlay CRN with SUs competing for the vacant primary channels and determining transmitting parameters in 
a cross-layer manner. In both works, with the resource allocation problem in the CRN being modeled as SGs, it is required that the state transition frequencies of the opponents' 
local states are tracked by each SU. In order to reduce the information exchange overhead about local state transitions, an SU abstracts the 
state space by classifying the opponent SUs' state space purely based on its local observation. Instead of learning the real state-transition frequencies, the transitions
of the abstracted state are recorded. The state value of the SG is updated based on the reduced states using the standard Bellman optimality equation (\ref{eq_value_fn}).  

\begin{figure}[!t]
  \centering
  \includegraphics[width=0.45\textwidth]{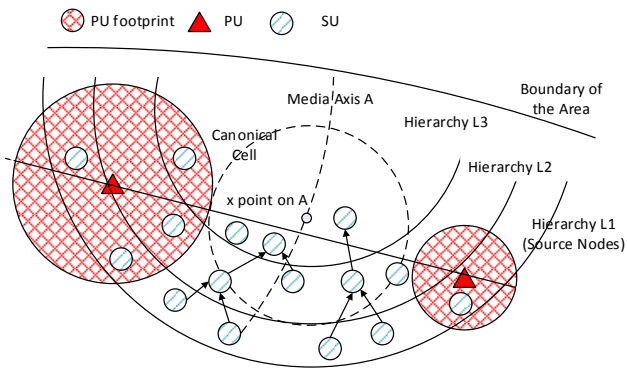}
  \caption{A snapshot of a hierarchical multi-hop CRN under the PU interference footprint (adapted from \cite{6331690}).}
  \label{fig_layered_routing}
\end{figure}

The special structure of some SGs can also be exploited to simplify the learning process for the FP-based learning mechanism. One example of such exploitation can be found in 
\cite{6331690}, which models the distributed dynamic routing in multi-hop CRNs as an SG (Figure \ref{fig_layered_routing}). Since the states of the routing SG in \cite{6331690} 
are defined as the state of channel availability in the CRN, the SG is featured by the state transitions which only depend on the PU activities. The SUs in the network attempt to 
find the route for minimizing the packet-forwarding delay due to queueing and channel collision while keeping their interference to the PUs as small as possible. Since the delay 
over a path is equal to the accumulated delay caused by each link in the path, and the state transition is independent of the SU's actions, the original SG in \cite{6331690} can 
be decomposed into a group of layered, stochastic subgames. Each subgame corresponds to a hierarchy level\footnote{According to \cite{6331690}, the hierarchy levels of the CRN 
are calculated along the ``media axis'', which is composed of a set of points. At these points, the lowest detection probability density of the PU's activities is (approximately) 
achieved.} in the routing path (see Figure \ref{fig_layered_routing}). The structure (i.e., 
the payoff matrix) of each subgame can only be determined when the cost (measured in delay) of the next-layer game is determined. A backward induction method is adopted in
\cite{6331690} to compute the equilibrium payoff in the layered routing game. The computation starts from the subgame of the layer which ends at the sink SU to the subgame of the 
layer which begins from the source SU. Since the state transition is independent of the SU's actions, the stochastic subgame in each layer can be reduced to a group of repeated 
games with fixed states. Therefore, the learning of state value becomes unnecessary and FP-based learning guarantees the convergence to the global NE, as long as the routing 
costs at the equilibrium point of each subgame are properly propagated to their lower layers.

In addition to learning algorithms that follow Algorithm \ref{alg_general_iter}, a number of miscellaneous learning mechanisms are applied to SG-based problems in wireless 
networks. In order to reduce the requirement of information exchange or to achieve convergence, most of these learning mechanisms exploit special properties from the SG. 
As we have discussed in Section \ref{subsec_app_dist}, for the Aloha-like spectrum access problem in CRNs \cite{Li:2010:MQA:1863662.1928710}, the near-NE policies of the 
stochastic access game can be obtained if all the SUs update their local policies with the Logit function (\ref{ea_la_softmax}), and the Q-value at state $\mathbf{s}$ is updated 
following (\ref{eq_myopic_q_learning}). In this specific scenario, the two-layer learning mechanism based on 
Q-value updating ensures the convergence to near-NE strategies of the SG without the need of any information exchange. In \cite{5397924, 5439936}, the structural property of a 
constrained SG is explored. Specifically, consider a utility-minimizing SG $G=\langle\mathcal{N}, \mathcal{S}, \times\mathcal{A}_i,\{c_i\}_{i\in\mathcal{N}},
\{d_i\}_{i\in\mathcal{N}},\Pr(\mathbf{s}'|\mathbf{s},\mathbf{a})\rangle$ with $c_i$ as the instantaneous local cost in the objective and $d_i$ as the instantaneous local cost in 
the constraint. If the following assumptions are satisfied with $G$:
\begin{itemize}
 \item [A1)] the set of policies that satisfy the constraint of the SG is non-empty,
 \item [A2)] the two cost functions $c_i$ and $d_i$ are multi-modular functions with respect to the actions and the state elements whose transition is a function of the joint
 local actions,
 \item [A3)] the transition probability $\Pr(\mathbf{s}'|\mathbf{s},\mathbf{a})$ is submodular with respect to the actions and the state elements whose transition is a function 
 of the joint local actions,
\end{itemize}
then $G$ has the following property in the structure of the NE:
\begin{theorem}
 \label{thm_NE_structure}
 Assume A1-A3 hold, then the NE policy of each player $i$, $\pi^*_i$, is a randomized mixture of two pure policies: $\pi_i^1$ and $\pi_i^2$. Each pure policy is 
 nondecreasing on the state elements whose transition is determined by the joint actions. 
\end{theorem}
Based on Theorem \ref{thm_NE_structure}, the search for NE policies $\pi^*_i$ can be reduced to finding a randomized mixture of discrete actions in the finite action
set. A policy-iteration-based strategy-learning algorithm can be developed based on the Simultaneous Perturbation Stochastic Approximation (SPSA) algorithm 
\cite{spall2005introduction}. In \cite{5397924}, the rate adaptation problem in a TDMA-based CRN is modeled as an SG with a latency constraint. In 
\cite{5439936}, the problem of joint source-channel rate adaptation in order to transmit layered video in a multi-user wireless local-area network is also formulated as an SG with 
the latency constraint. In both works, by showing that the assumptions A1-A3 hold in their respective SG-based model, the SPSA algorithm is applied for policy-learning. With the
SPSA algorithm, no explicit state value learning is needed, and the local policies are updated with a gradient-based method with random policy perturbation. Given that the 
assumptions A1-A3 holds in the SG, the SPSA algorithm is proved to converge in distribution to the Kuhn Tucker (KT) pair of the original constrained MDP (Theorem 3 in 
\cite{5397924}). 

In \cite{4539483}, another distributed learning algorithm is constructed based on the framework of $L_{R-I}$ learning in the team SGs. A team SG can be considered as a variation 
of potential games when all the players in a SG share the same payoff function (i.e., fully-cooperative SG). With the proposed learning scheme, an LA is maintained 
for every state of the underlying Markov chain by each player in the SG. At any time instance, only one LA is activated by each player to learn its optimal action probabilities 
in the corresponding state. The introduction of LA reformulates the stochastic game between the $|\mathcal{N}|$ players into a repeated game between the 
$|\mathcal{N}|\!\times\!|\mathcal{S}|$ automata. Extending from the special case of team SGs, the convergence condition of the LA-based learning scheme for SGs is generalized 
by the following theorem:
\begin{theorem}[\!\!\cite{4539483}]
 \label{them_conv_tem_sg}
 For SG $G\!=\!{\langle}{\mathcal{N},\mathcal{S}}, {\mathcal{A}}, \{u_i\}, \Pr(\mathbf{s}'|\mathbf{s},\mathbf{a}) {\rangle}$, assume that the multi-agent Markov chain 
 corresponding to each joint policy, $\pmb\pi(\mathbf{s})$, is ergodic. If $\pmb\pi^*(\mathbf{s})$ is a pure NE policy in the view of $|\mathcal{N}|$ players in $G$, 
 $\pmb\pi^*(\mathbf{s})$ is also a pure equilibrium for the reformulated game between the $|\mathcal{N}|\!\times\!|\mathcal{S}|$ LA, and vice versa.
\end{theorem}

According to Theorems \ref{thm_la_ne} and \ref{them_conv_tem_sg}, whenever an NE point in pure strategies exists in an SG (which is always the case for team SGs), the LA-based 
learning algorithm proposed in \cite{4539483} is guaranteed to find the NE. However, it is worth noting that only maintaining an independent, repeated-game-based learning process
(e.g., LA or SFP) for each state by the players may not necessarily produce the NE strategies for a general-case SG. Take the SFP learning scheme for example. In a general-case 
SG, the action-dependent state transition renders the Logit function in (\ref{eq_stoch_br}) no longer the solution to the perturbed best response. As a result, a Lyapunov 
function can not be found in the same way as for repeated games and the convergence property of the corresponding best-response dynamic in Robbins-Monro form is undermined.
Therefore, special structure is required for the SGs if the repeated-game-based learning processes are to be adopted. In \cite{ZhuBook2013}, a sufficient condition is given for 
the adoption of the CODIPAS-RL learning schemes (more specifically, LA and SFP-based learning) in the general-case two-player nonzero-sum SGs:
\begin{itemize}
 \item [C1)] the state transitions are independent of the player actions.
\end{itemize}

It is easy to prove that given condition C1, by fixing the state variable and solving for all the state-dependent NE with the repeated game-based learning algorithms 
discussed in Section \ref{subsec_app_game}, we are able to obtain the state-independent NE of the two-player nonzero-sum SGs. The conclusion can be further extended to $N$-player
games. When the state transitions are also independent of the current state, each player only needs to maintain a single learning process (see the examples in \cite{ZhuBook2013, 
hanif2012convergence}). However, due to the constraint on the state transition conditions, only a few applications of the SFP/GP/LA-based algorithms for the SGs-based network 
control problems can be found in the literature \cite{5991373, hanif2012convergence}.

\section{Challenges and Open Issues in Model-Free Learning for Cognitive Radio Networks}
\label{sec_open_issue}
In this section, we expand our discussion to the challenges and open issues that are yet to be addressed in the area of learning for distributed control and/or wireless 
networking. In Section \ref{subsec_optimality}, different aspects of the learning mechanism goals are reviewed, and the potential conflict between these aspects is discussed. 
In Section \ref{subsec_teaching}, we propose a problem to cope with the outlier agents who do not (necessarily) follow a given learning rule in a learner 
set. In Section \ref{subsec_transfer_learning}, the possibility of transferring experience from one learning scenario/process to a difference learning scenario/process is discussed.
In Section \ref{subsec_coordination}, we discuss a problem on the coordination among simultaneous learning modules over different protocol layers for the same network 
entity.

\subsection{The Goal of Learning: Self-Play, Stability and Optimality}
\label{subsec_optimality}
Generally, the goal of a perfect self-organized learning mechanism for multi-agent decision making processes is to achieve self-play (autonomy), stability and optimality at the same 
time. However, it has been well-recognized that for multi-agent learning (more frequently in a stochastic scenario), improving system performance typically incurs 
more signaling and coordination, thus undermining the self-play structure. Especially, when learning is implemented under the framework of games, achieving any two goals of 
self-play, stability and network optimality is usually at the cost of undermining the third goal. In recent years, the relationship between the three parties of the goals in 
multi-agent learning has been discussed in many works, but mostly from a high-level theoretical perspective \cite{4445757, Shoham2007365, Bowling00ananalysis}.

In regard to the applications of learning in wireless networks, the situations that have been discovered to keep consistence between a distributed solution and an optimal 
solution are limited within a small scope. One important case of these situations is the network control problems that is modeled as a potential game \cite{4814554}. For 
potential games, the following properties \cite{han2012game} make it possible to achieve convergence to the optimal operation point through adopting the learning algorithms 
that we discussed in Section \ref{subsec_app_game}:
\begin{itemize}
 \item Every potential game has at least one pure strategy NE.
 \item Any global or local maxima of the potential function defined in the game constitutes a pure strategy NE.
\end{itemize}
Based on the above properties, it is only necessary to prove the uniqueness of the NE in a repeated game for learning processes to achieve optimal operation point with 
sequential best-response play \cite{lasaulce2011game} or no-regret learning. Apart from the works discussed in Section \ref{subsec_app_game}, the applications of 
distributed learning in potential games in order to achieve global optimization can usually be found in a set of congestion-game-like problems such as \cite{6086561, 5783528}. 
However, the potential game requires that local users are able to (implicitly) perceive the utilities of the entire network in order to establish the correspondence between the 
local utility function and the constructed potential function \cite{4814554}. Since this requirement is at the cost of trading off the conditions for self-play, it significantly 
limits the applications for the potential-game-based learning algorithms.

For other model-free distributed learning mechanisms in a multi-device wireless network, how to coordinate the goal of optimality and self-organization when adopting a learning
scheme generally remains an open question. As a result, most current studies focus on ensuring convergence to the stable operation point in self-play by allowing a 
limited level of control signal exchange. Although there are a few already-known conditions that ensure the convergence of a learning algorithm, most of which are 
applicable to repeated games (e.g., Theorem \ref{th_fp_convergence} and \ref{thm_SFP_convergence}), for most current studies, whether a stability condition can be found for a learning scheme also remains an open issue. In the literature, the 
approaches to find the convergence condition of the learning algorithms generally fall into two major categories. For learning processes that can be approximated with a linear
system described as a set of ODEs in continuous time, the typical way of obtaining the convergence condition is to construct a Lyapunov function for the ODE-based dynamic and 
then prove that the strategy/utility updating mechanisms produce an asymptotic pseudo-trajectory of the flow defined by the ODE through the stochastic-approximation-based 
analysis (see the example in \cite{293490, leslie2003convergent}). The analysis of learning 
using the ODE-based approach can be found in \cite{Li:2010:MQA:1863662.1928710, 4278411, 6117762, 5462017, 6654861}. For the situations which cannot be easily modeled as an linear, 
ODE-based system, 
the contraction-map-based analysis (see the example in \cite{Hu:2003:NQG:945365.964288}) can be considered as an alternative. Usually, the contraction map is considered 
appropriate for the analysis of SG-based learning when modeling the problem is of high complexity \cite{6265055, 6779690}.
Table \ref{table_converge_conditions} summarizes convergence conditions for the multi-agent learning algorithms discussed in Sections \ref{sec_app_crn}-\ref{sec_app_game}.

\begin{table*}[!t]
  \caption{A summary of theoretical convergence conditions for the MAS-based learning algorithms}
  \centering
 \begin{tabular}{  p{65pt}|  p{70pt} | p{140pt} |  p{65pt} | p{75pt} }
  \hline
  Problem Formulation Category & Learning Scheme & Convergence Condition & Stable Operation Point & Required Signaling\\
  \hline
  Loosely Coupled MAS & Distributed (independent) Q-learning & Generally not known & Sub-optimal &  None \\
  \hline
  \multirow{5}{65pt} {Repeated Games} & Standard FP & Not guaranteed except in (a) two-player games and (b) multi-player game with common payoff
  \cite{1406126} & $\epsilon$-NE &  Exchange of local-action information \\  
  \cline{2-5}
  & Stochastic FP & Not guaranteed except in (a) potential games, (b) supermodular games (c) two-player zero-sum games and (d) two-player symmetric games \cite{ECTA:ECTA376} 
  & $\epsilon$-NE & None \\
  \cline{2-5}
  & Gradient play & Conditional convergence for strict NEs in multi-player games \cite{1406126} & NE & Exchange of local-action information \\
  \cline{2-5}
  & $L_{R-I}$& Conditional convergence for strict NEs in multi-player games (see Theorem \ref{thm_la_ne})  & $\epsilon$-NE & None \\
  \cline{2-5}
  & No-external-regret learning (Hedge) & Potential games & NE & None \\
  \cline{2-5}
  & No-internal-regret learning & CE in multi-player games & Non-social-optimal CE \cite{ECTA:ECTA153} & None \\
  \hline
  \multirow{4}{65pt} {Stochastic Games} & Minimax Q-learning & Not known & NE & Knowing the structure of local payoff function\\
  \cline{2-5}
  & Nash Q-learning/R-learning & Each matrix game has a unique NE \cite{Hu:2003:NQG:945365.964288, li2007reinforcement} & NE & Exchange of local action/payoff information \\
  \cline{2-5}
  & Conjecture learning & Conditional convergence & Conjecture equilibrium & Knowing the reference point\\
  \cline{2-5} 
  & FP-based policy updating & Generally not known & NE & Exchange of local-action information \\
  \hline  
\end{tabular}
  \label{table_converge_conditions}
\end{table*}

In addition to the issues associated to finding the convergence condition for a learning scheme, another concern when applying model-free learning in wireless networks is the 
convergence rate of learning algorithms. Although analytical results for the convergence rate of learning algorithms are highly desired, most of the existing studies are only 
able to show empirical results for the learning convergence rate through numerical simulations (see the examples in \cite{6265055, 4581648}). The reason for this is partly due to 
the asymptotic convergence condition (if there is any), which requires for most of existing learning algorithms that the states and actions are visited infinitely to ensure the 
convergence. Given such a limitation, one known approach to analyze the convergence speed of a learning scheme is to view the learning process itself as a discrete time Markov 
chain. In this approach, the standard Markov chain analysis can be applied to obtain the expected time (number of iterations) to learn before reaching the chain's absorbing 
state (e.g., the equilibrium point of a repeated game). Such a technique 
can be found in the recent studies \cite{6620914, 6678685}. In \cite{6620914}, the Markov-chain-based analysis is used to measure the lower bound of the iterations needed for 
the Logit-function-based learning scheme to leave a sub-optimal NE in a potential game for gateway selection \cite{6620914}. In \cite{6678685}, the same method is employed to 
track the average iterations that a trial-and-error-based learning method needs for reaching the NE of a joint channel-power selection game for the first time. However, such an 
approach could be computationally intractable when the system/learning scheme is too complicated, and it is yet to be found applicable to the more complex learning 
algorithms such as those in the SGs.

\subsection{Heterogeneous Learning and Strategic Teaching in the Context of Games}
\label{subsec_teaching}
For the existing studies of strategy learning in wireless networks, one most important assumption is that each individual agent abides by the same learning rule (or just uses
variable parameters for the same learning scheme). Only with such an assumption, the convergence properties of the learning scheme can be mathematically tracked. However, in many 
practical scenarios, especially in the scenarios when malicious nodes exist in the network, such an assumption may not be applicable and the malicious nodes may intentionally 
deviate from the given learning rule. One possible scenario of such a case can be 
found in a selective-forwarding-based attack-defense game, in which a sophisticated attacker with the ability of selectively forwarding the received packets may wait and abide by
the normal packet forwarding rule until some critical packets are sent to it before dropping. To the best of our knowledge, currently there are few (if not any) works 
discussing this situation. 

To further demonstrate the situation in which a learner may benefit by deviating from a common learning rule, we introduce the concept of ``strategic teaching'', which is first
discussed in the studies of economic games \cite{Camerer2002137}. With strategic teaching, it is assumed that the game is composed of a number of adaptive players and 
sophisticated players. An adaptive player learns its strategy following the learning scheme that it is assigned to. By contrast, the sophisticated players are able to adopt a 
non-myopically optimal strategy and afford a certain short-term loss. Since the adaptive learners will finally learn the best response to a pre-committed strategy by the 
sophisticated player under the given learning scheme, the sophisticated players will be able to induce the adaptive players to expect some specific patterns of strategies from 
them in the future \cite{Camerer2002137}. Then, the sophisticated players will be able to take advantage of the behavior patterns that they ``teach'' the adaptive players. 
It has been found that a sufficiently patient strategic teacher can achieve as much utility as from first-play in a Stackelberg game\footnote{About the difference of a 
Stackelberg equilibrium and an NE, the readers are referred to \cite{han2012game} for more details.} \cite{Camerer2002137}. Thus, the sophisticated play may become a favorable 
way of strategy adoption for a noncooperative or a malicious node in the wireless network compared with the way of strictly following the same learning rule.

In \cite{Camerer2002137}, a heuristic, model-free learning method known as Experience-Weighted Attraction Learning (EWAL) \cite{ECTA:ECTA054} is applied to a repeated trust 
game (i.e., lender-borrower game) as the basis of both adaptive learning and sophisticated learning. In that game, $M$ borrowers try to borrow money from each of a series 
of $N$ lenders. A lender only makes a one-time binary decision on either \emph{Loan} or \emph{No Loan} in a single round out of a $N$-round game. A borrower makes a series of $N$
binary decisions on \emph{Repay} or \emph{Default} regarding each lender that it borrows money from after observing the lender's decision. The sequences of the $N$-round stage-games 
(also known as supergames) are repeated for many times with a random order of lenders to make decisions with each sequence. In one sequence, one borrower is picked as the common 
borrower in the game. All the lenders and some of the borrowers play as adaptive players and learn their strategies with EWAL. The rest of borrowers are assumed to be dishonest 
and adopt sophisticated play. It is assumed that the actions and instantaneous payoffs of one player are observable by the other players. For the adaptive players, EWAL uses the 
Logit-function-based rule as in (\ref{ea_la_softmax}) for strategy updating. Instead of directly using the instantaneous/accumulated payoff as the argument of operator 
$\exp(\cdot)$ in the Logit function, EWAL introduces the concept of experience accumulation through reinforcement and
employs two new measurements to build local experience: the observation-equivalents of the past experience and the attraction to a specific strategy \cite{ECTA:ECTA054}. The former is 
similar to the action-frequency estimation in FP and the latter is used as the argument of the Logit function.
In the game, the adaptive players apply EWAL twice to build 
their attraction first within a lending-borrowing sequence (i.e., supergame) and then across the consequent sequences. For the sophisticated borrowers, the learning process does 
not differentiate between attraction building within a supergame and across different supergames. A sophisticated borrower guesses how the lender learns according to the attraction 
value of the adaptive lender that it observes. Then, the policies of default and repay are sought by incorporating estimated policies of the lenders into the computation of its own 
sophisticated attraction function (see Section 4.1 of \cite{Camerer2002137} for the details). It has been demonstrated in \cite{Camerer2002137} that by adopting sophisticated 
play with the attraction updating mechanism based on lender policy estimation, the dishonest borrowers are able to outperform the adaptive borrowers which follow the same EWAL 
learning rule as the lenders. For simplicity, the mechanism of sophisticated play can be interpreted as playing additional tricks to the adaptive lenders by repaying frequently 
enough so if the dishonest borrowers do default, it won't lower the belief probability of the lenders about the trustworthiness of these borrowers below a 
critical level. Such an example provides an important insight into the possible strength of sophisticated play in repeated noncooperative games. However, few studies discuss such
an issue in the context of wireless networks. Also, it is generally not clear how strategic teaching with sophisticated play in other forms can be enforced or avoided in the 
current framework of learning and in what ways it will affect the equilibria that can be reached. 

\subsection{Experience Transferring between Heterogeneous Learners}
\label{subsec_transfer_learning}
As we note from Sections \ref{sec_app_crn}-\ref{sec_app_game}, one of the significant benefits of model-free learning is to allow the decision-making entities to learn the 
strategies from scratch without the a-priori knowledge of the wireless network. However, since model-free learning is based on trial-and-error, when the network 
environment has dramatically changed, the learners generally need to start the same learning process from the very beginning. One example of such scenarios can be found in 
interference mitigation problem for cellular networks, in which mobile stations may enter or leave the network frequently. For most of the existing model-free learning algorithms,
such changes in the network topology mean the changes in the MDP model of the network with new dimension of states/actions, if MDP-based learning is adopted, or the transition 
from an old network-control game to a new one since the set of players is different, if game-based learning is adopted. As a result, when it is required that the 
decision-making agents swiftly switch from an old scenario to a new one, the existing learning methods will face great challenges if they can only restart the learning process
in the new scenario.

In order to address such a challenge, a natural consideration is to utilize the acquired experience of strategy taking which is obtained from the old scenario. We note 
that such a process is fundamentally different from the experience sharing process discussed in Section \ref{subsec_app_docition}, since for the experience-sharing 
framework such as docitive networks, the parallel and homogeneous learning processes are assumed so the expert agent is able to share its better experience of the same stochastic
process with the newcomers. In the scenarios of dramatical environmental changes, the experience transferring paradigm, Transfer Learning (TL) 
\cite{Taylor:2009:TLR:1577069.1755839}, is considered more appropriate for the tasks of sharing experiences of strategy taking between heterogeneous learning processes. Compared 
with the experience transferring between homogeneous learners, the motivation of TL is to transfer knowledge (i.e., experience) from the well-established learning processes 
(known as the source tasks) to the newly established learning processes (known as the target tasks) in a different situation. It is worth noting that under the framework of 
MDP-based learning, TL allows the difference in state spaces, state variables/transition, reward functions and/or sets of actions \cite{Taylor:2009:TLR:1577069.1755839}.

\begin{figure}[!t]
  \centering
  \includegraphics[width=0.45\textwidth]{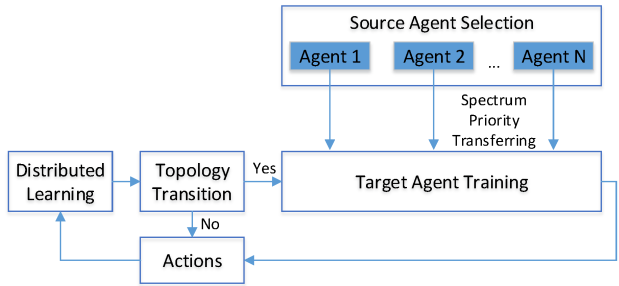}
  \caption{Architecture of the policy-transfer mechanism in the LTE-A based opportunistic network \cite{6692444}.}
  \label{fig_TL_policy}
\end{figure}

TL has been considered difficult to implement for learning in wireless networks. This is mainly due to the fact that it is difficult to find a proper mapping 
(either in value-function representation or directly in policy transferring \cite{Taylor:2009:TLR:1577069.1755839}) to transfer between learning tasks with different action-state 
representations. For the applications in wireless networks, one example of policy-transferring TL can be found in \cite{6692444}. In \cite{6692444}, a highly dynamic 
opportunistic network which is based on LTE-A is studied. The network topology is assumed to change with time, and the eNodeBs (eNBs) are supposed to be responsible for learning 
channel allocation under the conditions of mutual interference among the user equipments. The mechanism of policy transferring is adopted on the basis of two model-free learning 
algorithms: the linear reinforcement learning and the single-state Q-learning. The former employs a simple, linear updating function for state-value updating, while the latter 
applies Q-learning to update a state-less Q-table.
For TL, one shot of the changing network topology is considered as a learning phase, then the objective of TL is to apply the experience learned 
in previous phases (sources) to the similar phases (targets) in the future. The eNBs which attempt to assign channels to the user devices for interference 
coordination work as the learning agents and obtain the spectrum priority through sorting the Q-table obtained in the current phase in a descent order. A policy function is
designed to transfer the Q-table learned in a previous phase to the new phase through assigning weights to the source priority table to the target priority table in the new phase.
Such a procedure of associating the channel priority in the target task with the channel priority in the source target can be considered as initializing the learning process 
in the new phase with the transferring knowledge from the old phase. Thereby, the information from transfer learning and distributed learning is combined through weighting the 
values of channel priorities. The Q-table in the new phase is learned with the given reinforcement learning methods. The policy transferring process in \cite{6692444} is 
demonstrated in Figure \ref{fig_TL_policy}.

\begin{figure}[!t]
  \centering
  \includegraphics[width=0.47\textwidth]{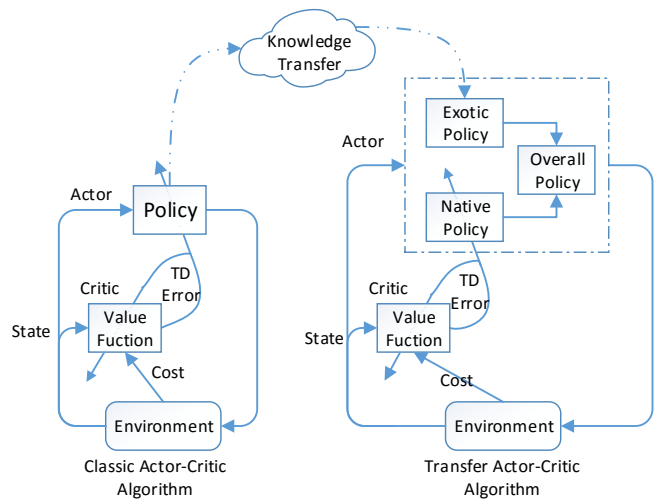}
  \caption{Architecture of the transfer-actor-critic algorithm \cite{6747280}.}
  \label{fig_TL}
\end{figure}

A different approach of applying TL to the wireless networking problems can be found in \cite{6747280}, where the authors apply TL to a series of actor-critic learning processes 
to coordinate BS switching/sleeping in a cellular network. In \cite{6747280}, the possibility of 
improper guidelines provided by transferred knowledge of the old task to the new task is considered. The actor-critic learning scheme is performed by a BS-operation controller, 
and is based on a multi-state MDP model for the traffic load of the serving BSs. Compared with \cite{6692444}, the difference of the TL mechanism in \cite{6747280} lies in the 
way of adopting the transferred policies. Instead of using the static transferred knowledge for the initialization of the new learning phase, the experience in the new learning phase
is divided into two sources: the ``native policies'' obtained through actor-critic learning and the ``exotic policies'' obtained as transferred policies from old tasks. The 
weight of the exotic policies contributing to the overall strategy selection decreases as the native learning process progresses. The learning-knowledge-transferring process is demonstrated 
in Figure \ref{fig_TL}. It is mathematically proved that regardless of the initial value of the overall policies and the transferred policies, the actor-critic-learning-based 
algorithm is guaranteed to converge. Also, numerical simulations show that TL does improve the learning speed when compared with the reinforcement learning methods without TL.

In the literature, most of the applications of TL in wireless networks are set in the scenarios which can be modeled as MDP-based MAS.  
With all the existing effort for establishing a general framework of applying TL to learning in wireless networks, the following questions are to be answered:
\begin{itemize}
 \item [1)] Whether and how can TL be applied between related games (e.g., symmetric games with the same structures of payoffs and actions, but with different sets of players) for 
 accelerating the convergence speed to the equilibrium?
 \item [2)] How can we measure the efficiency of knowledge transferring?
 \item [3)] Apart from policy transferring and value-function transferring, can TL also be applied to heterogeneous learning processes using different learning schemes?
\end{itemize}
In the literature, few studies in wireless networks are found discussing the aforementioned topics. However, discussions on cross-game learning or cross-mechanism 
learning have already begun in the area of economic games \cite{CooperTL2008} and automatic control \cite{taylor2009transfer}. Although the detailed discussion on these topics is 
beyond the scope of this survey, it is believed that addressing these issues will bring great improvement to the existing learning mechanisms in the CRNs.

\subsection{Coordination of Learning Modules: Integration vs. Decomposition}
\label{subsec_coordination}
In addition to the problems in heterogeneous learning processes, handling experience sharing or transferring knowledge among different network devices, 
the coordination of simultaneously learning modules may still be a challenging issue even within a single network device. As shown in our previous discussion, learning processes
targeting at various functionalities (which may or may not involve the interactions with other users) can happen in any layer of the protocol stack (see Figures
\ref{fig_layered_MDP_System}, \ref{fig_cr_mesh_network} and \ref{fig_layered_routing} for example). Although many existing works have succeeded in applying
the learning-based solution to their dedicated functionalities, a systematic discussion on coordinating these learning processes for different functionalities
generally remain untouched in the current research progress. In the seminal work \cite{7143333}, it is pointed out that different functionalities across the protocol layers may 
exhibit a range of conflicts and/or dependence when working concurrently in the same network. Thereby, it becomes a natural idea to consider the solutions to the learning 
module coordination by first identifying the conflicts or dependence in practical scenarios.

Based on the work in \cite{7143333,6666642}, we consider the following major conflicts and/or dependence among different network functionalities: 
\begin{itemize}
 \item [1)] Logical dependence: this kind of dependence may arise when there is a logical dependence between the objectives of different network functions.
 \item [2)] Parameter conflict/dependence: this kind of conflicts or dependence is triggered either when different networking functions try to modify the same configuration parameters
 or when the parameters of one function depends on some other network parameters.
 \item [3)] Measurement conflict: measurement conflicts exist if a learning module depends on the state of the other learning modules.
\end{itemize}

Logical dependence happens when different learning modules exhibit a hierarchical dependence on the output of each other. In this sense, the relationship between different learning 
modules in a CR device shares a lot of similarity with the relationship between the subtasks of a hierarchical reinforcement learning mechanism \cite{Barto2003, Ghavamzadeh2006}.
The major difference is that in (MDP-based) hierarchical reinforcement learning, a single learning process is decomposed into a number of subtasks with their own sub-states, 
actions, transition functions and rewards in a top-down manner with the help of recursive value function decomposition\footnote{A general principle for a hierarchical value 
function decomposition is that the reward function of a parent task is the state-value function of the child task \cite{Ghavamzadeh2006}.} \cite{Ghavamzadeh2006}. Since
hierarchical learning requires to finish each child learning task before starting its parent task, it extends the MDP-based system model into a semi-MDP-based system model, 
in which the amount of time for the transition from one action to the next is a random variable due to the existence of the subtask sequences. By adopting the general idea of 
hierarchical learning, learning coordination with logical dependence can be considered as a reverse process of hierarchical learning by integrating the existing learning modules
according to their dependence and forming a macro learning task. Practically, such an operation of module concatenation may be extended to the non-MDP-based learning mechanisms.
For example, in \cite{JCM0410790802, 6554824} a hybrid structure of both MDP-based Q-learning and repeated game-based no-regret learning is formed to approximate the equilibrium 
strategy of an SG. In those two cases, the expected utility based on the learned equilibrium of the repeated game can be considered as the instantaneous utility of the parent-level
Q-learning process. However, the major difficulty in applying a hierarchical learning-based coordination mechanism lies in the uncertainty of convergence, 
as we have highlighted in \ref{subsec_game_SG}. Unlike the well-established examples of hierarchical learning in the domain of robot control \cite{Barto2003}, For the applications
in CRNs there usually exists no terminal state for a subtask to determine when to stop its execution. As a result, when to start and terminate a task in the framework of 
hierarchical learning are usually determined empirically, and the convergence conditions of such a learning process still remains an open issue. 

Unlike logical dependence, parameter conflict/dependence and measurement conflict are caused by the conflicts of the actions and states in different learning modules, respectively.
For example, parameter conflict may happen between the inter-cell interference control and the coverage/capacity optimization modules of a cellular network.
With respect to downlink transmit power control, the interference control module may want to decrease the transmit power in order to reduce the inter-cell interference,
while the coverage/capacity optimization modules may want to increase the transmit power to improve the local link quality at the same time. For those two kind of conflicts, 
one traditional solution is to build a decision tree to activate different decision modules according to the pre-determined conditions, which is also called trigger-condition-action 
points \cite{7143333}. However, the trigger-condition-action based solution is a typical model-based method, and thus cannot be directly incorporated into the coordination 
process of learning modules.

Although no prototypical solution has been proposed to resolve conflicts 2) and 3), it is still possible to address these conflicts by imitating the existing model-based methods
when some certain property can be found in the learning modules. Consider a general case where a number of learning modules share a subset of network states, and try to learn 
the strategy on the same action parameters to achieve different goals. To resolve the conflicts, we can adopt the idea of layering by decomposition in 
\cite{4118456} to coordinate the learning modules. One typical way of doing so is to pick the objective of one network functionality as the major goal and treat the goals of all 
the other functionalities as constraints. It is worth noting that such an operation can be also considered as a way of integration. However, the ultimate goal of it 
is to create a structure of optimization which suits the further operation of decomposing it into interrelated but layered learning processes. 
A revisit to the work on layered Q-learning for video compression \cite{5299114} helps to exemplify such an idea in details.
In \cite{5299114}, a multimedia processing system considers three different concurrent objective functions, which are the video distortion at the codec level,
the queueing delay for video frame processing in the pre-encoding buffer, and the energy cost in the OS/hardware layer. The distortion and queueing delay can be treated as two 
objective functions in the application layer of the system sharing the same system state, while the configuration that defines the energy cost (the operating frequency in this 
case) also determines the distortion of the compressed video. In \cite{5299114}, minimizing the queuing delay is considered as the main objective, and the rest two objective
functions are treated as constraints. Conflicts between different functionalities can be easily found in this case, 
since increasing the operating frequency will lead to a better video quality but result in more energy consumption. By creating such a constrained optimization problem, 
a layered Q-learning mechanism is designed in a way that is similar to the procedure of dual decomposition. As briefly discussed in Section \ref{sec_app_crn}, 
a two-layer learning framework is created in the following way. In the application layer, the Q-learning module receives the signaling from the OS/hardware layer about its action (frequency 
selection) information, and learns the local state value. In the OS/hardware layer, the local learning process receives the estimated Q-value of the 
application layer as part of its instantaneous utility, and then learns its own state value. Unlike the hierarchical learning based integration method, layered learning based 
on decomposition does not require that one learning process to be finished first before another learning process starts. 

Like integration-based learning, the mathematical proof of convergence for decomposition-based learning is still rarely discussed in the existing literature. In the meanwhile, although considered more 
autonomous than the model-based coordination methods such as trigger-condition-action, decomposition-based learning needs a pre-determined constrained-optimization structure for 
layering of the learning processes. Such a requirement may limit the ability 
of decomposition-based learning in quickly responding to the requests of a certain network functionality that cannot be reached in the given constrained-optimization structure.
From this point of view, finding a satisfying tradeoff between different functionalities still remains an open question for decomposition-based learning coordination.

\section{Conclusion}
\label{conclusion}
Owing to the distributive nature of cognitive wireless networks, model-free learning is especially appropriate for the wireless nodes to adaptively choose their
transmission strategies in a self-organized manner without much requirement for knowing the network conditions. In this paper, we have provided a comprehensive survey on 
the applications of the state-of-the-art learning mechanisms in a wide range of scenarios of network modeling.
With a broad-scope analysis and comparisons of the literature, we have focused on learning algorithms that can be categorized with a set of 
prototypical schemes. Briefly, these prototypical schemes includes MDP-based learning and experience sharing, conjecture-based learning, FP/GP-based learning, LA-based learning 
and no-regret learning. We have classified the various scenarios for the applications of learning into three major categories, namely, the SAS-based network control, the 
loosely-coupled MAS-based network 
control and the game-based network control. We have mainly focused on the following characteristics of the selected learning algorithms: (i) the ability of the learning schemes 
to achieve optimality/equilibria without knowing an a-priori model for the environment, (ii) the ability of the learning schemes to achieve optimality/equilibria 
without obtaining the information that is not locally available and (iii) the ability of the learning schemes to quickly adapt by exchanging experience.
In addition to detailed reviews of the existing applications of learning in wireless networks, we have also discussed a variety of open issues that need to be addressed in future 
research. We hope this survey will serve as an important guideline for future research directions 
to further understand model-free learning mechanisms and expand their applications in cognitive wireless networks.

\bibliographystyle{IEEEtran}
\bibliography{Reference}
\end{document}